\documentclass[fleqn,usenatbib]{mnras}

\usepackage{newtxtext,newtxmath}

\usepackage[T1]{fontenc}
\usepackage{ae,aecompl}

\usepackage{graphicx}	
\usepackage[export]{adjustbox}
\usepackage{amsmath}	

\usepackage{xcolor}
\usepackage{comment}

\newcommand{\srg}{\textit{SRG}}

\defcitealias{2021A&A...651A..41C}{Paper~I}

\usepackage{hyperref}

\title[WHIM filaments]{Prospects of detecting soft X-ray emission from typical WHIM filaments around massive clusters and the Coma cluster soft excess}

\author[Churazov et al.]{E.~Churazov,$^{1,2}$ I.I.~Khabibullin,$^{3,2,1}$, K.~Dolag,$^{3,2}$, N.~Lyskova$^{1,4}$, R.A.~Sunyaev,$^{1,2}$ 
\\
\\
$^1$~Space Research Institute (IKI), Profsoyuznaya 84/32, Moscow 117997, Russia \\
$^2$~Max Planck Institute for Astrophysics, Karl-Schwarzschild-Str. 1, D-85741 Garching, Germany  \\
$^3$~Universitäts-Sternwarte, Fakultät für Physik, Ludwig-Maximilians-Universität München, Scheinerstr.1, 81679 München, Germany\\
$^4$~Astro Space Center, P. N. Lebedev Physical Institute of RAS, Profsojuznaya 84/32, Moscow 117997, Russia
}

\begin{document}
\label{firstpage}
\pagerange{\pageref{firstpage}--\pageref{lastpage}}
\maketitle

\begin{abstract}
 While hot ICM in galaxy clusters makes these objects powerful X-ray sources, the cluster's outskirts and overdense gaseous filaments might give rise to much fainter sub-keV emission. Cosmological simulations show a prominent "focusing" effect of rich clusters on the space density of the Warm-Hot Intergalactic Medium (WHIM) filaments up to a distance of $\sim 10\,{\rm Mpc}$ ($\sim$ turnaround radius, $r_{ta}$) and beyond. Here, we use \texttt{Magneticum} simulations to characterize their properties in terms of integrated emission measure for a given temperature and overdensity cut and the level of contamination by the more dense gas. We suggest that the annuli $(\sim 0.5-1)\times \,r_{ta}$ around massive clusters might be the most promising sites for the search of the gas with overdensity $\lesssim 50$. 
We model spectral signatures of the WHIM in the X-ray band and identify two distinct regimes for the gas at temperatures  below and above $\sim 10^6\,{\rm K}$. Using this model, we estimate the sensitivity of X-ray telescopes to the WHIM spectral signatures. We found that the WHIM structures are within reach of future high spectral resolution missions, provided that the low-density gas is not extremely metal-poor. We then consider the Coma cluster observed by SRG/eROSITA during the CalPV phase as an example of a nearby massive object. 
We found that beyond the central $r\sim 40'$ ($\sim 1100\,{\rm kpc}$) circle, where calibration uncertainties preclude clean separation of the extremely bright cluster emission from a possible softer component, the conservative upper limits are about an order of magnitude larger than the levels expected from simulations.
\end{abstract}




\section{Introduction}
Warm-hot intergalactic medium with temperature from $10^{5}$~K to a few times $10^{6}$~K and overdensity (relative to the mean baryonic density of the Universe) from a few to a few tens is believed to contain almost a half of all baryons in the modern Universe \citep{1999ApJ...514....1C,2001ApJ...552..473D,2010MNRAS.407..544B,2019MNRAS.486.3766M,2021A&A...646A.156T}. On the low temperature and density end, it is located in sheets of matter collapsed in only one direction, while on the high temperature and density end, it corresponds to the hot and tenuous outskirts of virialized objects, i.e. galaxies, groups, and clusters.

The relatively high temperature and low density of this gas make its direct observation very hard. Indeed, on the one hand, thermally-excited emission is very faint and falls into UV and soft X-ray bands, suffering from high line-of-sight absorption and foreground emission of our own Galaxy. On the other hand, the degree of hydrogen and helium ionization in this medium is extremely high (due to both collisional and photoionization in the radiation field of cosmic X-ray and UV background) prohibiting any detectable absorption signal akin to Lyman forest absorption readily observed in the spectra of quasars at redshifts above two. 

This picture changes dramatically if WHIM is allowed to be significantly enriched with metals, which are capable of staying only partially ionized under WHIM-relevant physical conditions. Both emission \citep[][]{2008SSRv..134..405P} and absorption \citep[][]{2008SSRv..134...25R} in lines of the metals have long been considered as the most promising venue for the direct observation of this medium, in particular, thanks to the exquisite redshift-sensitivity of the signal enabling the possibility of cross-correlation of the X-ray signal with galaxies overdensities inferred from the optical surveys \citep[][for recent reviews]{2019BAAS...51c.450B,2022arXiv220315666N}. 

An alternative approach is to focus on the overdense regions which are naturally present in the vicinity of massive galaxy clusters. Namely, the expected overdensity of all structures near
a turn-around radius of the cluster is $\sim 2-3$ (see, e.g. Fig. 2 in \citealt{2014ApJ...789....1D} or Fig.5 in \citealt{2021MNRAS.504.4649O}), and one can expect WHIM gas to be concentrated there as well, in particular in the form of the matter filaments with an overdensity of few tens. As a result, the excess X-ray emission from such regions would naturally contain the contribution of WHIM, and this can be used even without the selection of individual filaments traced by the overdensity of galaxies at the redshift of the central cluster. 

Given that the surface brightness of the WHIM emission does not depend on the distance to the object (for small redshifts), it is primarily the sky area covered by the object that determines the total integrated X-ray signal. In this regard, deep observations of nearby massive clusters are well suited for the purposes of WHIM detection with X-ray telescopes featuring large Field of View (FoV).

For future X-ray missions with microcalorimetric spectral resolution in the soft X-ray band ($\lesssim$ a few eV near energies of the most prominent oxygen lines at 0.5-0.7 keV), the situation is different thanks to the possibility of selecting redshift windows which are less affected by the Galactic foreground emission, effectively reducing the astrophysical background level from the bright foreground emission lines by two orders of magnitude. In this case, deep observations of $\sim$5-10 Mpc regions around massive clusters at redshift $\sim0.1$ offer best opportunities.  

The Coma cluster (Abell 1656) is particularly interesting in the context of the WHIM detection in X-rays since evidence for such emission has been reported based on \textit{EUVE} and \textit{ROSAT} data \cite[e.g.][]{1996Sci...274.1335L,1999ApJ...510L..25L,2022MNRAS.514..416B}. 
This object is one of the few X-ray brightest nearby ($z=0.0231$) massive ($M_{500}\sim 6\times 10^{14}\,M_\odot$; \citealt{2013A&A...554A.140P}) clusters, which subtends several square degrees on the sky. Here we revisit the question of the so-called "soft-excess", i.e. a distinct spectral component at sub-keV energies, which might be associated with $\sim 10^6\,{\rm K}$ gas in the vicinity of Coma.

The paper is organized as follows. In Section~\ref{sec:mass} we introduce quantities that we use to characterize the amount of gas (hydrogen and helium) associated with the WHIM. In Section~\ref{sec:exp} we use the data of numerical simulations in order to extract these quantities from the data of numerical simulations. We then analyze the excess of the WHIM gas within a turn-around radius near massive clusters (Section \ref{sec:where}). The level of the WHIM signal contamination by denser gas present in the simulation is estimated in Section~\ref{sec:cont}.
The expected spectral signatures of the WHIM gas are calculated in Section~\ref{sec:smodel}. The sensitivity of X-ray telescopes to this signal is estimated in the same section. In Section~\ref{sec:coma} the observations of the Coma cluster with SRG/eROSITA are discussed and compared with the predictions of numerical simulations (see Figs.~\ref{fig:ring} and \ref{fig:eml_erosita}).  In Section ~\ref{sec:discussion} we discuss the implications of our results in the context of WHIM search strategies and mention a possible boost of the WHIM signatures by the "side illuminations". Finally, Section~\ref{sec:conclusions} summarizes our findings. 

Throughout the paper, we assume a flat Lambda Cold Dark Matter ($\Lambda$CDM) cosmology with $\Omega_m = 0.3$, $\Omega_{\Lambda}$=0.7, $H_0 = 70$ km/s/Mpc. For $\Omega_b$ we use Planck-2015 value 0.0486 \citep{2016A&A...594A..13P}.


\section{Measures of line-of-sight emission and Thomson optical depth}
\label{sec:mass}
As we discuss later in the text, the detectable X-ray signal associated with the WHIM largely comes from spectral signatures of metals (and their amplitude scales  essentially linearly with the total gas metallicity), except for a pure Thomson scattering and the part of the bremsstrahlung emission due to hydrogen and helium. The latter two components do not have sharp spectral features and it is not easy to differentiate them from other backgrounds and foregrounds. On the other hand, parameters like gas density and temperature of the WHIM appear to be more robustly predicted by numerical simulations than metal abundances. It therefore makes sense to use densities and temperatures from simulations and treat the metal abundance as an additional parameter. In this section, we discuss only the density-related quantities, which determine the amplitude of the X-ray signals, either in absorption or in emission.  

A convenient and commonly used way of characterizing the density-related dependence of the intensity of an optically thin plasma emission (when collisions are the dominant excitation mechanism) is via the line-of-sight emission measure 
\begin{eqnarray}
{\rm EM_{\it l,cgs}}=\int n_e n_H dl=\int \left ( \frac{\rho_c\Omega_b}{\mu m_p}\right )^2 \left (\frac{n_e}{n_t}\right )  \left (\frac{n_H}{n_t}\right ) (1+\delta)^2 dl, 
\end{eqnarray}
which we have reformulated in terms of the critical\footnote{Since we are interested in the baryonic density, a more logical approach would be to specify the physical density of the gas via the parameter $\Omega_b\,h^2$. We, however, write the mean baryonic density as $\rho_c\Omega_b$, which appears more "transparent" when  used in the above equation.} density of the Universe $\rho_c\approx 9.21\times 10^{-30}\,{\rm g\,cm^{-3}}$ and the baryon fraction $\Omega_b=0.0486$ at $z=0$. $\mu\approx 0.608$ is the mean atomic weight for fully ionized plasma with the abundance of heavy elements relative to the Solar photosphere $Z/Z_\odot=0.2$, $m_p$ is the proton mass, $\left (\frac{n_e}{n_t}\right )\approx 0.52$ and $\left (\frac{n_H}{n_t}\right )\approx 0.44$ are the fractions of electrons and protons relative to the total number of particles (including helium and heavier elements), $(1+\delta)$ is the gas density relative to the mean baryon density and $L$ is the length of the region occupied by the gas along the line of sight.

For $z=0$, we can define a factor 
\begin{eqnarray}
\eta_{\,0}=\left ( \frac{\rho_c\Omega_b}{\mu m_p}\right )^2 \left (\frac{n_e}{n_t}\right )  \left (\frac{n_H}{n_t}\right )\times {\rm Mpc}=1.36\times 10^{11}\, {\rm cm^{-5}},
\end{eqnarray}
and $\langle n_e\rangle=2.3 \times 10^{-7}\,{\rm cm^{-3}}$ , $\langle n_H\rangle=1.9\times10^{-7}\,{\rm cm^{-3}}$.
As a function of $z$, $\eta$ obviously changes as $\eta=\eta_{\,0}(1+z)^6$. 

When dealing with objects, which size is small compared to cosmological scales, it is convenient to move $\eta$ outside of the integral and characterize the line-of-sight emission measure by the quantity
\begin{eqnarray}
{\rm EM_{\it l}}=\int (1+\delta)^2 dl, 
\end{eqnarray}
where $l$ is in units of Mpc. We use this approach below and often quote ${\rm EM_{\it l}}$ in units of Mpc. With this definition, for a 10~Mpc size object with overdensity $\delta=9$, ${\rm EM_{\it l}}=(9+1)^2\times 10=10^3 \,{\rm Mpc}$.

For comparison with the emission measure normalization convention in a widely used \texttt{XSPEC} package \citep{1996ASPC..101...17A}, it is convenient to provide conversion coefficients from  ${\rm EM_{\it l}}$ to ${\rm K_{XSPEC}}=\frac{10^{-14}}{4\pi\left[D_A (1+z)\right ]^2}\int n_e n_H dV$. Since were are dealing with the diffuse emission, we give the conversion factors per unit of solid angle:
for ${\rm EM_{\it l}=1\, Mpc}$,  ${\rm K_{XSPEC}}=1.08\times 10^{-4}(1+z),~3.30\times 10^{-8}(1+z)$, and $9.16\times 10^{-12}(1+z)$ per steradian, square degree, and square arcminute, respectively. 

Another, similarly useful, quantity is the Thomson optical depth of a gas lump, which for a uniform density at $z\sim 0$ is
\begin{eqnarray}
\tau_T=n_e l \sigma_T=\langle n_e \rangle (1+\delta) l\sigma_T= \nonumber \\
4.7\times10^{-7} (1+\delta) \frac{l}{\rm Mpc}= 
4.7\times10^{-7} \frac{{\rm EM_{\it l}}}{(1+\delta)}.
\label{eq:tau2eml}
\end{eqnarray}
Below, we use both quantities to characterize the expected signal from the low-density gas. It turns out that for the WHIM gas in the relevant range of overdensities,  ${\rm EM_{\it l}}$ is a more suitable measure for temperatures $T\gtrsim 10^6\,{\rm K}$, while at low temperature $\tau_T$ is more appropriate, because of photoionization effect of the CXB photons.

\section{Expected levels of the WHIM emission measure from cosmological simulations}

\label{sec:exp}


A number of cosmological hydro simulations have been used to predict density, temperature, and spatial distribution of the WHIM, either using Eulerian \citep[][]{1999ApJ...514....1C,2011ApJ...731....6S,2012ApJ...759...23S,2018MNRAS.476.4629P,2019A&A...627A...5V,2021ApJ...920....2Z}, or Lagrangian \citep[][]{2001ApJ...552..473D,2003MNRAS.339..289S,2003PASJ...55..879Y,2010MNRAS.407..544B,2016MNRAS.459..310R,2019MNRAS.482.4972K,2019MNRAS.486.3766M,2021A&A...649A.117G,2021A&A...646A.156T} approaches.

To estimate the X-ray signal that can be associated with the WHIM gas, we used two cubical cut-outs from the \texttt{Magneticum}\footnote{\href{www.magneticum.org}{www.magneticum.org}} suite of cosmological simulations \citep[][]{2016MNRAS.463.1797D}. 

Specifically, we extracted two boxes of 57.5 Mpc on each side from the full simulation \textit{Box2b/hr} at redshift $z=0.252$ (the final redshift for this simulation). The \textit{Box2b/hr} simulation has the total comoving volume of (640 $h^{-1}$ cMpc)$^3$  with dark matter m$_\mathrm{DM}=6.9\times 10^8 \ \rm{M_\odot}$ and gas m$_\mathrm{gas}=1.4\times 10^8 \ \rm{M_\odot}$ mass resolutions across it, simulated with 2880$^3$ particles by an improved version \citep{2016MNRAS.455.2110B} of the N-body code \textsc{Gadget~3}, which is an updated version of the code \textsc{Gadget~2} \citep{2005MNRAS.364.1105S} involving a Lagrangian method for solving smoothed particle hydrodynamics (SPH).

In addition to solving for the cosmological hydrodynamical evolution of the gas, the simulation follows the evolution of various metal species and their relative composition in the interstellar, circumgalactic, intracluster, and intergalactic media via computing continuous enrichment by supernovae of type Ia and type II, and asymptotic giant branch star winds in the fully cosmological context. The enrichment is computed self-consistently with the underlying evolution of the stellar populations \citep[for details, see][]{2007MNRAS.382.1050T} and matter outflows from the galaxies initiated by star formation and AGN feedback \citep{2014MNRAS.442.2304H}. The simulations were capable of reproducing detailed metal distributions within galaxies and galaxy clusters at redshift $z\sim 0$ \citep{2017Galax...5...35D}, while achieving  significant enrichment of the intergalactic medium already at $z\sim 2-3$ \citep{2017MNRAS.468..531B}.


The first cut-out box was centered on the second most massive cluster in the simulated volume, while the second one was taken at a random position to serve as a reference box.
Namely, the cluster has the total mass of $M_{500,c}\approx7.5\times10^{14}h^{-1}M_\odot\approx10^{15}M_\odot$ and $R_{500,c}=1.3 h^{-1}$ Mpc$\approx1.8$Mpc. 

For a massive galaxy cluster at $z\sim 0$, the turn-around radius (the position of the non-expanding shell which is furthest away from the cluster center) is equal to $\sim 3R_{200m}\sim 10$ Mpc, where $R_{200m}$ is the radius for which mean enclosed matter density exceeds the mean matter density $\rho_m$ of the Universe by a factor of 200. This relation can be obtained from the self-similar spherical collapse model \citep{1984ApJ...281....1F,1985ApJS...58...39B} specialized for $\Lambda$CDM \citep[e.g.][]{2016MNRAS.459.3711S} or derived from numerical simulations \citep[e.g.][]{2020A&A...639A.122K,2020JCAP...01..048H}. The mean enclosed density within the turn-around radius is $\rho(<R_{ta})/{\rho_{m}} \sim 10$ \citep[e.g.][]{2016MNRAS.459.3711S, 2016arXiv160103740T}. At the turn-around radius, numerical simulations predict local density  $\displaystyle \rho(R_{ta})/\rho_m \simeq 2-3$ (see, e.g. Fig. 2 in \citealt{2014ApJ...789....1D} or Fig.5 in \citealt{2021MNRAS.504.4649O}). Clearly, our cut-outs from the simulations are few times bigger than the turn-around radius of the cluster, and we expect convergence to the "mean" properties in the outer parts of the cluster box.

The particle-based data of the SPH simulation were remapped onto a uniform cartesian grid with the voxel size of (100~kpc)$^3$, where we used a so-called gather approach to compute the SPH smoothed quantities at the center of the grid cells, resulting in 575$^3$ boxes with physical parameters of the gas density, temperature and metallicity within them. Although this size of the grid cell is significantly bigger than the smoothing length size of the particles within strongly overdense structures, e.g. in the inner parts of galaxies, groups, and clusters, the primary focus of the current study is on the diffuse intergalactic medium, where such gridding is fully adequate. The consistency of the grid-based and particle-based predictions for emission and absorption by the intergalactic medium has been confirmed by both statistical and point-by-point direct comparisons between them.


\begin{figure*}
\centering
\includegraphics[angle=0,trim=1cm 10cm 2cm 8.8cm,width=1.95\columnwidth]{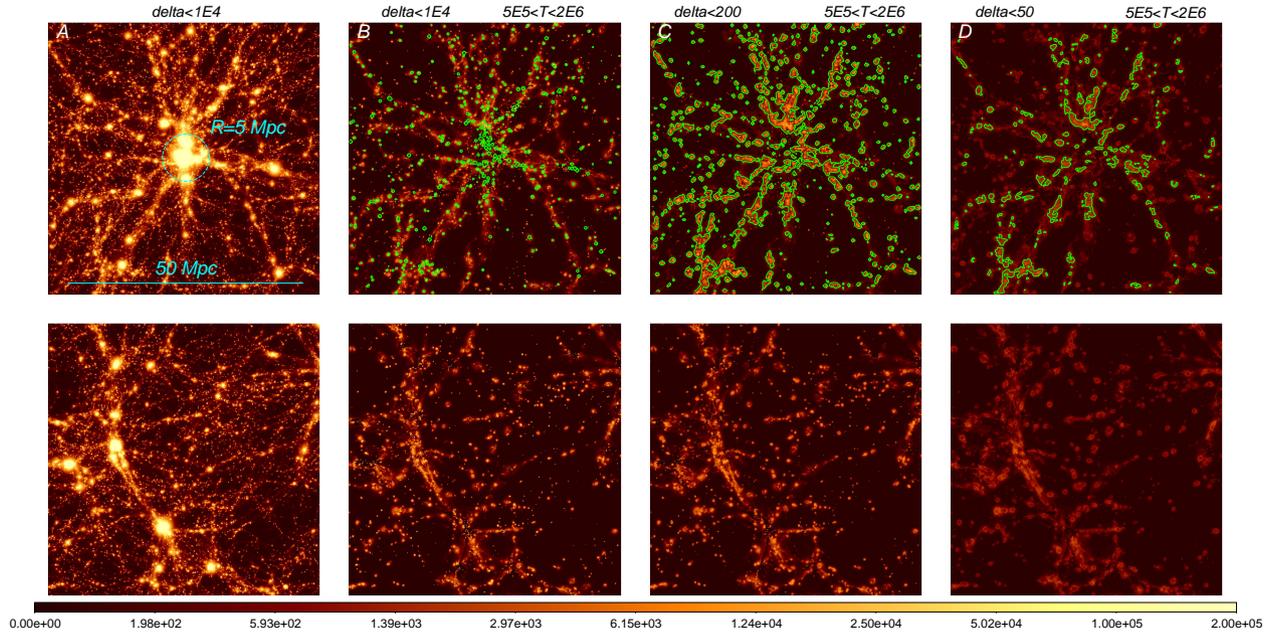}
\caption{Maps of the line-of-sight emission measure extracted from Magneticum simulations. The two rows show $\int (1+\delta)^2 dl$ for two boxes, one centered on a massive cluster (top) and another one, representing a more "typical" (randomly selected) volume. Both boxes are 57.5~Mpc on the side. From left to right, progressively stronger constraints on the temperature and overdensity were applied to gas cells when calculating the emission measure. Namely, for column $A$, the only constraint was overdensity $\delta < 10^4$. For the remaining columns, the temperature is in the range $0.5\times10^6{\rm K}<T<2\times10^6 {\rm K}$, while the overdensity is  $\delta < 10^4$, $\delta < 200$, and $\delta < 50 $ for $B,C$, and $D$, respectively.  With these definitions, the left-most maps show the total emission measure (any temperature or overdensity), while the right-most maps are associated with the low-density gas with the temperature $\sim 10^6\,{\rm K}$ (within a factor of 2), which encapsulates a fraction of the WHIM emission. 
As expected, these maps show clearly that the fraction of area subtended by filaments increases towards the cluster. In panel B (top row), the green contours show the regions for which $\int (1+\delta)^2 dl$ is larger than $10^4$~Mpc. Clearly, these regions correspond to individual halos. In panels C and D (top row) the contours correspond to ${\rm EM_{\it l}}=10^3$~Mpc. At this level, the contours trace a more diffuse gas in the most prominent filaments. While in the azimuthally-averaged data, the mean level will drop below $10^3$, individual filaments are clearly above this threshold (up to $(2-3)\times10^3$) even when a cut $\delta < 50 $  is used. We reiterate here that $\delta$ is the local (rather than mean enclosed) overdensity of baryons. }  
\label{fig:eml_maps}
\end{figure*}

In Fig.~\ref{fig:eml_maps}, we show maps of ${\rm EM_{\it l}}$ projected along one of the axes of the box. The upper and lower rows show such maps for the "cluster" and "reference" boxes, respectively. 
The maps shown also differ by the selection of cells contributing to  ${\rm EM_{\it l}}$. In particular, the left column (A) shows the total emission measure including all cells with overdensity $\delta < 10^4$. On the contrary,  the right column (D) corresponds to the most stringent selection criteria: only the low overdensity ($\delta<50$) cells with the gas temperature in the range $0.5\times10^6\,{\rm K}<T<2\times10^6 \,{\rm K}$. 
The middle two panels illustrate the importance of selecting only the gas with the temperature $\sim 10^6\,{\rm K}$ and excluding dense regions, which likely correspond to cores of virialized halos.
In particular, columns B and C use the same temperature range, but different cuts in overdensity: $\delta<10^4$ and $\delta<200$ for B and C, respectively.  
We note in passing, that  $(1+\delta)$ refers to the local density, rather than to the enclosed density, which for virialized halos is always larger. Therefore, for halos, the above overdensity cuts effectively remove cores that correspond to a factor of 1.5-2 larger enclosed overdensity.

\begin{figure*}
\centering
\includegraphics[angle=0,trim=1cm 10cm 2cm 9cm,width=1.3\columnwidth]{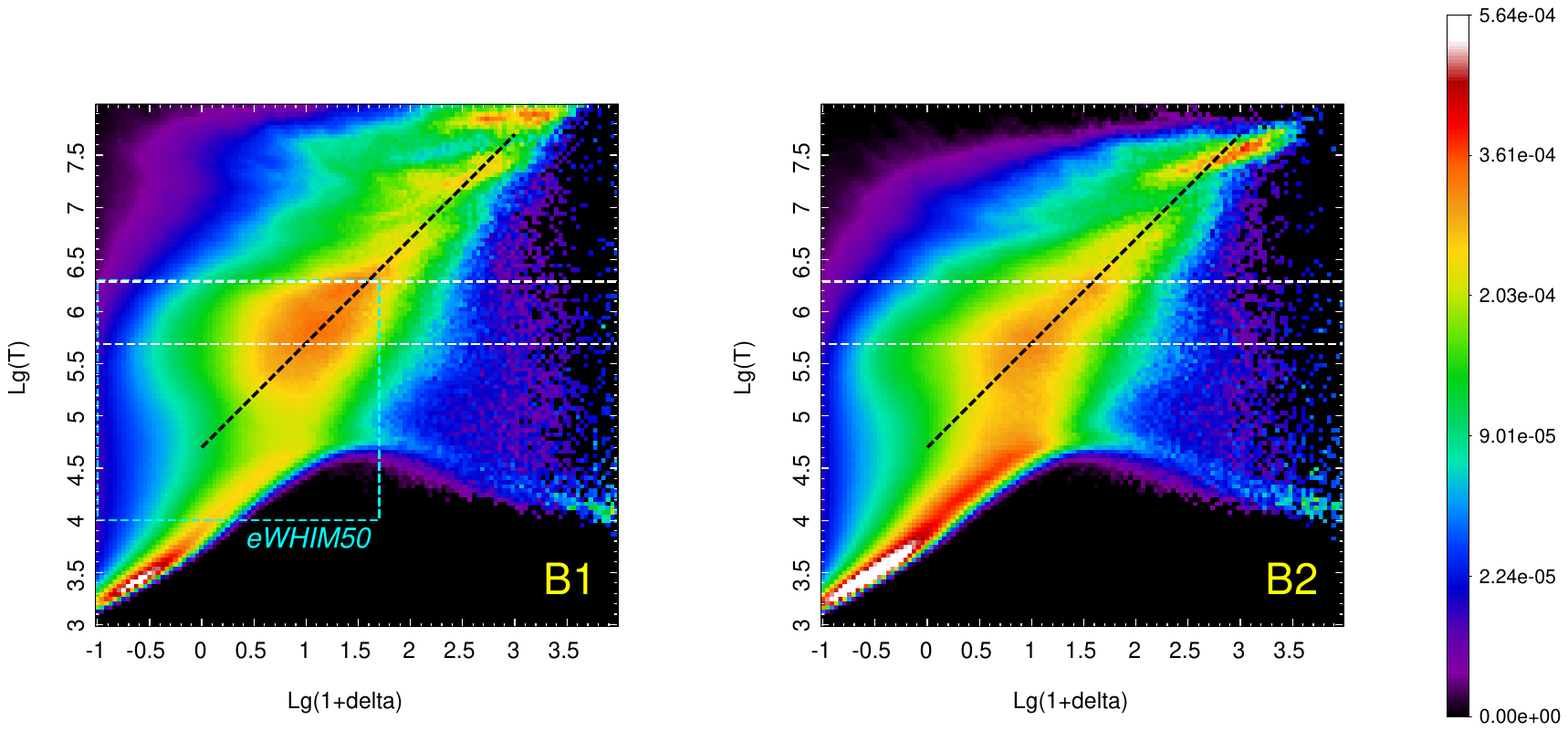}
\includegraphics[angle=0,trim=1cm 6cm 1.5cm 5.5cm,width=0.73\columnwidth]{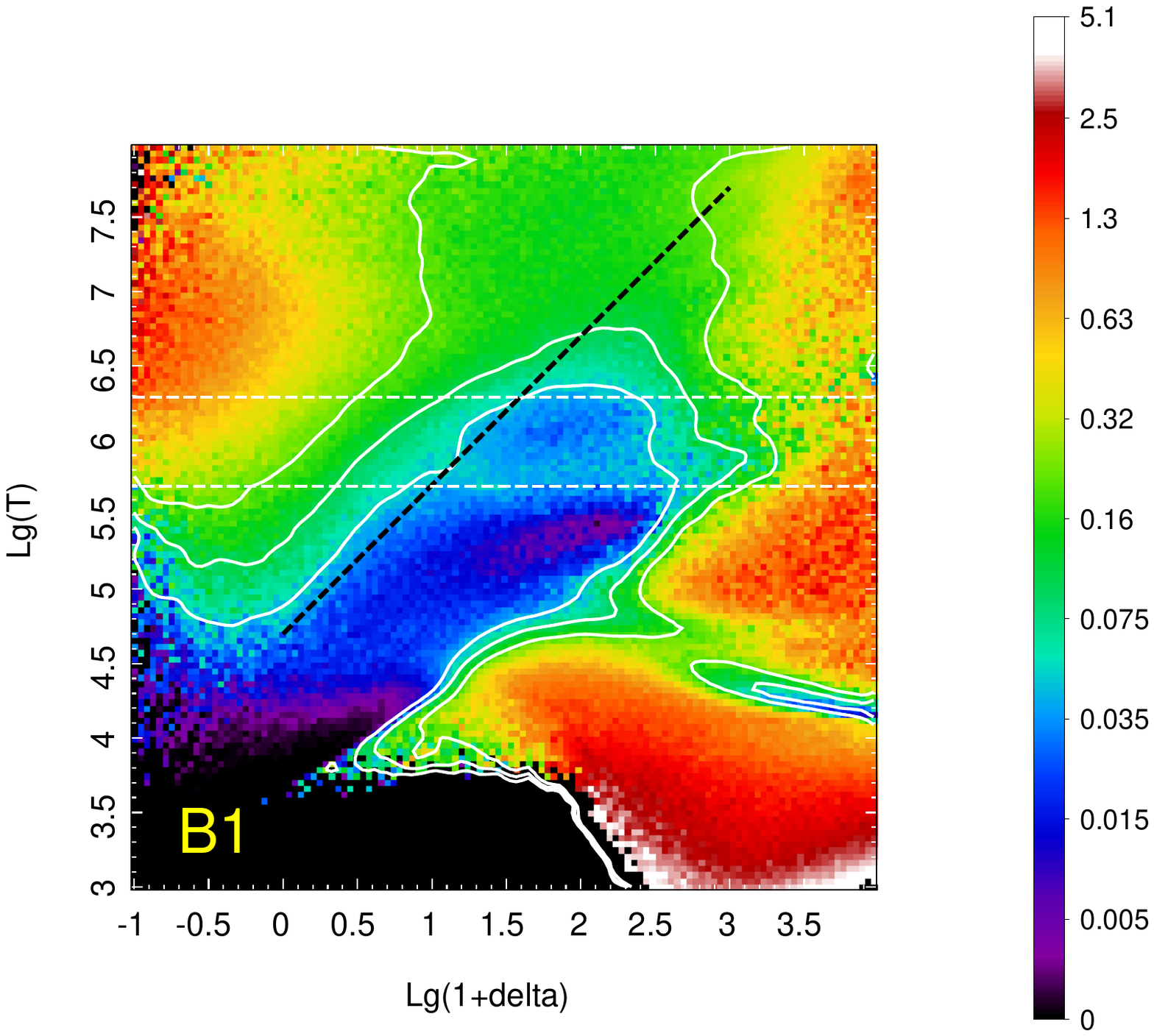}
\caption{{\bf Left and middle: } Mass-weighted gas distribution in the overdensity-temperature plane for the "cluster" (B1) and "reference" (B2) cubes shown in Fig.~\ref{fig:eml_maps}. The temperature is in Kelvin (vertical axes) and the density $(1+\delta)$ is in units of the mean baryon density of the Universe (horizontal axes). The fraction of gas mass at a given  density and temperature is color-coded. The white dashed box encompasses the gas with temperature $0.5\times10^6\,{\rm K} < T <  2\times 10^6\,{\rm K}$. The gas in this temperature range accounts for 20\% of the baryons in the simulated volume, with $\sim 43$\% and $\sim 33$\% in colder and hotter phases, respectively. About $\sim 4$\% of baryons are in stars outside the considered temperature and density ranges.  The dashed black line shows the locus of the points that make an important contribution to the total mass budget of the WHIM. In this paper, we focus on the gas with the density $(1+\delta)<50$ and the temperature in the range $10^4\,{\rm K}<T<2\times10^6\, {\rm K}$.  To indicate the extended range of temperatures and the overdensity cut, we will call this phase "eWHIM50". 
{\bf Right}: Metal abundance as a function of overdensity and temperature in the simulated box. The abundance was calculated by averaging it over all particles with similar densities and temperatures (see also Appendix~\ref{app:metals}). 
The abundance is relative to the Solar photospheric abundance. The contours are  at 0.05, 0.1, 0.2. In these simulations, the  abundance in a typical WHIM region 
($\delta \lesssim 10^2$ and  $T\sim 10^6 \,{\rm K}$) is very low, $\sim 2-5$\%. These pictures are similar to the outputs of other cosmological simulations \citep[e.g.][]{2001ApJ...552..473D,2005MNRAS.364.1105S,2019MNRAS.486.3766M,2021A&A...646A.156T} and previous analysis of \texttt{Magneticum} simulations \citep[][]{2019MNRAS.482.4972K,2022A&A...663L...6A}.
}
\label{fig:n_t_boxes}
\end{figure*}

The selection cuts used above are further illustrated in Fig.~\ref{fig:n_t_boxes}, which show the mass-weighted gas distribution in the overdensity-temperature plane for the "cluster" and "reference" cubes. The white dashed line delineates the selection criteria corresponding to column B in Fig.~\ref{fig:eml_maps}, namely  $0.5\times10^6\,{\rm K}<T<2\times 10^6\, {\rm K}$ and $\delta < 10^4$. In the simulated data, these cuts account for $\sim 20$\% of all baryons in the box with $\sim 43$\% and $\sim 33$\% in colder and hotter phases, respectively.  As we show below, low-density and cool gas is still able to produce observable signatures in the X-ray spectra. We, therefore, defined as "WHIM" the gas with the density $(1+\delta)<50$ and the temperature in the range $10^4\,{\rm K}<T<2\times10^6\, {\rm K}$.  To indicate the extended range of temperatures and the overdensity cut, we will call this phase "eWHIM50". As a further justification of the density and temperature cuts used here, we note that for a massive halo, a local matter overdensity  $\sim 40-50$ is found near the radius $r_{200m}$ where the enclosed matter overdensity if a factor of 200 larger than the mean matter density of the Universe, \citep[e.g.][]{2014ApJ...789....1D,2021MNRAS.504.4649O}.
In simulations, the radial density profile is the steepest in the vicinity of $r_{200m}$, suggesting that this radius can be used as one of the definitions of a boundary separating the cluster gas from a more diffuse WHIM. Our  $(1+\delta)<50$ cut, therefore, broadly corresponds to this definition. We have also completely ignored the gas with $T\lesssim10^4\,{\rm K}$ when considering the WHIM. Such gas might be present either as cold and dense clumps in galaxies or a low-density gas outside halos. In the former case, no X-ray emission is expected and the absorption signatures are confined to galaxies. In the latter case, the density and the mass of such gas are so small that it is safe to ignore this phase.

The abundance of metals predicted by simulation for different gas phases is shown in Fig.~\ref{fig:n_t_boxes}.  
The abundance is relative to the Solar photospheric abundance and is estimated as the ratio of the total mass of elements heavier than helium to the hydrogen mass and scaled by the Solar value of $\approx 2.6\times10^{-2}$  \citep[e.g.][]{1989GeCoA..53..197A}. We note here, that the  scatter of abundances inside each overdensity/temperature bin can be substantial (see Appendix~\ref{app:metals}) and the mean value does not necessarily characterize the bulk of the gas with a given density and temperature. However, it is this value that one would get if the components with different abundances are physically co-spatial and mixing between them is allowed to take place (the enrichment model implemented in \texttt{Magneticum} does not allow metals exchange between the particles). 

As expected, ${\rm EM_{\it l}}$ maps (Fig.~\ref{fig:eml_maps}) show a network of filaments converging on the massive cluster. These filaments are often a few tens of Mpc long and $\sim$4-6~Mpc thick. Once the selection based on the gas temperature (or temperature and overdensity) is applied, the inner parts of the filaments (inside the hot and dense cluster region) become suppressed. In the reference box maps, similar filaments are present, but their number is significantly smaller, and in the absence of a prominent cluster at the center of this box, the distribution of filaments does not show clear convergence of filaments on a single center. Instead, there is one very prominent and largely one-dimensional filament that is traced by a chain of lower-mass halos. At the edges of the "cluster" and "reference" boxes their appearance is similar, suggesting that at these distances ($\gtrsim 20\,{\rm Mpc}$) the influence of the overdensity around the main cluster on the distribution of matter is not very strong. This, in particular, reflects the higher density of all types of structures near the turn-around radius of a massive cluster.

\section{Radial profiles}
\label{sec:where}

\begin{figure*}
\centering
\includegraphics[angle=0,trim=1cm 5cm 0cm 2cm,width=0.9\columnwidth]{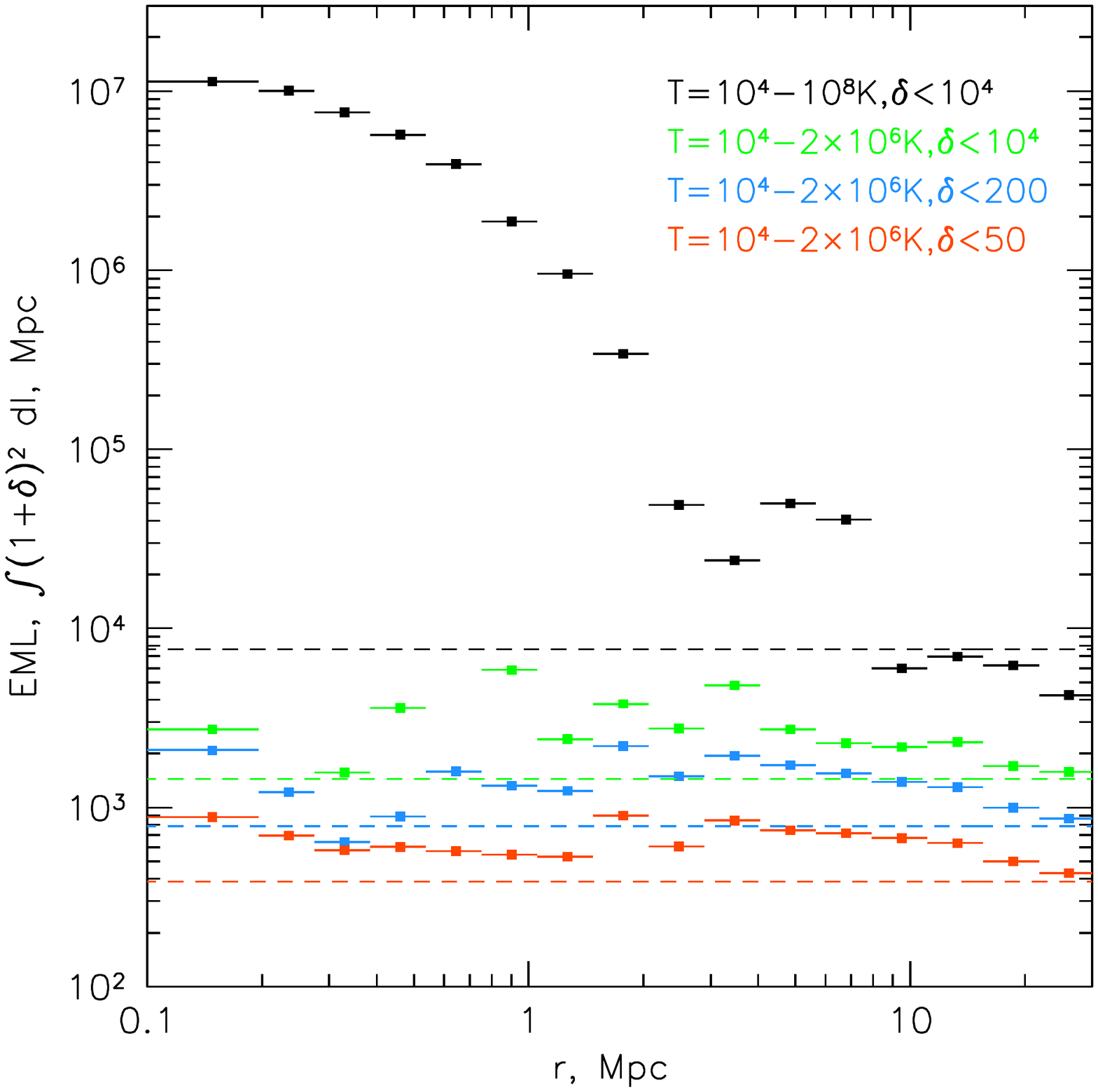}
\includegraphics[angle=0,trim=1cm 5cm 0cm 2cm,width=0.9\columnwidth]{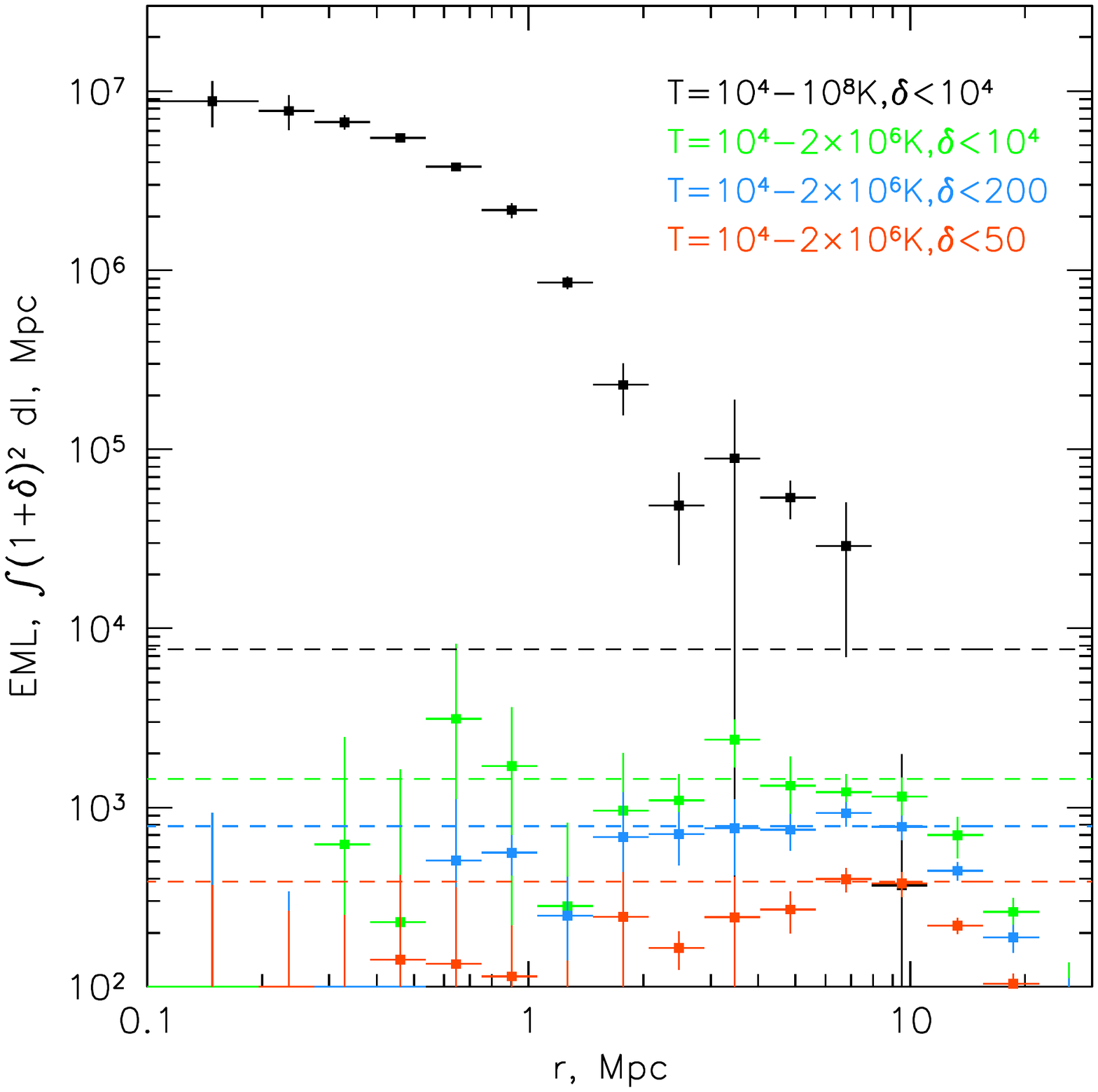}
\caption{{\bf Left: }Radial profiles of ${\rm EM}_l$ extracted from the data cube shown in the top row of Fig.~\ref{fig:eml_maps}. The black, green, blue, and red colors correspond to the same temperature and density cuts shown in the Figure legend.  The least stringent cut (black color) is intended to encompass the bulk of the gas associated with the cluster. Hence, the broad range of temperatures from $10^4$ to $10^8\,{\rm K}$. Selected patches in the simulated cluster have temperature up to $2\times 10^8\,{\rm K}$, but their contribution to the plotted points is subdominant. The same is true for the gas below $10^4\,{\rm K}$ since the requirement of $\delta<10^4$ excludes the very dense cold clumps that might be present in galaxies. For comparison, the dashed horizontal lines show the mean levels of  ${\rm EM}_l$ for the reference data cube (bottom panels in Fig.~\ref{fig:eml_maps}). For the "cluster cube", the center corresponds to the position of the massive cluster in the Magneticum halo catalog. For the reference data cube, the center is simply the center of the cube, not associated with any particular halo. 
Clearly, beyond 15-20~Mpc there is no big difference between the projected emission measures in the cluster and reference cubes. 
{\bf Right:} The same radial profiles from which the mean level obtained from the  reference box have been subtracted.  The profiles were averaged over three projections to assign error bars that reflect the scatter between these projections. These two plots suggest that the most robust constraints on the WHIM emission could be obtained by studying distant outskirts ($\sim 10$~Mpc) of massive clusters. It is noteworthy that  the excess emission is of the same order as the mean level found in the reference simulated cube, as one can expect within the turnaround radius of a massive cluster (but far enough from the cluster itself).
} 
\label{fig:eml_radial}
\end{figure*}

\begin{figure}
\centering
\includegraphics[angle=0,trim=1cm 5cm 0cm 2cm,width=0.95\columnwidth]{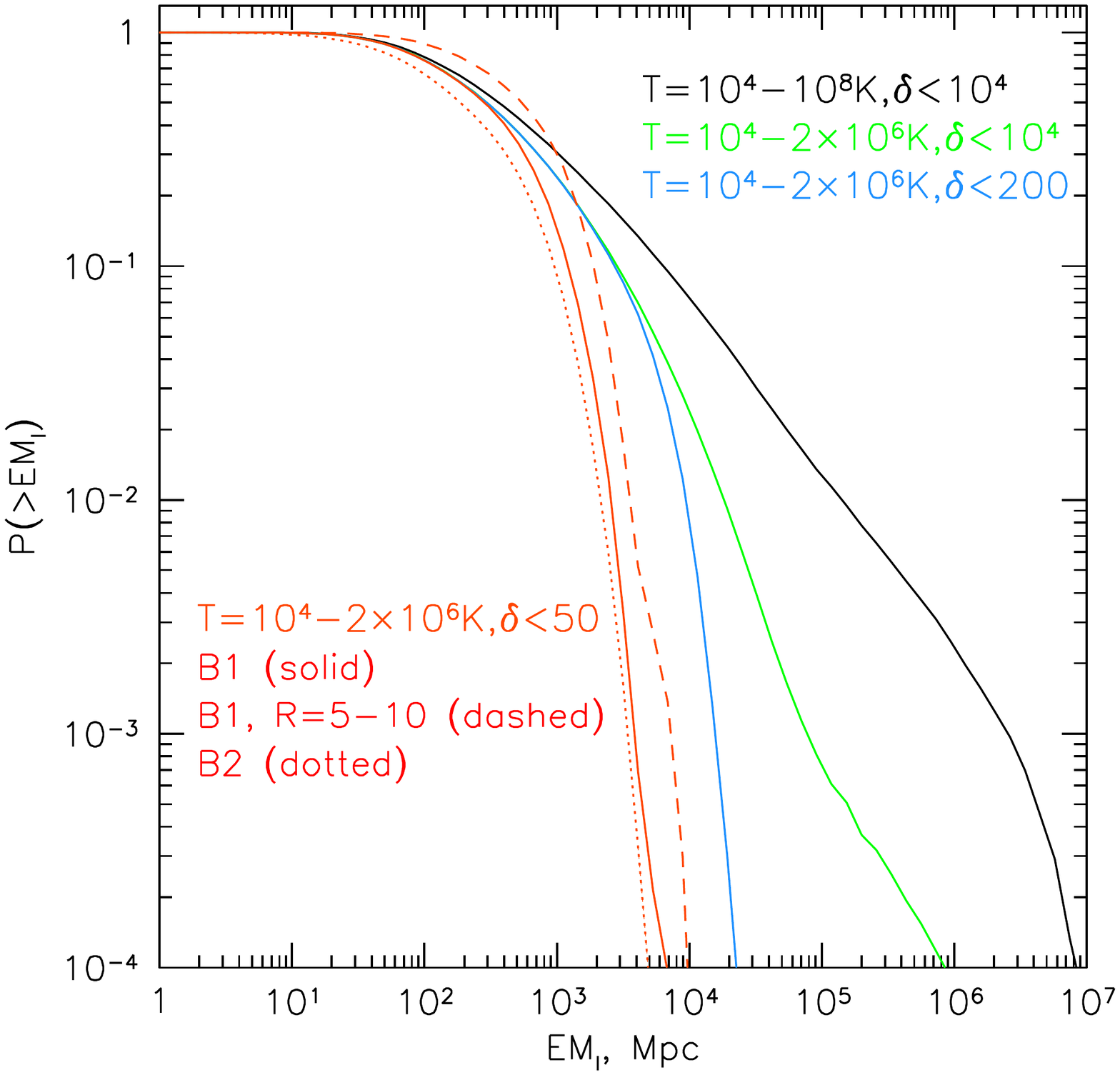}
\caption{Probability of finding  ${\rm EM_{\it l}}$ above a given value in a projected image of the simulated "cluster" box  (57~Mpc along the line-of-sight). Different colors (see legend) correspond to different cuts in the 3D density and temperature made before generating a projected image. The solid red curve corresponds to our our provisional eWHIM50 definition with $10^4\,{\rm K}<T<2\times10^6 \,{\rm K}$ and $(1+\delta)<50$. For these parameters the median value of ${\rm EM_{\it l}}$ is $\sim 240\,{\rm Mpc}$ and exceeds $\sim 1200\,{\rm Mpc}$ with the probability of $\sim 10$\%. The dashed red line (labeled as "R5-10") shows the same probability calculated for $r=5-10$~Mpc annulus centered at the main cluster. In this region both the median  ($\sim 520$~Mpc)  and the top 10\% quantile of the ${\rm EM_{\it l}}$  values ($\sim 1900$~Mpc) are higher than in other portions of the image.
For comparison, the dotted red line (labeled "B2") shows  $P(>{\rm EM_{\it l}})$  for the projected image generated for the second "reference" box, which is comparable to the  $P(>{\rm EM_{\it l}})$ calculated for the entire cluster box.}
\label{fig:pdf_eml}
\end{figure}

A more quantitative characterization of  ${\rm EM_{\it l}}$ maps is shown in Fig.~\ref{fig:eml_radial} left panel, where the radial profiles for concentric rings around the cluster are shown. The horizontal lines show the mean levels of ${\rm EM_{\it l}}$ in the reference box with the same temperature and overdensity cuts. For a homogeneous box having the mean baryonic density of the Universe, the expected value of ${\rm EM_{\it l}}$ is, obviously, the size of the box, i.e. 57.5~Mpc. 

These radial profiles illustrate several characteristic properties:
\begin{itemize}
\item The level of ${\rm EM_{\it l}}$ is a strong function of the applied gas phase selection criteria.
\item For all temperature and overdensity cuts, the cluster box profiles converge to the mean level of the reference box beyond $\sim 20$~Mpc, i.e.  $\sim 5-7 R_{200m}$. This further strengthens the suggestion that the chosen size of the box is large enough and the impact of the cluster on the matter density at larger distances is subdominant, at least for the purpose of this study.  
\item Inside $\sim 10-20$~Mpc, the projected emission measure due to "focused" filaments is significantly higher than for the reference box.   
\item In the inner region ($<1\,{\rm Mpc}$), the scatter in the ${\rm EM_{\it l}}$ radial profiles for the warm $10^4T<2\times10^6\,{\rm K}$ gas is large, as illustrated by error bars shown in the right panel of Fig.~\ref{fig:eml_radial}. These error bars were crudely evaluated as the root-mean-square variations among radial profiles calculated for 3 different projections (along the X, Y, and Z axes of the simulated box). The level of the projected emission measure can be higher or lower than the corresponding values for the reference box. This partly reflects the fact, that not much of the $10^6$~K gas is present inside the cluster and, more importantly, the stochastic fluctuations due to a highly inhomogeneous gas density distribution in cluster outskirts are dominated by a rather small number of the most prominent filaments.
\item There is a range of projected radii  ($\sim 1 - 10 \,{\rm Mpc}$), where the scatter is moderate. 
\end{itemize}

Given that both the cluster and reference boxes have similar levels of ${\rm EM_{\it l}}$ beyond $\sim 20$~Mpc, it makes sense to calculate the "excess" emission associated with the focusing effect of the cluster on the distribution of filaments. We do this by subtracting the mean levels obtained for the reference box from the radial profiles. The results are shown in the right panel of Fig.~\ref{fig:eml_radial}.  There we average 3 projections to reduce the scatter and we assign an error bar to each radial bin, which characterizes this scatter\footnote{This is, of course, a very crude and incomplete way of characterizing the scatter. We do this only for illustration since we do not use this information here.}. Unlike the left panel of Fig.~\ref{fig:eml_radial} the excess emission should be independent of the size of the box (once it is larger than the correlation length in the density distribution). Fig.~\ref{fig:eml_radial} (right panel) also corroborates the conjecture that the range of radii of $\sim$5-10~Mpc is well suited for the search of $T\sim 10^6~K\,{\rm gas}$ in the vicinity of a massive cluster (apart from possible observational limitations).

Instead of calculating the excess emission of the WHIM in radial bins, one can search for bright localized spots associated with the WHIM within 5-10~Mpc of the cluster. The probability of finding ${\rm EM_{\it l}}$ above a given value in a projected image of the simulated "cluster" box  can be characterized by the probability distribution function PDF(${\rm EM_{\it l}}$), which is shown in Fig.~\ref{fig:pdf_eml} for the entire box and the 5-10~Mpc ring. This plot shows that the probability of finding excess values of ${\rm EM_{\it l}}\sim 10^3\,{\rm Mpc}$ associated with eWHIM50 is $\sim 10$\% in the 5-10~Mpc ring. We repeated the same procedure for two other massive clusters from the Magneticum and found very similar probability distribution functions of the ${\rm EM_{\it l}}$.

\section{Contamination of the WHIM signal by high-density regions}
\label{sec:cont}


One of the issues that might hamper the detection of the WHIM signal, is its contamination by higher-density gaseous lumps seen in projection. In Figs.~\ref{fig:eml_maps},~\ref{fig:eml_radial}, and \ref{fig:pdf_eml},  the regions of high-overdensity have been removed from the original (3D) data. Of course, this can only be done for the  simulated data. The discussion on what approach should be used in application to real data \citep[see, e.g.][]{2007ApJ...661L.117H} is beyond the scope of this paper. However, we can simply estimate the probability of such contamination by using the same simulation data. The size of the cube, corresponding to the redshift bin of $\Delta z\sim 0.01$ is well suited for this purpose. The chances of a random superposition of two structures can already be easily estimated from Fig.\ref{fig:pdf_eml}. 

\begin{figure*}
\centering
\includegraphics[angle=0,trim=1cm 12cm 1cm 9.8cm,width=2.1\columnwidth]{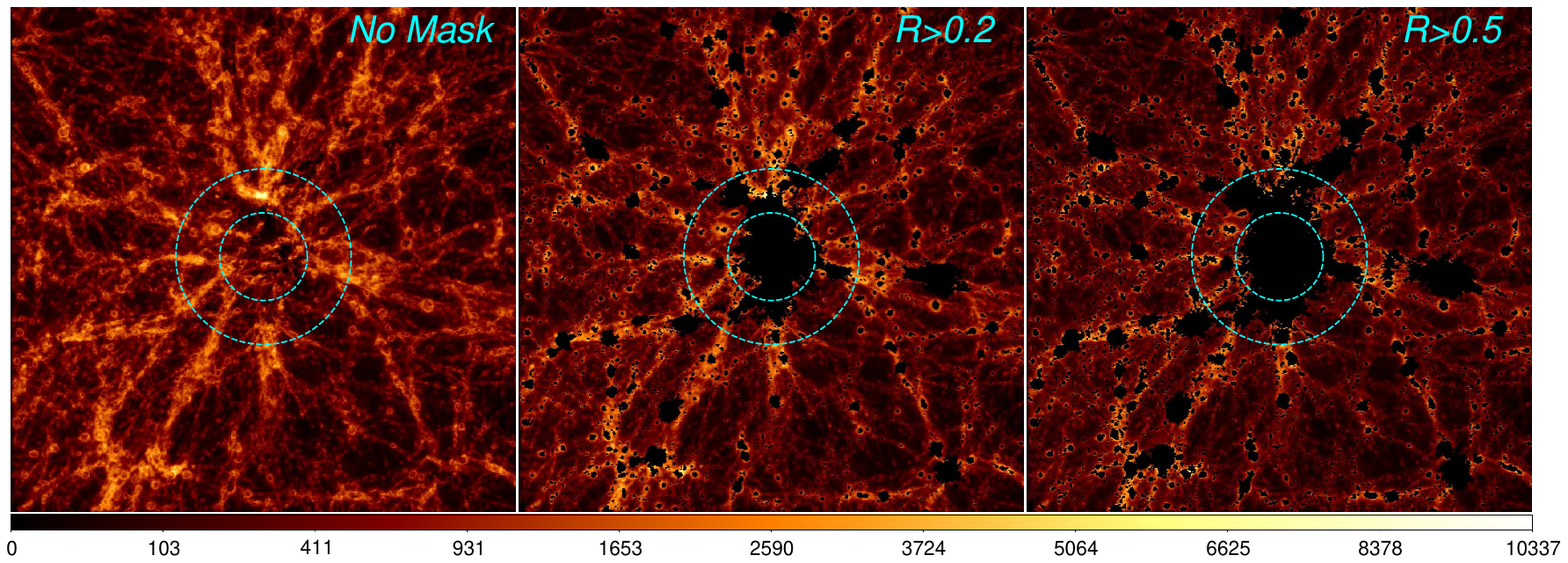}
\caption{Contamination of the eWHIM50 signal by high-density regions seen in projection. The images show three ${\rm EM_{\it l}}$ maps for $\delta < 50$ gas and different masks. The mask is defined by the level of the "contamination" of a given region, which is evaluated by comparing the value of ${\rm EM_{\it l}}$ due to $\delta < 50$ gas and the total ${\rm EM_{\it l}}$ for the gas with any overdensity (see Eq.~\ref{eq:r}).  
Masks that keep regions with $R>0.2$ and  $R>0.5$ are shown in the middle and right panels, respectively. Two dashed circles correspond to  5 and 10 Mpc distances from the cluster center. The requirement of having $R>0.5$ is a rather strong one. However, even with this very stringent cut less than half of the area is affected beyond 5~Mpc from the cluster, and filaments are clearly visible in the remaining data, although the brightest regions of these filaments are masked out. } 
\label{fig:eml_masked}
\end{figure*}

\begin{figure}
\centering
\includegraphics[angle=0,trim=1cm 5cm 0cm 2cm,width=0.95\columnwidth]{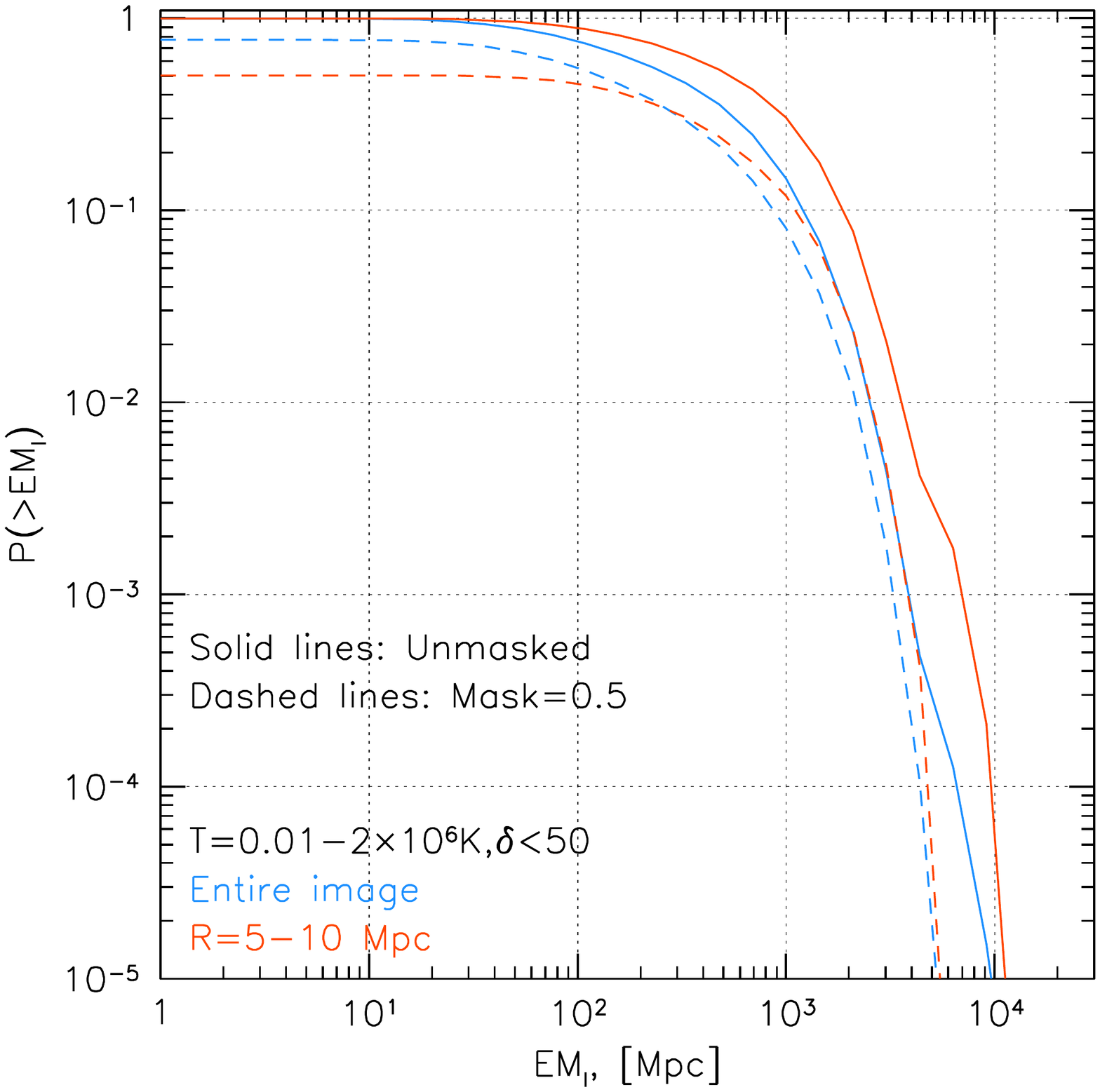}
\caption{Impact of the $R>0.5$ mask (see Eq.~\ref{eq:r}) on the probability of finding a given value of the emission measure in the projected (and masked) images.  The solid lines show ${\rm P(>EM_{\it l})}$ for the entire unmasked image (blue) and the $r=5-10$~Mpc annulus (red), respectively. The dashed lines show the same quantity for the masked images. The masked regions are counted as zero-value pixels, so that for small positive ${\rm EM_{\it l}}$ the values of $P<1$ reflects the fraction of the image area that is contaminated by gas lumps with overdensity higher than 50.
}
\label{fig:pdf_masked}
\end{figure}

However, the largest signal is coming from collapsed halos and from the regions adjacent to them. It is therefore likely that a random superposition approach severely underestimates the level of the contamination. To characterize this effect we have calculated the ratio of the two ${\rm EM_{\it l}}$ maps   
\begin{equation}
R=\frac{{\rm EM_{\it l}}(T<2\times10^6; \delta<50)}{{\rm EM_{\it l}}(\forall \, T; \forall \delta)},
\label{eq:r}
\end{equation}
where the map in the numerator accounts for the WHIM only, while the denominator includes all gas phases. The larger the value of $R$, the higher the contribution of the WHIM-like gas to the total signal. Therefore, masking the regions with low $R$ retains a cleaner WHIM signal. The impact of the masking is illustrated in Fig.~\ref{fig:eml_masked} for $R>0.2$ and 0.5. 
In principle, the low value of $R$ does not mean that it is not possible to distinguish the contributions of the low and high-density components, provided that their spectra are sufficiently different. On the other hand, $R>0.5$ ensures that the signal is indeed dominated by the low-density gas. All in all,
this exercise suggests that if real WHIM properties are reproduced by simulations used here, then (i) about half of the area must be excluded (for the 5-10~Mpc annulus around the most massive clusters) and (ii) the maximum amplitude of the WHIM signal in the "uncontaminated" regions reduces by a factor of $\sim 2$ compared to the unmasked case. This is further illustrated by  Fig.~\ref{fig:pdf_masked} that shows the probability of finding ${\rm EM_{\it l}}$ above a given value for the projected images without contaminated regions. 


\section{Expected spectra and detection sensitivity}
\label{sec:smodel}

\begin{figure}
\centering
\includegraphics[angle=0,trim=1cm 5cm 0cm 2cm,width=0.9\columnwidth]{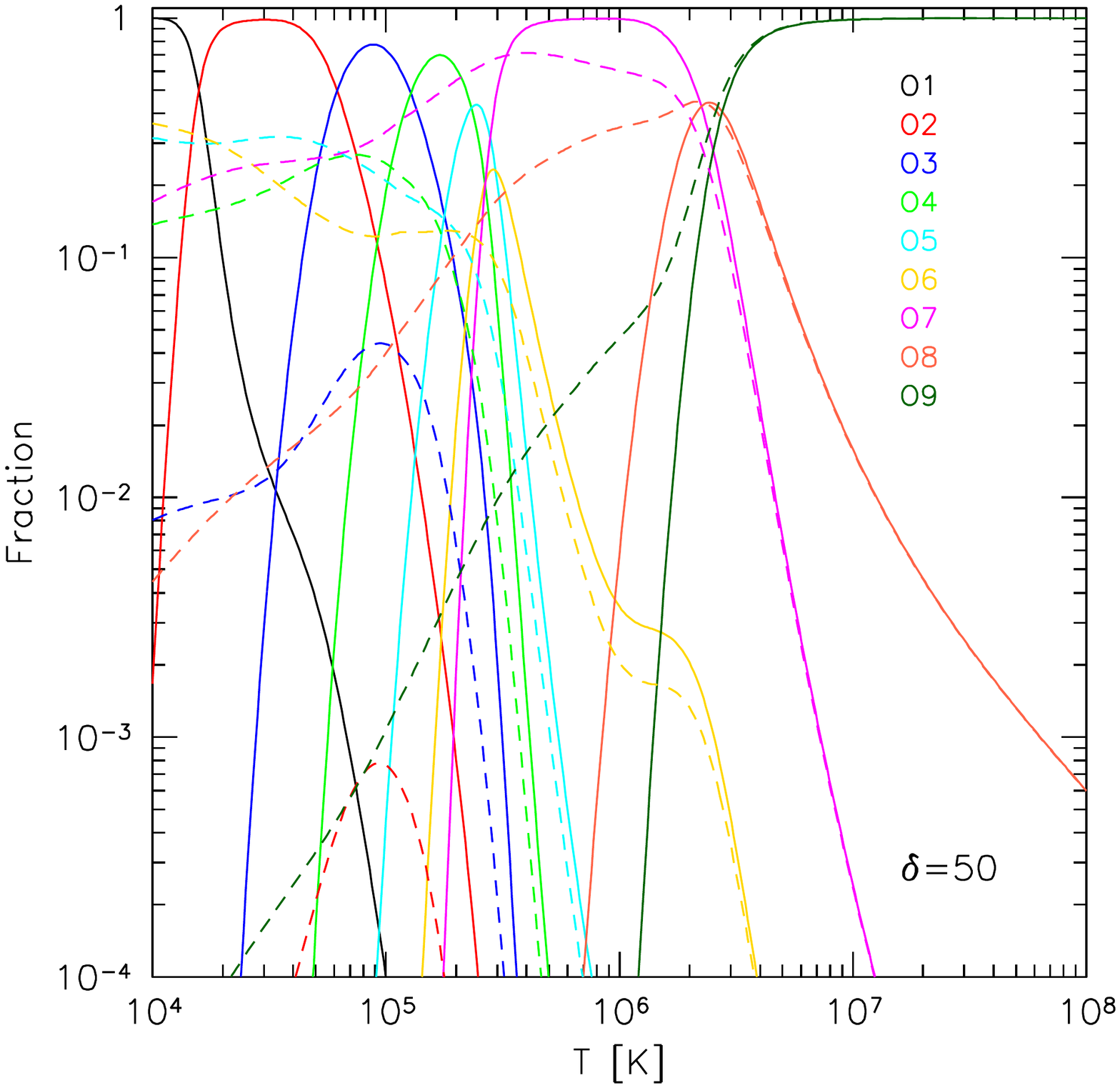}
\caption{Comparison of the oxygen ion fractions for CIE (solid lines) and photoionized plasma (dashed lines) with overdensity $\delta=50$  as a function of electron temperature. The colors correspond to different degrees of ionization (spectroscopic symbols) shown in the plot. For the X-ray signatures of WHIM, the most important effect produced by photoionization is the presence of He-like oxygen ions across the entire range of temperatures shown in the plot. 
}
\label{fig:cie_vs_photo}
\end{figure}

Given the density and temperature of the WHIM gas, we calculate the ionization balance of heavy elements by taking into account photoionization by cosmic UV and X-ray background radiation fields at $z\sim0$ \citep[as we have done in ][]{2001MNRAS.323...93C,2019MNRAS.482.4972K}. For the thermal emission, \verb#MEKAL# model \citep{1985A&AS...62..197M,1995ApJ...438L.115L} as implemented in the \verb#XSPEC# package \citep{1996ASPC..101...17A} was used.  A comparison of the predicted spectra with the results from \cite{2019MNRAS.482.4972K} that uses \texttt{Cloudy} \citep[see][]{2017RMxAA..53..385F} shows good consistency that is fully sufficient for the purposes of this study.
The role of photoionization is particularly important for low temperatures and small overdensities. Indeed, as illustrated in Fig.~\ref{fig:cie_vs_photo}, photoionization by CXB photons leads to the appearance of OVII and OVIII ions over a broad range of (low) temperatures and overdensities, which are characteristic of the WHIM component. We note, that while the time needed to establish ionization equilibrium in the low-density WHIM phase can be long \cite[see, e.g., Fig.3 in][]{2006PASJ...58..641Y}, this mostly concerns the case of recombining hot plasma. Here, we are more focused on the case of photoionized plasma, when the CXB photons provide relatively fast channels of reaching approximate ionization equilibrium for warm gas. We, therefore, assume that the ionization equilibrium has been reached.

The following processes can contribute to the potentially observable X-ray spectral signatures (either in emission or absorption) of a given gas lump:
\begin{itemize}
\item Photoelectric absorption, Thomson, and resonant scattering of the background X-ray emission, dominated by the CXB. For the WHIM case, the role of the Thomson scattering is not particularly important. These processes remove CXB photons from the line of sight going through the WHIM, resulting in a decrement of the CXB emission in direction of the gas lump compared to an "empty" field.
\item Thermal emission of the gas, where due to the strong contribution of photoionization, the role of Radiative Recombination Continuum (RRC) is especially prominent. We also note here, that the photoelectric absorption and RRC can partly cancel each other - these are two reciprocal processes that affect the spectrum above the ionization edges of corresponding ions. 
\item Fluorescent and resonantly scattered photons. Fluorescent photons are not important for the WHIM emission unlike the resonantly scattered photons at the energies of the most prominent lines, principally of O~VII and O~VIII.  As discussed in \cite{2001MNRAS.323...93C,2019MNRAS.482.4972K}, this component is visible only if a significant fraction of the X-ray background is resolved or there is an additional bright source that illuminates the gas from a side (an AGN or a galaxy cluster). 
\end{itemize}

\begin{figure*}
\centering
\includegraphics[angle=0,trim=1cm 5cm 0cm 2cm,width=0.99\columnwidth]{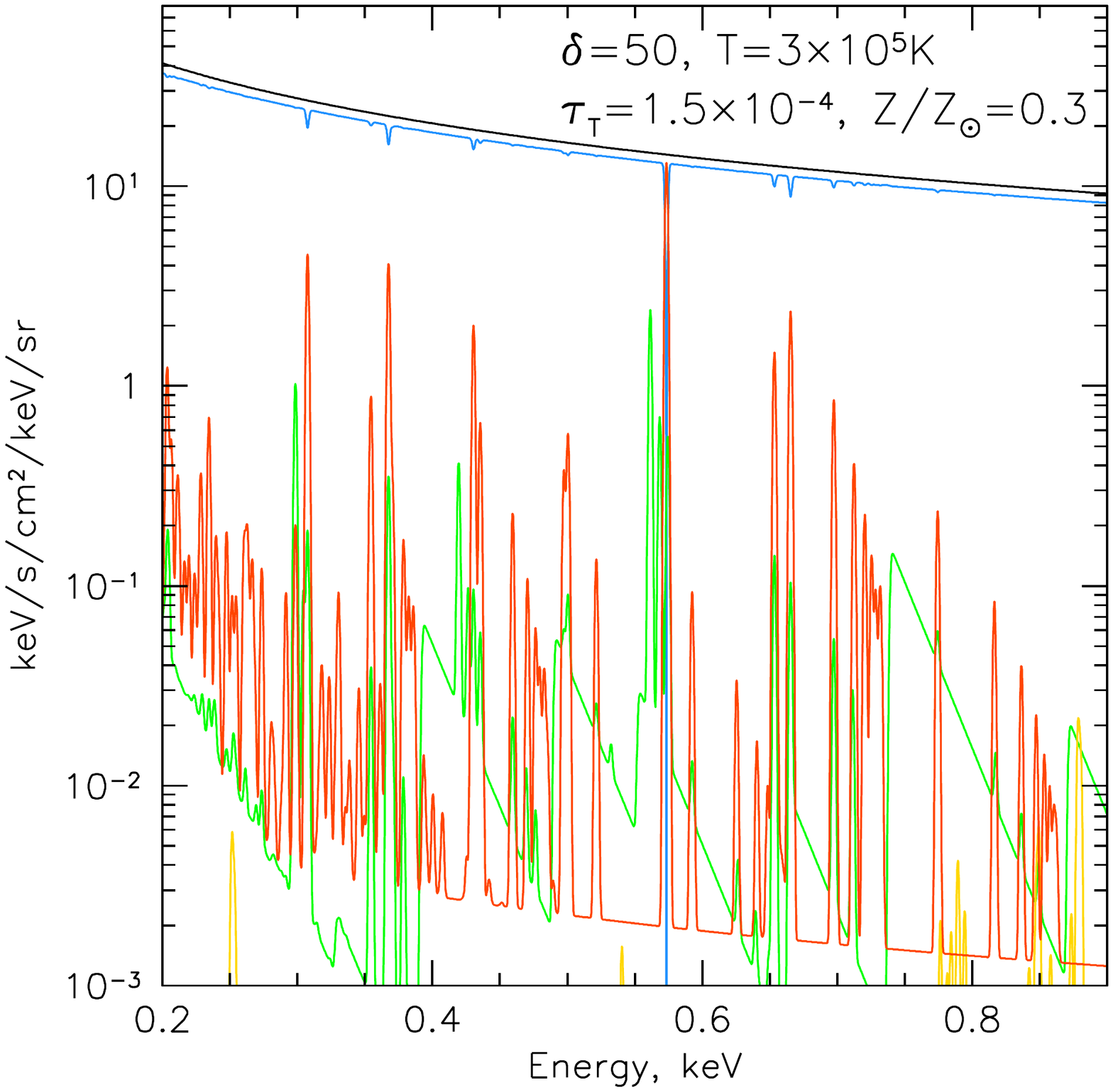}
\includegraphics[angle=0,trim=1cm 5cm 0cm 2cm,width=0.99\columnwidth]{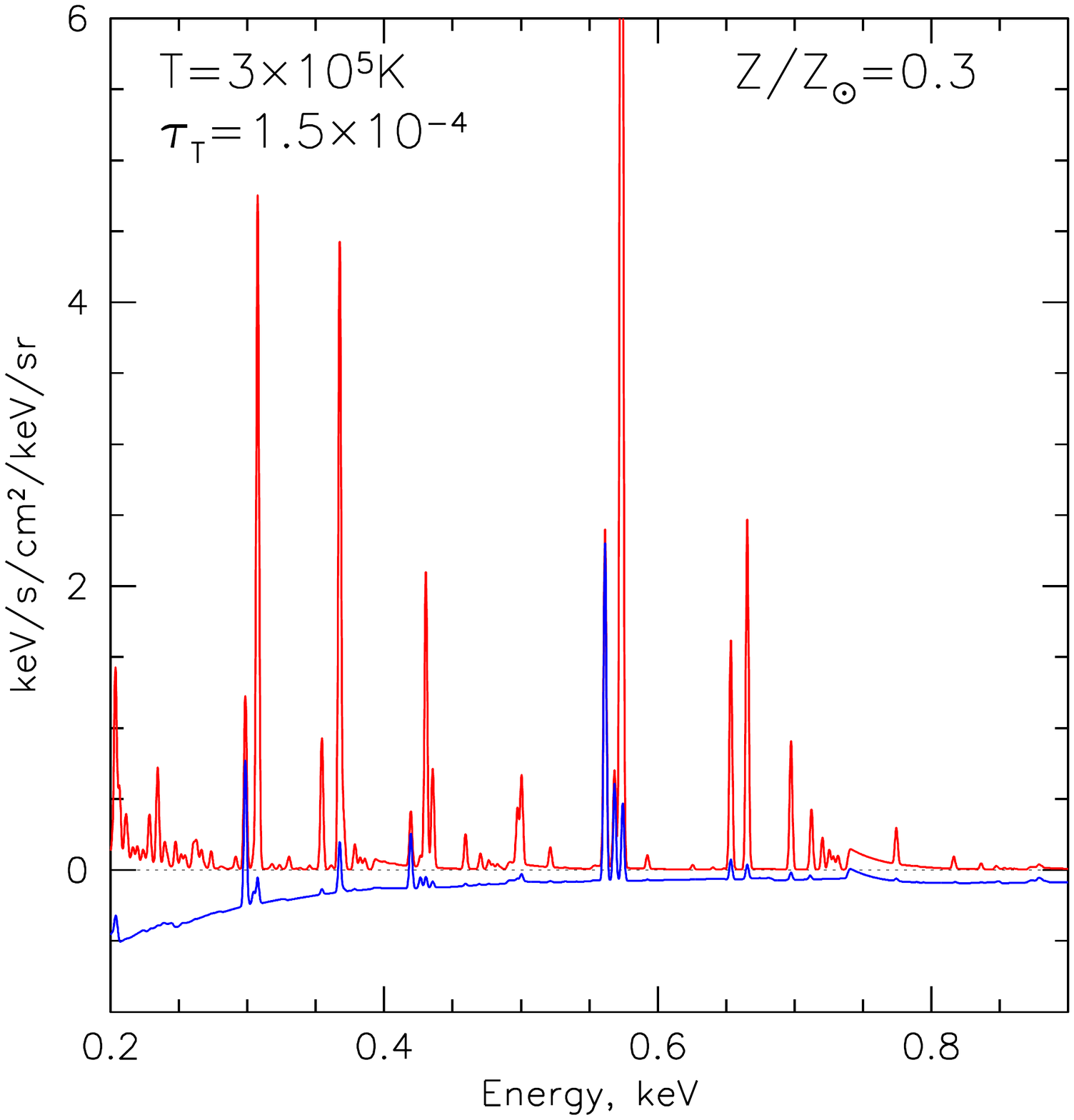}
\caption{{\bf Left:} Illustration of various spectral components in the direction of a filament with temperature $T=3\times10^5\,{\rm K}$ and overdensity $\delta=50$. The blue curve is the CXB spectrum attenuated due to resonant scattering, photoelectric absorption, and Thomson scattering. For comparison, the black curve shows the CXB spectrum (multiplied by 1.1 for clarity). The green curve shows the thermal emission of the photoionized plasma, consisting of emission lines and recombination radiation. The red line is the scattered radiation (resonant and Thomson scatterings). The yellow curve shows the fluorescent emission. All spectra were convolved with $\sigma=1\,{\rm eV}$ Gaussian. The resonant lines are assumed to be unsaturated and all effects second-order in the optical depth have been ignored. {\bf Right:} Difference between the spectrum coming from a direction towards the filament ("on") and that of a "blank" region ("off"). The blue curve shows the case when CXB is unresolved, i.e. it is contributing to the on and off spectra. In this case, a negative signal is seen due to photoelectric absorption plus (mostly) recombination radiation. The red curve, on the contrary, shows the hypothetical case when the background CXB is fully resolved. In this case, the resonantly scattered lines become very prominent in the on-off spectrum.} 
\label{fig:whim_model}
\end{figure*}

All these spectral components are shown in the left panel 
of Fig.~\ref{fig:whim_model}. In this plot, the absorbed and emitted spectra are shown separately. In reality, only the sum of all components can be seen in the direction of a filament, with the additional possibility to resolve some fraction of compact sources that compose CXB. 

The presence or absence of a filament can be tested by comparing the spectrum toward the filament and the spectrum toward an "empty" field. Here we assume that the spectrum of the "empty" field consists of a pure CXB and that the fraction of the resolved CXB is the same as in the filament data. Of course, there are other foreground/background components contributing to the observed spectra, but those will cancel once the difference "filament" minus "empty" is calculated. This difference is essentially the signal that we are looking for. 

The corresponding spectrum is shown in the right panel of Fig.~\ref{fig:whim_model} for some fiducial parameters of a WHIM filament. One of the spectra corresponds to the case when the CXB is not resolved (blue line). In this case, a negative signal at low energies is due to photoelectric absorption, while the positive signal is due to fluorescence, recombination, and collisionally excited lines. For comparison, the red line shows the difference spectrum (hereafter on-off spectrum) when the CXB is fully resolved (hypothetical case). In this limit, the photoelectric absorption signal is gone, while the resonant lines become very prominent. 

Apart from the difference spectrum, the spectrum of resolved CXB sources collected from the filament region will feature absorption lines that collectively account for the same number of photons that are seen as resonantly scattered lines in the difference spectrum. This is of course true only on average. As discussed in \cite{2019MNRAS.482.4972K} comparing the absorbed and emitted flux one can determine the properties of the gas that constitutes the filament.

\begin{figure*}
\centering
\includegraphics[angle=0,trim=1cm 5cm 0cm 2cm,width=0.65\columnwidth]{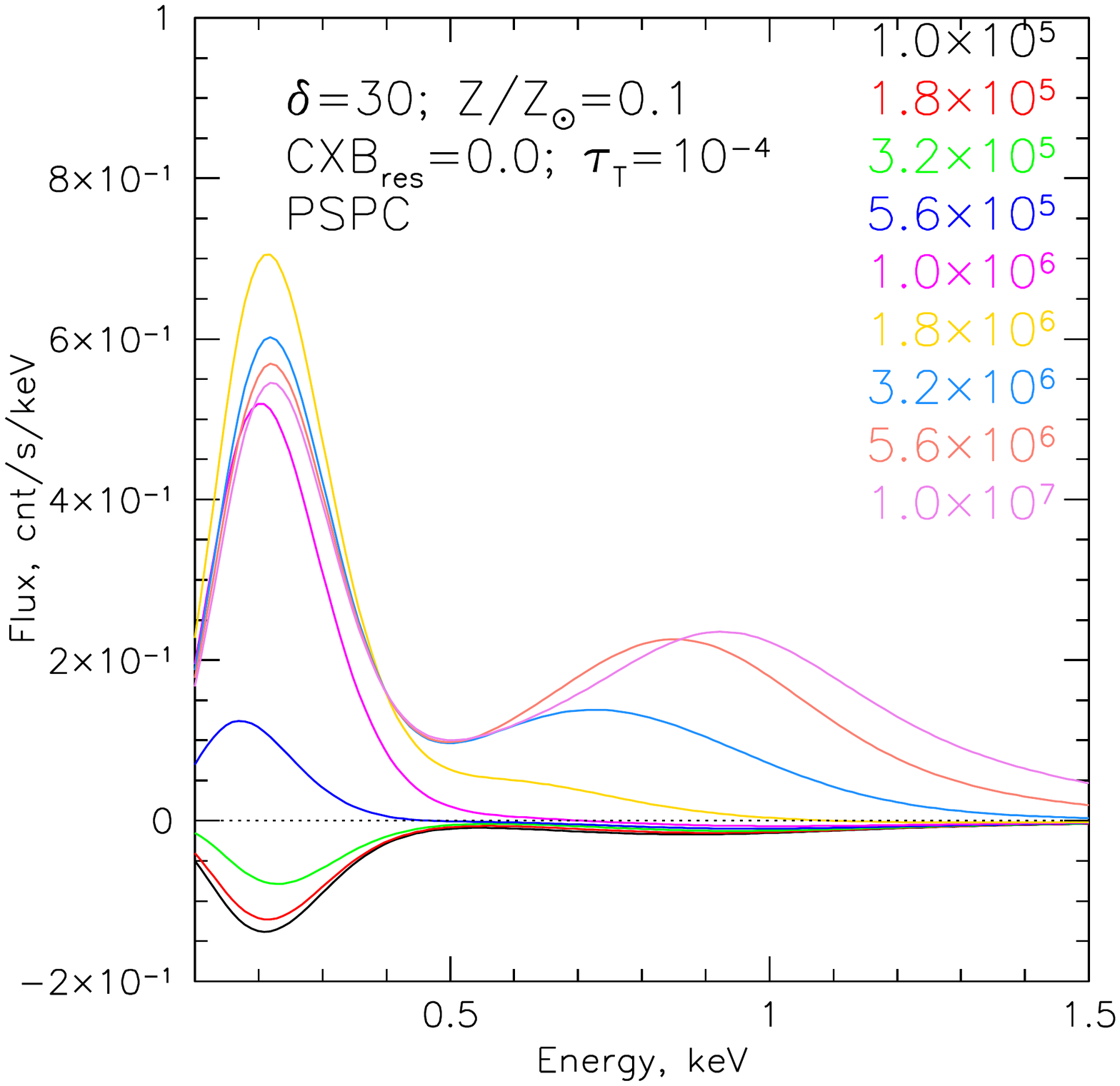}
\includegraphics[angle=0,trim=1cm 5cm 0cm 2cm,width=0.65\columnwidth]{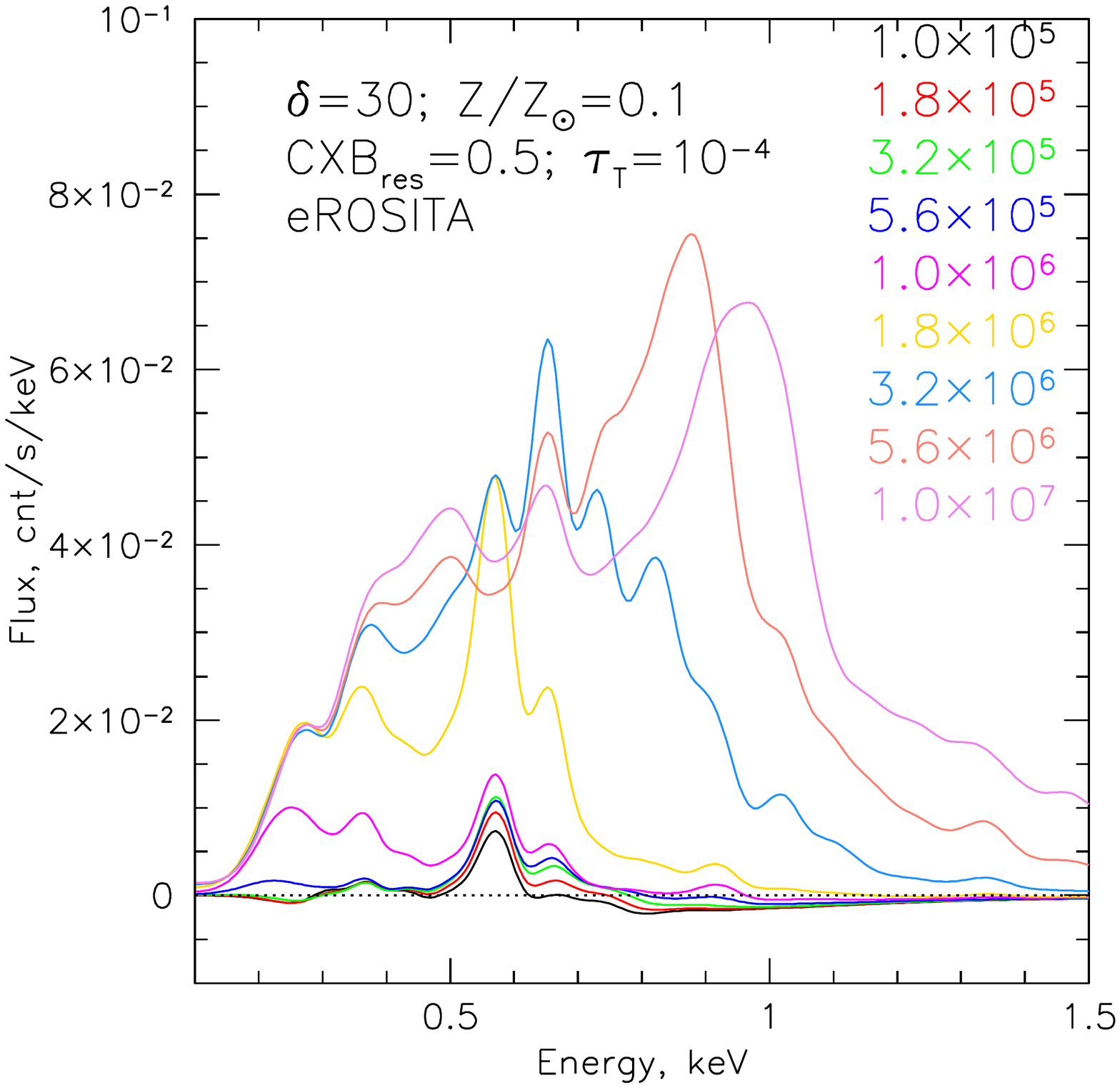}
\includegraphics[angle=0,trim=1cm 5cm 0cm 2cm,width=0.65\columnwidth]{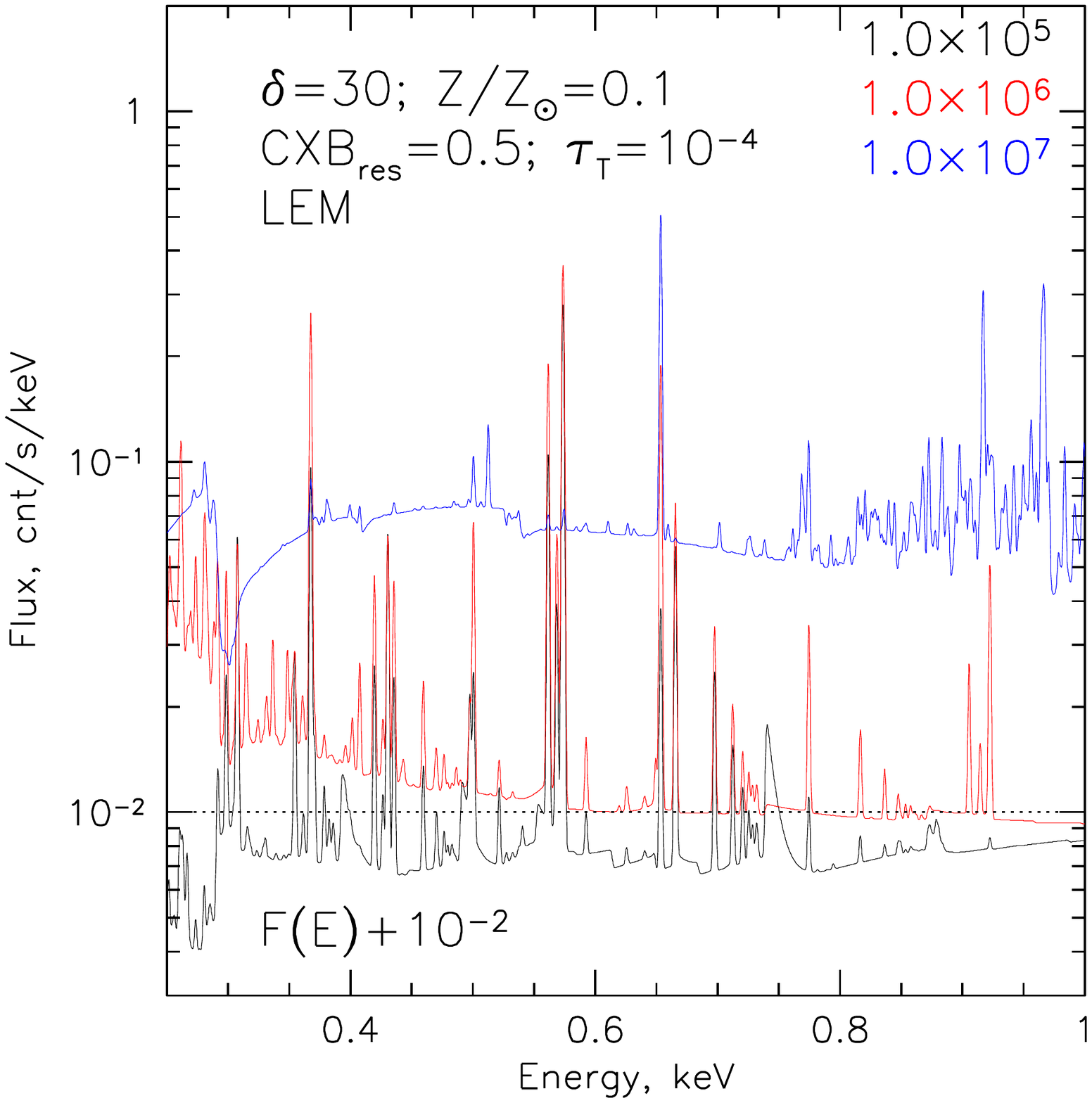}
\caption{Illustrative examples of the on-off spectra for ROSAT, eROSITA, and perspective LEM mission for several values of the gas temperature. The spectra have been convolved with the default responses of the three instruments. It is assumed that the WHIM signal subtends the entire FoV of each instrument.  For the adopted overdensity, the transition between the two regimes (at low and high temperatures) is very clear. For the LEM plot, a constant ($10^{-2}$) was added to the $T=10^5$~K spectrum (black line), to see the negative parts of the spectrum in the log plot.} 
\label{fig:whim_model_3inst}
\end{figure*}

Here, we focus more on the detection of the WHIM via the on-off spectra. Fig.~\ref{fig:whim_model_3inst} shows how the on-off spectra look like when convolved with the spectral response (a combination of the effective area and the energy redistribution matrix) of different instruments, namely, \textit{ROSAT}/PSPC \citep{1982AdSpR...2d.241T}, \textit{SRG}/eROSITA \citep{2021A&A...647A...1P} and proposed soft X-ray microcalorimeter \textit{LEM} \citep{2022arXiv221109827K}. For all plots, the gas metallicity is set to 0.1, overdensity $\delta=30$, the Thomson optical depth $\tau_T=10^{-4}$, and foreground absorbing column density $N_H=10^{20}\,{\rm cm^{-2}}$. The resolved CXB fraction is set to zero for \textit{ROSAT}/PSPC and to 0.5 for eROSITA and \textit{LEM}. It is assumed that the filament fills the entire FoV, which is a strong assumption, especially for PSPC which has the largest FoV. 

The spectra, shown in Fig.~\ref{fig:whim_model_3inst} have been convolved with the standard spectral responses of these instruments. At low temperatures, the on-off spectrum can be negative. This is especially clear in the PSPC and LEM spectra. As discussed above, the negative features are caused by the photo-electric absorption of the CXB signal in the direction of the filament. This negative signal would diminish if most of the CXB signal is resolved, although this is not feasible with these three instruments. At higher temperatures, negative parts go away and the spectra become more close to the thermal emission plus resonantly scattered emission in the lines of elements like oxygen or neon. The latter signal depends linearly on the CXB resolved fraction.

One can use the on-off model spectra to estimate the sensitivity of a given instrument to the presence of the filament. Since the spectra shown in Fig.~\ref{fig:whim_model_3inst} represent the difference between the on-filament and off-filament spectra, one has to take into account the inevitable Poissonian noise of the signal that comes from the full spectrum. To this end, we model the total spectrum in the filament direction as a combination of five components: the Local Hot Bubble (LHB), the Milky Way diffuse emission, the unresolved part of the CXB, the instrumental background, and the filament emission itself. We assume that the "off-filament" spectrum is known and it does not introduce additional errors. Furthermore, we assume that the abundance, temperature, and overdensity of the gas inside the filament are known. With these assumptions, the only unknown quantity is the normalization of the filament signal. We further assume that the exposure is long enough so that the $\chi^2$ statistic is applicable. With these assumptions, the sensitivity was calculated for several values of overdensity, temperature, and abundance.

\begin{figure*}
\centering
\includegraphics[angle=0,trim=1cm 5cm 0cm 2cm,width=0.99\columnwidth]{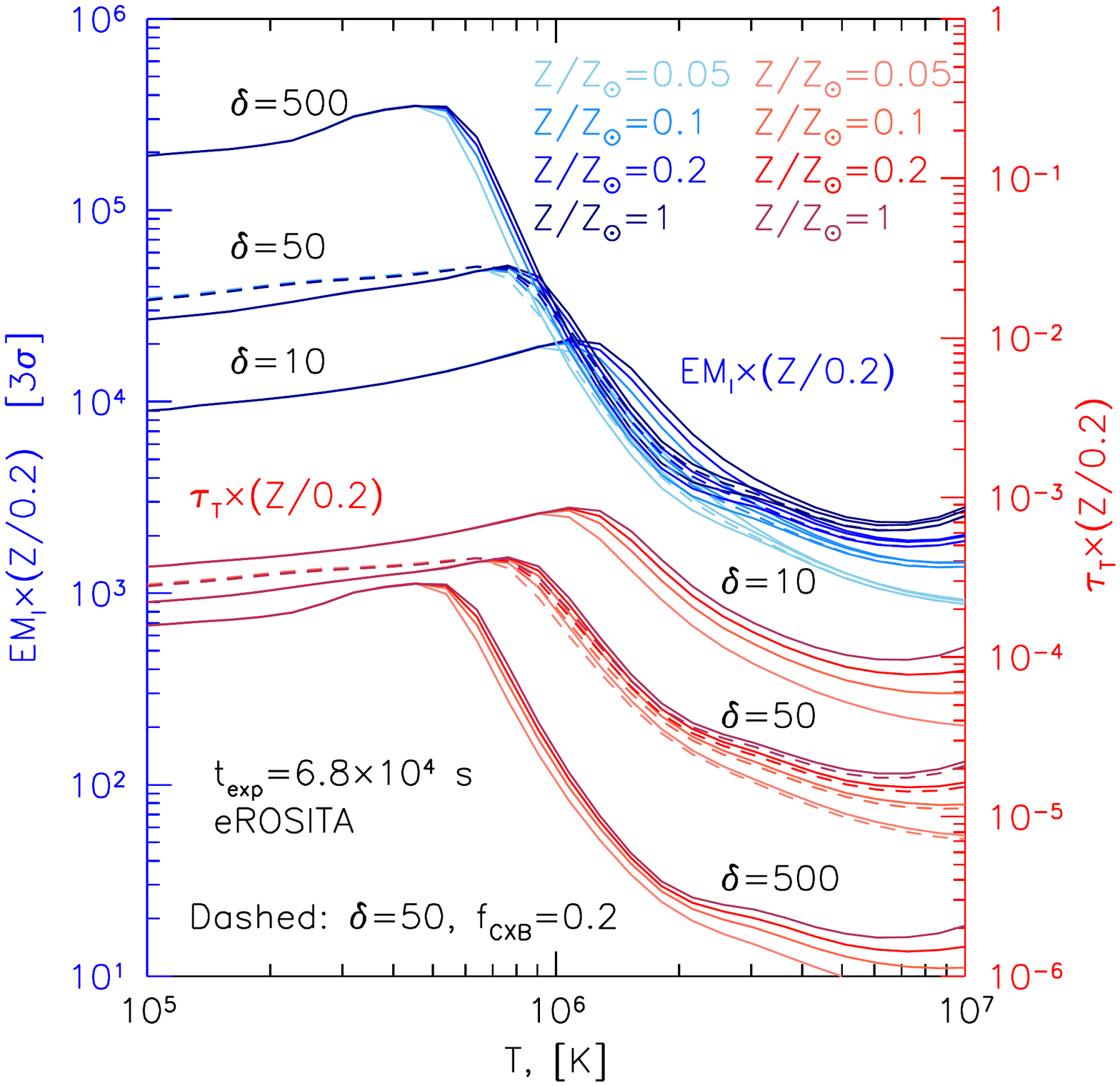}
\includegraphics[angle=0,trim=1cm 5cm 0cm 2cm,width=0.99\columnwidth]{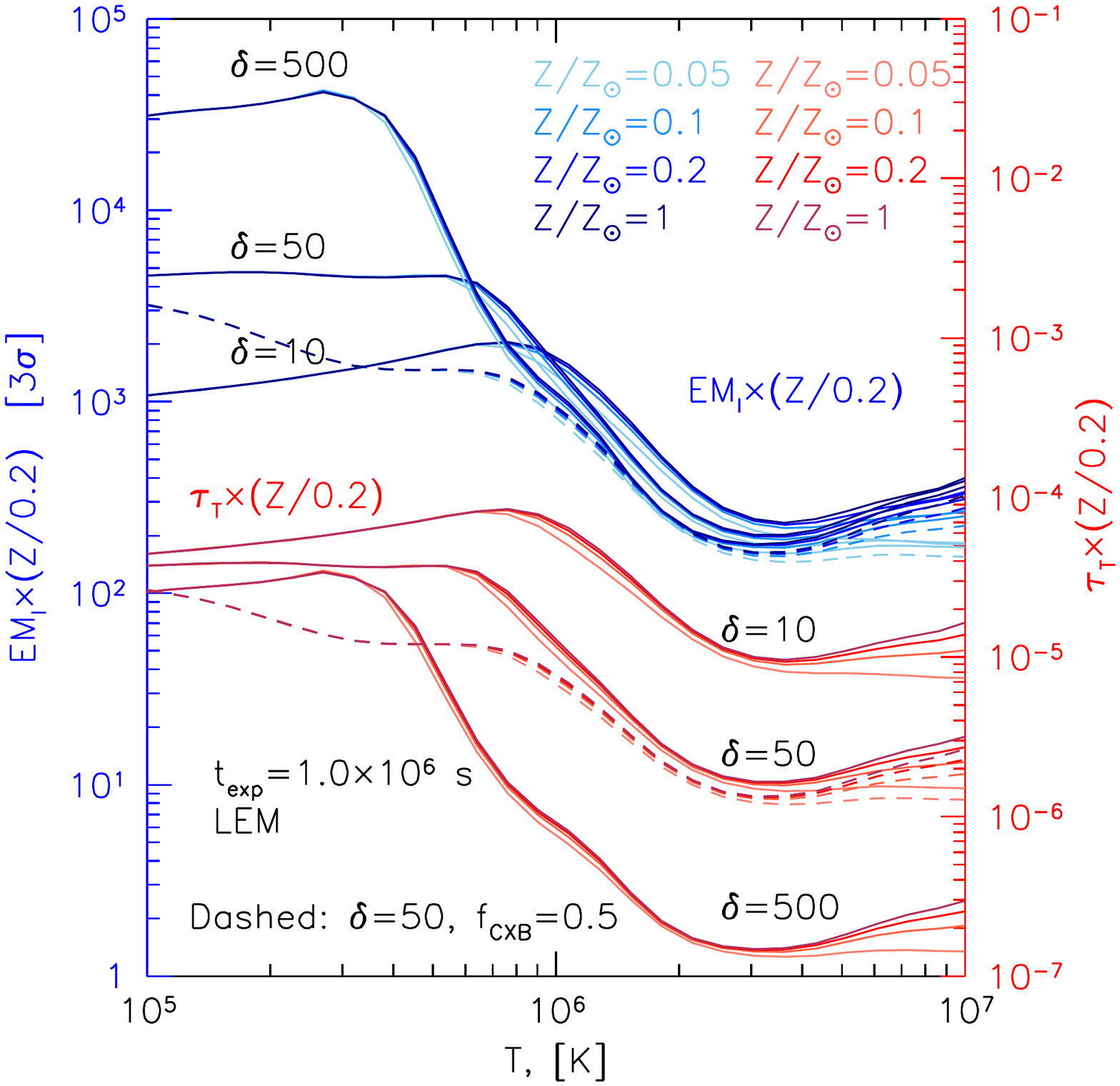}
\caption{Expected sensitivity of eROSITA and LEM for detection of a filament at a known redshift as a function of temperature. The sensitivity is expressed in terms of the ${\rm EM}_l$ (left axis) and the Thomson optical depth $\tau_T$ (right axis). In this highly idealized exercise, it is assumed that all backgrounds and foregrounds are known and they only contribute to the Poissonian errors. The only free parameter in the model is the normalization of the on-off spectrum. For eROSITA the filament redshift is set to 0.0231 (= Coma cluster), while for LEM it is 0.08. In terms of ${\rm EM}_l$, typical values are a few $10^4$~Mpc for eROSITA and a few $10^3$~Mpc for LEM at $T\lesssim 10^6\,{\rm K}$. In terms of $\tau_T$, the sensitivity is a few $10^{-4}$ and $10^{-5}$ for eROSITA and LEM, respectively. Allowing for freedom in the background and/or foreground parameters would shift these curves up. Therefore, quoted sensitivity can be viewed as the best possible sensitivity that might be achieved. The dashed line shows the curves when a certain fraction of CXB is resolved.
The range of temperatures shown in the plots covers the most relevant regime for the WHIM of $T\sim 10^6\,{\rm K}$ (see Fig.~\ref{fig:n_t_boxes}). At temperatures lower than  $\sim 10^5\,{\rm K}$ (not shown), the sensitivity is essentially constant. On the high-temperature end, the plots show the transition region to hotter ICM in groups and clusters of galaxies. There, the sensitivity increases as the temperature rises to $\sim (2-3)\times 10^6\,{\rm K}$  and then levels off. } 
\label{fig:sens}
\end{figure*}

In Fig.~\ref{fig:sens} we show the $3\sigma$ values expressed in terms of ${\rm EM_{\it l}}$ (blue curves) and the Thomson optical depth (red curves). All curves with different metalicities are scaled to $Z/Z_\odot=0.2$. Clearly, the error on the filament strength scales inversely with $Z$. This result is a natural consequence of the fact that much of the signal is associated with spectral signatures of metals. 
It is also clear that there are two different regimes and low and high temperatures, respectively. 

At low temperatures ($T\lesssim 10^6\,{\rm K}$), the production of photons via excitation of important transition by electron collisions is subdominant and the expected signal is dominated by CXB photons which are "transformed" by interactions with the WHIM, namely absorbed photons are re-emitted as the recombination radiation and fluorescent lines, while the resonant scattering changes the direction of photons in the lines of heavy elements, principally - oxygen. In this regime, the strength of the signal scales approximately linearly with $\tau_T$. For instance, for $T=10^5\,{\rm K}$ and  $(1+\delta)\gtrsim 10$ the fraction of OVII ions decreases with density
\citep[see, e.g.][]{2001MNRAS.323...93C,2003PASJ...55..879Y,2019MNRAS.482.4972K,2019MNRAS.488.2947W}
and the column density of these ions remains weakly sensitive to the actual gas density (see also Appendix~\ref{app:smodel}).

In the opposite limit ($T\gtrsim 10^6\,{\rm K}$), the collisional excitation/ionization can be much more efficient than the photoionization, provided that the gas density is not too low. As a result, even for moderately high densities, the dominant signal is largely due to thermal emission that scales as the density squared. The same scaling remains valid in the low-density limit when photoionization strongly dominates. As a result, the fraction of O~VII and O~VIII ions becomes a linear function of density, and, therefore, the total signal scales with the density squared.  Fig.~\ref{fig:sens} clearly illustrates this behavior: in terms of ${\rm EM_{\it l}}$, all sensitivity curves overlap with each other (once scaled with the abundance).

From Fig.~\ref{fig:sens} it follows that for eROSITA (with the exposure time accumulated in the CalPV phase towards the Coma cluster) the levels of the detectable signal corresponds to  ${\rm EM_{\it l}} \approx 10^{4}(Z/Z_\odot/0.2)\,{\rm Mpc}$ for $T\lesssim 2\times10^6\,{\rm K}$. This value exceeds expectations by $\sim$ order of magnitude. For LEM, exposure times $\sim 1$~Ms would bring the sensitivity down to   $\approx 10^{3}(Z/Z_\odot/0.2)\,{\rm Mpc}$, i.e. close to expectations. From Fig.~\ref{fig:sens} it is also clear that detecting $T\gtrsim 3\times10^6\,{\rm K}$ gas is much easier to detect than the cooler gas (see also Fig.~\ref{fig:diag} in the Appendix). Given the linear scaling of the signal with $Z$, the abundance of metals at the level of a few \% of the Solar values would make the detection of this gas extremely difficult.

\section{Coma cluster}
\label{sec:coma}
We now proceed with the comparison of the aforementioned spectral model and the estimates of ${\rm EM_{\it l}}$ from the \textit{SRG}/eROSITA observations of the Coma cluster.

\subsection{Observational data}

The \srg~ X-ray observatory \citep{2021A&A...656A.132S}  was launched on July 13, 2019,  from the Baikonur cosmodrome. It carries two wide-angle grazing-incidence X-ray telescopes, eROSITA \citep{2021A&A...647A...1P} and the Mikhail Pavlinsky ART-XC telescope \citep{2021A&A...650A..42P}, which operate in the overlapping energy bands of 0.2–8 and 4–30 keV, respectively.

The main dataset used here is similar to the one described in \citep[][hereafter Paper I]{2021A&A...651A..41C}. Briefly, dedicated \textit{SRG} observations of the Coma cluster were performed in two parts, on December 4-6, 2019, and June 16-17, 2020 in the course of Calibration and Performance Verification phases (hereafter CPV). The observatory was operating in a scanning mode when a rectangular region of the sky is scanned multiple times to ensure uniform exposure of the region at the level of $\sim 20\,{\rm ks}$ per point. For the spectral analysis, we use the data of five eROSITA telescopes equipped with the on-chip filter. These telescopes have a very similar response and are not prone to optical light leakage that might complicate the analysis of the soft part of the spectrum \citep[see][]{2021A&A...647A...1P}. 
 In addition, we used the data from the SRG/eROSITA All-sky Survey in order to determine the typical sky background level around the cluster.

\subsection{Results}
\begin{figure}
\centering
\includegraphics[angle=0,bb = 70 190 550 680,width=0.999\columnwidth]{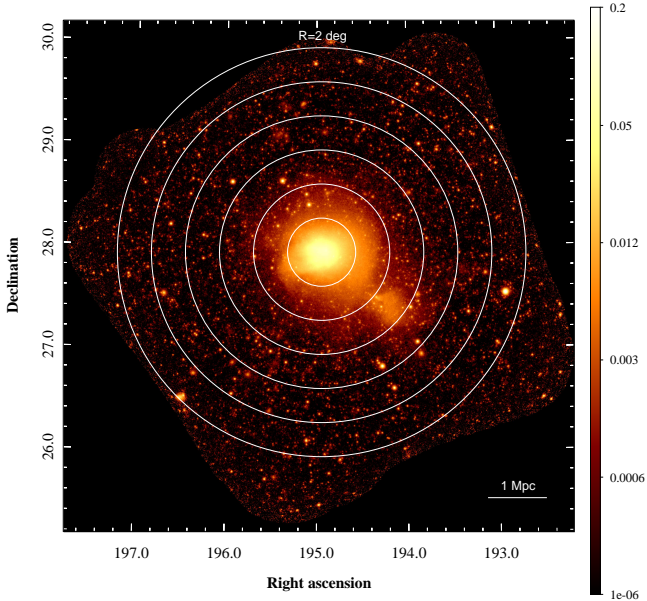}
\caption{X-rays image of the Coma cluster in the 0.4-2 keV band obtained by \textit{SRG}/eROSITA during Calibration and Performance Verification (CalPV) phase \citepalias{2021A&A...651A..41C}. A set of concentric annuli ($20'$-wide $\sim 560\,{\rm kpc}$ are used to extract spectra. The surface brightness is in units of ${\rm counts\,s^{-1}\,arcmin^{-2}}$ normalized per single module of seven eROSITA telescopes.
}
\label{fig:ximage}
\end{figure}

The spectra were extracted from a set of concentric annuli centered on the Coma cluster (see Fig.~\ref{fig:ximage}). The width of each annulus is $20'$ ($\sim 560\,{\rm kpc}$), so that the outermost annulus goes up to $\sim 3.4\,{\rm Mpc}$. 

\begin{figure}
\centering
\includegraphics[angle=0,bb = 70 240 540 640,width=1.\columnwidth]{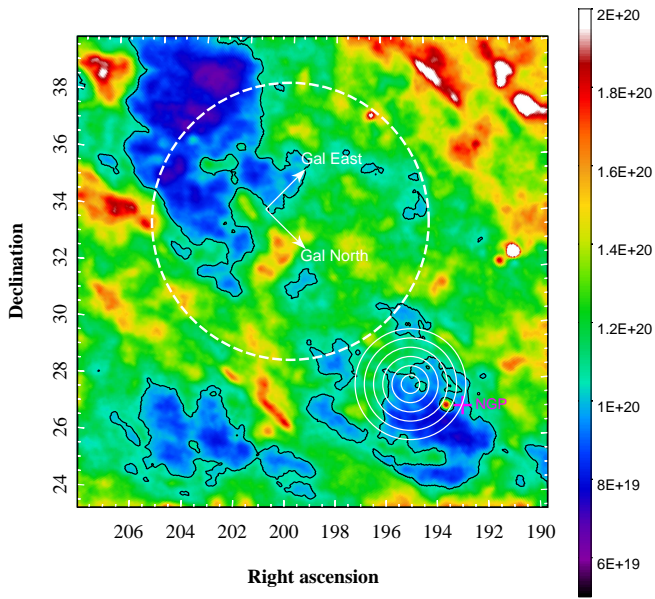}
\caption{Effective Galactic  hydrogen column density map with the Coma annuli (solid circles) and the background region (the dashed line, radius $5^\circ$) superposed. This map was derived from a combination of the HI 21~cm emission and the extinction maps, following the procedure described in \citep{2013MNRAS.431..394W}.   In the area covered by the Coma CalPV observations (concentric circles) and in the background region (big circle) the total effective column densities are very similar. The position of the Galactic North Pole is marked with the red cross, direction of the Galactic coordinates is indicated with the white compass.
}
\label{fig:nht}
\end{figure}

Since we are looking for signatures of faint emission from the $T\sim 10^6\,{\rm K}$ gas, the magnitude of Galactic interstellar absorption is very important in order to properly estimate the flux. Shown in Fig.~\ref{fig:nht} is the total effective column density (NH$_t$) of the Milky Way in the region $\sim 16 \times 16$ square degrees in size that includes the Coma cluster. The value of NH$_t$ was evaluated by combining the HI data from the HI4PI survey \citep{2016A&A...594A.116H} and $E(B-V)$ map  \cite{2015ApJ...798...88M} following the recipe described in  \cite{2013MNRAS.431..394W}.


The eROSITA background can be schematically decomposed into three components: (1) intrinsic detector  background, (2) X-ray background due to distant sources well beyond the boundaries of the Milky Way, e.g. AGN, clusters, galaxies, etc., and (3) X-ray background associated with the compact and diffuse sources in the Milky Way. The first component (detector background) is, of course, not affected by extinction. For the second component (distant objects) the total ${\rm NH}_t$ determines the absorption. The third component is the foreground (for Coma) and is mostly associated with the Milky Way soft diffuse emission, which itself consists of several components. This makes the background (foreground for Coma) model somewhat uncertain.

In the direction of Coma, the mean value of ${\rm NH}_t$ is $1.0\times 10^{21}\,{\rm cm^{-2}}$. We have selected a large region (radius = $5^\circ$) close to the Coma field (see  Fig.~\ref{fig:nht}), where the mean ${\rm NH}_t$ is essentially identical to the Coma field. Of course, the agreement in the total column density between two regions does not guarantee that the distribution of the absorbing material is the same, and, consequently, the impact of the diffuse Milky Way emission might be different. However, selecting the background region in the vicinity of the source region should minimize this effect.

This choice is also not without a flaw, however - according to Fig.~\ref{fig:eml_radial}, the excess surface brightness is rather flat between 1 and 10~Mpc from the cluster center. This means, that the background region is expected to contain some "excess" related to the Coma cluster too. Therefore, the signal in the difference between the source and background regions might be attenuated. We decided that for the purposes of this paper, the advantages of using the background region close to Coma outweigh the disadvantages. The size of the background region is large enough so that the cumulative exposure (of the all-sky survey data) is comparable with that of the Coma rings and, therefore, the noise introduced by the background does not increase the uncertainties in measured net Coma fluxes significantly.   

Apart from the problem of selecting the background field, it is desirable to remove bright X-ray sources unrelated to the soft diffuse emission. The main issue here is not resolving a significant fraction of the background, but rather suppressing the level of fluctuations caused by fluctuations of the number of objects (i.e. the shot noise and even their intrinsic variability) in the Coma and background regions. Since the background region is covered in the survey, where the exposure time is much smaller than in the CalPV data in the Coma field, we selected a rather high threshold for excising compact sources, namely $F_X>10^{-13}\,{\rm erg\,cm^{-2}\,s^{-1}}$ in the 0.5-2~keV band to ensure a uniform level of the background resolved fraction. 

From the LogN-LogS curve of AGN obtained by \textit{Chandra} \citep[see, e.g.][]{2017ApJS..228....2L}, we estimate that about 16\% of the flux, associated with AGN, will be resolved, and the level of fluctuations of the remaining AGN flux is $\sim$5\% on scales of order 1~${\rm deg^2}$. This estimate corresponds to pure Poissonian fluctuations of the number of sources and does not include other terms. In addition, we have removed four regions manually. Namely, a bright and variable star 41~Com, two highly variable AGNs, and a $20'$ circle (radius) covering a compact (in projection) group of AGNs. None of these excised regions have a strong impact on the results.



\begin{figure}
\centering
\includegraphics[angle=0,trim=1cm 5cm 0cm 2cm,width=0.99\columnwidth]{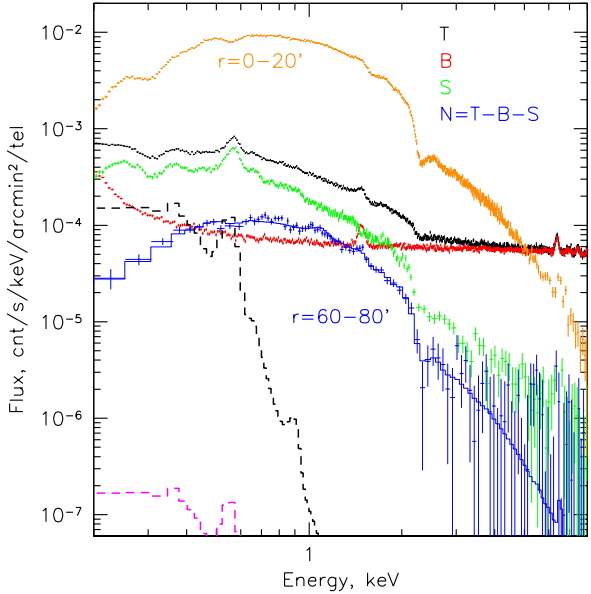}
\caption{eROSITA spectrum of the $60-80'$ ring around the Coma cluster. The spectra are presented as count rate per square arcminute and per one (out of seven) eROSITA telescope module, hence the units are ${\rm cnt/s/keV/arcmin}^2/{\rm tel}$.
Different spectra show the total measured spectrum (T, black), detector internal background spectrum (B, red), sky spectrum without detector background, extracted from the circular "background" region (S, green), net spectrum in the direction of the Coma cluster (N=T-B-S, blue). For comparison, the orange points show the background-subtracted spectrum extracted from the innermost ring $0-20'$. 
The blue histogram shows the best fitting single temperature APEC model with fixed absorption ${\rm NH}=1.03\times10^{20}\,{\rm cm^{-2}}$ and $kT=3.3\pm0.2\,{\rm keV}$. In terms of projected emission measure, the normalization of the observed Coma spectrum in the $60-80'$ ring corresponds to ${\rm EM_{\it l}}=2.1\times 10^5\,{\rm Mpc}$.
For comparison, the black dashed line shows the APEC spectrum with $T=10^6\,{\rm K}$, abundance $Z/Z_\odot=0.2$ and ${\rm EM_{\it l}}=9 \times10^5$~Mpc. This would correspond to the normalization suggested in \citep{2022MNRAS.514..416B}. The magenta dashed line shows the same spectrum with ${\rm EM_{\it l}}=10^3$~Mpc motivated by the Magneticum simulations. We note here that WHIM signatures, especially for low overdensities and temperatures can be substantially different from the APEC model. Also, the true WHIM metallicity can be much lower than  $Z=0.2$ adopted for this plot.
}
\label{fig:ring}
\end{figure}

The spectra were extracted for each of the annuli shown in Fig.~\ref{fig:ximage} and corrected for the particle and sky background. The errors were calculated following the procedure outlined in \cite{1996ApJ...471..673C}. Namely, to avoid the bias due to the low number of counts per bin when minimizing $\chi^2$ for unbinned spectra  (provided that the total number of counts associated with every independent spectral component is large), the errors were calculated from the smoothed spectra.  An example of various components (including detector intrinsic background and sky background) contributing to the total observed spectrum is shown in Fig.~\ref{fig:ring}. The spectra for the  annulus between 60$'$ and 80$'$ ($\sim$1.7-2.2~Mpc) were used.  

Clearly, for this annulus, both particle and sky backgrounds make very large contributions. The "cleaned" spectrum, i.e. corrected for all backgrounds, is shown with blue crosses. For comparison, the histogram (blue) shows the best-fitting single-temperature APEC model. In this model, the interstellar absorption, redshift, and abundance of heavy elements were fixed at $NH=1.0\times 10^{20}\,{\rm cm^{-2}}$, $z=0.0231$ and $Z/Z_\odot=0.2$, respectively. The best-fitting temperature is $\sim 3.3\,{\rm keV}$. Clearly, at the level of accuracy provided by these data, the one-temperature model appears to be sufficient to describe the data in this annulus. 

For comparison, the black and magenta dashed lines show the APEC model with $T=10^6\,{\rm K}$, $Z/Z_\odot=0.2$ and normalizations corresponding to ${\rm EM_{\it l}}\approx 10^6$ and $10^3$ Mpc, respectively. The former corresponds to the normalization suggested in \citep{2022MNRAS.514..416B}, i.e. $\delta\sim300$ and $L\sim 10$~Mpc, while the latter with ${\rm EM_{\it l}}=10^3$~Mpc is motivated by the \texttt{Magneticum} simulations.

Fig.~\ref{fig:ring} illustrates the difficulties of measuring the emission of the filaments in X-rays. In the inner regions (exemplified by the orange data points), where the Coma emission is bright, the expected signal is very small compared to the flux from the hot and dense Coma core. The response of the instrument (in particular, its off-diagonal part) has to be extremely well-calibrated to place reliable constraints on the WHIM signal. At larger radii, the Coma contribution becomes subdominant, but the sky background dominates (Fig.~\ref{fig:ring}). In this case, the uncertainties in response are less important, but now the background level has to be calibrated to high accuracy. These requirements make the whole exercise very challenging. As we see below, both issues are important.

To estimate a possible contribution of the WHIM signal to the observed Coma spectra, we used a model consisting of two components. The first component is the APEC model which represents the emission associated with the cluster itself. The second component is the "on-off" WHIM spectrum considered in Section~\ref{sec:smodel}. We fixed all major parameters of the WHIM model: metal abundance $Z=0.2$ \citep[abundance table of][]{2003ApJ...591.1220L}, temperature $T=1.5\times 10^6\,{\rm K}$, gas density $(1+\delta)=30$, so that the temperature and density follow the correlation of the most widespread WHIM phases (see the diagonal line in Fig.~\ref{fig:n_t_boxes}). This choice of $T$ and $(1+\delta)$ is rather arbitrary, but it serves the purpose of illustrating the characteristic sensitivity of the present data set to the potential WHIM emission.
The same metal abundance $Z=0.2$  was assumed for the \texttt{APEC} model. The effective column density of the Galactic absorption is fixed at  ${\rm NH}=1.0\times 10^{20}\,{\rm cm^{-2}}$.

\begin{figure}
\centering
\includegraphics[angle=0,trim=2cm 5cm 0cm 2cm,width=0.99\columnwidth]{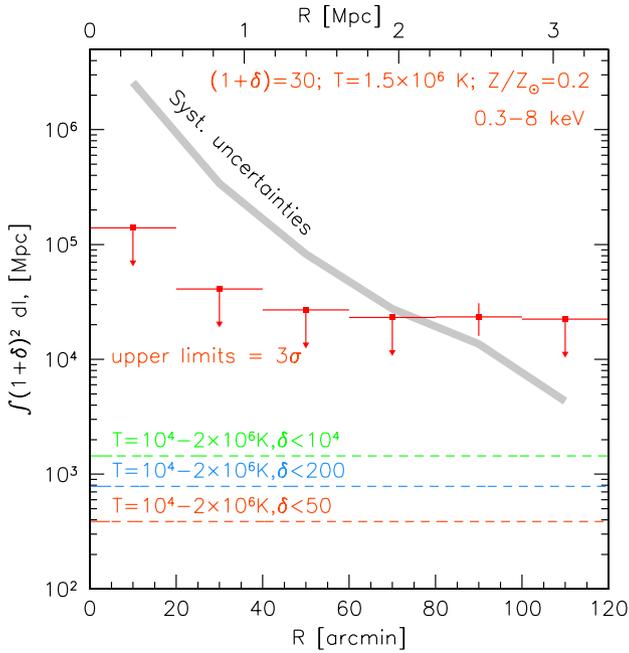}
\caption{Constraints on the WHIM signal in a set of concentric annuli around the Coma cluster (red points are mostly $3\sigma$ upper limits, except for one point is 3.1$\sigma$ level). In the Coma center, present-day systematic uncertainties in the spectral response (gray curve) exceed the statistical upper limits for the central 60' by a large factor, while beyond $R_{200c}\sim74'$ statistical uncertainties dominate. For comparison, the colored horizontal lines show the typical level of the WHIM signal expected from the \texttt{Magneticum} simulations.
} 
\label{fig:eml_erosita}
\end{figure}

The resulting model has three free parameters: the normalizations of both components and the temperature of the cluster emission. The resulting constraints on the emission measure associated with the WHIM component are shown as $3\sigma$ upper limits (red crosses)  in Fig.~\ref{fig:eml_erosita}. For one radial bin, the best-fitting normalization of the WHIM component is just above the $3\sigma$ level, but it could well be the result of a too simplistic model rather than the evidence for the WHIM signal. For comparison, the colored dashed lines show levels of the excess emission measure associated with the WHIM predicted by simulations (see \S\ref{sec:smodel} and Appendix). It is clear that even for a very optimistic value of the WHIM metallicity $Z=0.2$, the upper limits are more than one order of magnitude above the expectations. 

The thick gray line in Fig.~\ref{fig:eml_erosita} shows the (crudely estimated) level of systematic uncertainties associated with existing calibration uncertainties of the low-energy tail of the spectral response. These systematic uncertainties scale with the brightness of the cluster emission and, therefore, increase dramatically in the inner region.  Other sources of systematic errors include the uncertainties in the stray light contribution \cite[see, e.g.][]{2022arXiv220507511C} and time variable Solar Wind Charge Exchange emission. However, given that the pure statistical noise is much larger than the anticipated signal, the additional noise introduced by these uncertainties is not crucially important. 

We reiterate here that given the discussion of the contaminating emission from the cluster itself (and its immediate outskirts) in \S\ref{sec:cont}, the \textit{SRG}/eROSITA CalPV observations of the Coma cluster may not be the most promising data for the WHIM search since they cover only the region within $\sim 3$~Mpc from the cluster center. In reality parts of the region that we use here as the "background" field (see Fig.~\ref{fig:nht}) might be better suited for this purpose. Nevertheless, typical filaments are clearly too faint for a secure detection with eROSITA as illustrated by Fig.~\ref{fig:eml_erosita}.

\section{Discussion}
\label{sec:discussion}
The detection of WHIM signatures is particularly interesting on two grounds. First of all, a significant fraction of baryons can be "hidden" in a low-density warm gas \citep[e.g.][]{1999ApJ...514....1C}. Secondly, the enrichment of this gas with metals is sensitive to the physical processes that control the escape of metals from halos where most of the star formation takes place \citep[e.g.][]{2017MNRAS.468..531B}. Since the WHIM signal in X-rays is primarily sensitive to the spectral features induced by metals, the detection of WHIM would help answer these questions.  

The metal distribution predicted in the \texttt{Magneticum} simulations is rather complex (see Appendix~\ref{app:metals}). In particular, all metal-rich particles (i.e. gas mass elements) are confined to regions within a few virial radii around halos, while it is "zero-metallicity" gas that forms a network of filaments. This does not necessarily mean that the filaments are made of pure hydrogen and helium. Rather it shows that the feedback and enrichment models implemented in \texttt{Magneticum} and shown to be able to reproduce observed properties of galaxies, groups, and clusters (including their stellar masses and metal contents) might produce close-to-zero metallicity of the diffuse intergalactic medium with eWHIM properties. Therefore,  the detection of the WHIM metal signatures with a sensitive X-ray instrument, providing a lower limit on the gas metallicity, can be used as a diagnostic of the actual enrichment process. Similarly, upper limits on the gas metallicity can be used to constrain models characterized by stronger feedback spreading the metals over larger IGM volumes.

From the observational point of view, traditionally, two approaches for the search for the WHIM signal in X-rays are considered: either via absorption lines in the spectra of a bright distant AGN \citep[see, e.g.][for the recent review]{2022arXiv220315666N} or via diffuse emission \citep[][]{2001MNRAS.323...93C, 2008SSRv..134..405P,2010MNRAS.407..544B,2011ApJ...734...91T,2018MNRAS.476.4629P, 2019MNRAS.482.4972K,2019A&A...627A...5V,2022arXiv220900657P}, including stacking over large sky areas \citep[][]{2020A&A...643L...2T,2022A&A...667A.161T}. In the former case, a serendipitous intersection of a WHIM structure by the line of sight towards the bright source is used to probe the WHIM. In this case, the brightness of the background objects is the main criterion.
In the latter case, the region selected by the overdensity of galaxies (or of other tracer objects) is used as a proxy for a possible WHIM patch. In particular, the identification of a long and massive filament oriented approximately along the line of sight is a viable strategy \citep[e.g.][]{1999ApJ...522L..13M}. 

In this study we mostly advocate for another approach, namely using regions enclosed by the turn-around radius of massive clusters, where, by definition, the density of WHIM gas should be higher than in other locations. This approach has a few attractive features: (i) the redshift of the expected signal is known and (ii) the size of the affected region is large enough so that the emission from the cluster itself does not contaminate the WHIM signal, and (iii) any cluster can be used as a target and (iv) typical rather exceptional WHIM patches are probed.

Along the same line, the analysis of regions around massive clusters can be used to generate two spectra: the combined spectrum of all resolved AGN more distant than the cluster and the remaining "unresolved" spectrum of the diffuse emission (see Appendix~\ref{app:smodel}). While the former spectrum contains absorption signatures
Joint analysis of the two spectra not only increases the significance of the WHIM signal detection but also allows one to better constrain the temperature and density of the gas constituting the filament. As in other cases, removing the contaminating signal from smaller halos is a serious issue.

This  approach implies specific requirements on the instrument characteristics. Since the aim is to detect faint diffuse emission spread over substantial areas of the sky at low redshifts, the best-suited future X-ray mission would need to have a large grasp (i.e. the product of the effective area and size of the field of view), high spectral resolution and stable instrumental background (at least in terms of spurious spectral features). While the large grasp allows obtaining sensitive maps of large areas in a reasonable amount of time, the latter two requirements ensure good control over systematic uncertainties in the estimation of the astrophysical and instrumental background signals. 

Examples of such missions are \textit{ATHENA} \citep{2013arXiv1306.2307N} and the wide-field X-ray microcalorimeter \textit{LEM} \citep[][]{2022arXiv221109827K}. For nearby objects that subtend large solid angles in the sky, the grasp, i.e. the product of the effective area and the size of the field of view matters most, while for distant objects, the effective area is more important. Therefore, \textit{LEM} with a factor $\sim 10$ larger grasp is more promising for $z\sim 0$ objects, while  \textit{ATHENA} with $\sim 6$ times larger effective area is more suitable for the more distant ones \citep[e.g.][]{2019A&A...627A...5V,2022arXiv220900657P}.

In principle, yet another effect that might boost the WHIM signal in the vicinity of a cluster is the illumination of the gas by an additional X-ray flux (on top of the CXB) that might come from an AGN \citep[e.g.][]{2001MNRAS.323...93C} or the cluster itself \citep[e.g.][]{2022MNRAS.515.3162S}. This affects the ionization fractions, but more importantly, the extra illumination boosts the resonant scattering signal. The most advantageous configuration is when the illumination is coming from the side rather than from behind the WHIM patch so that the illuminating flux does not contaminate the WHIM signal. 
One can consider two flavors of this boost factor. In the first case, the illuminating flux contains a narrow resonant emission line scattered in the WHIM region. This is a standard case of the resonant scattering that can boost the signal from cluster outskirts \citep[e.g.][]{1987SvAL...13....3G,2013MNRAS.435.3111Z}. 
In the second case, the illuminating spectrum is featureless like the CXB. The only difference between these two cases is that a fraction of the total illuminating flux that can be scattered is higher in the former case.

The obvious condition for the importance of the resonant scattering effects with the "side illumination" is that the flux density of the illuminating source across the width of the line is comparable (or exceeds) that of the CXB. In the case of clusters like Coma, that lack prominent emission lines below a few keV, it is sufficient to compare the cluster surface brightness within a given radius $I_{cl} (<R_{cl})$ with that of the CXB ($I_{CXB}$) at the same energy, i.e.
\begin{eqnarray}
r\lesssim \left ( \frac{I_{cl}}{I_{CXB}}\right )^{1/2} \frac{R_{cl}}{2}\sim 2.7\,{\rm Mpc}.
\end{eqnarray}
Clearly, this is still a close vicinity of the cluster, but it might be important when studying the outskirts. In this case, resonant lines will be present in emission even if CXB sources are not resolved. 

If the illumination is provided by an AGN, a more convenient estimate is via its luminosity at ($\nu L_\nu$) at the line energy \citep{2001MNRAS.323...93C}
\begin{eqnarray}
r\lesssim 11\,{\rm Mpc} \left ( \frac{L}{10^{45}\,{\rm erg\,s^{-1}}} \right )^{1/2}.
\end{eqnarray}
A further variation of the same type is a beamed source, e.g. a blazar, that is pointing away from the observer and could illuminate a conical region in the WHIM.

\section{Conclusions}
\label{sec:conclusions}

We describe in full detail spectral signatures associated with the Warm-Hot Intergalactic Medium (WHIM) in the soft X-ray band and assess the prospects of its detection and diagnostic.

We factorize the strength of the WHIM signal into several components. The amount of the gas is quantified via the quantities (i) ${\rm EM_{\it l}}=\int (1+\delta)^2 dl$ (in units of length), where $(1+\delta)$ is the ratio of baryon density to the cosmic mean value and/or (ii) the Thomson optical depth $\tau_T=\int (1+\delta)\langle n_e\rangle\sigma_T dl$. The density and temperature of the gas determine the shape of the spectral distortions, while the net signal amplitude is proportional to the abundance of metals in the WHIM (Section~\ref{sec:mass}), except for a small contribution from a pure H+He bremsstrahlung and the Thomson scattering.

We further extend our previous models \cite{2001MNRAS.323...93C,2019MNRAS.482.4972K} of spectral distortions associated with WHIM. The model takes into account collisional and photo-ionisation (by the CXB photons), photoelectric absorption, resonant scattering, fluorescence, and local production of X-ray photons. The model predicts the difference of the spectrum in the direction of a "WHIM patch" and from the "empty" region. It takes into account the fraction of the CXB resolved into individual sources and can simultaneously describe the absorption signal in the resolved spectrum and the total signal in the diffuse part (Section~\ref{sec:smodel} and Appendix~\ref{app:smodel} ). 

At low temperatures ($T\lesssim 10^6\,{\rm K}$), the production of photons via excitation of important transition by electron collisions is subdominant and the expected signal is dominated by CXB photons which are "transformed" by interactions with the WHIM, namely absorbed photons are re-emitted as the recombination radiation and fluorescent lines, while the resonant scatterings change the direction of photons in the lines of heavy elements, principally - oxygen. In this regime, the strength of the signal scales linearly with $\tau_T$ and gas metallicity. 

At higher temperatures and moderate overdensities ($\sim 10$), the excitation of major transitions by electrons becomes significant and the strength of the signal scales with  ${\rm EM_{\it l}}$. As before, the scaling with metallicity is linear as well.

Motivated by the results of the \texttt{Magneticum} simulations and the analysis of expected spectral distortions, we define a specific region in the density-temperature diagram, which is the focus of our studies. Namely, we selected the gas with an extended (compared to other definitions of WHIM) range of temperatures
$10^4\,{\rm K}<T<2\times10^6 \,{\rm K}$ and the density $(1+\delta)\lesssim 50$. To distinguish it from more restrictive WHIM definitions, we call this gas eWHIM50 (Section~\ref{sec:exp}). 

One promising strategy for finding WHIM (and, in particular, eWHIM50) is by focusing on the region within the turnaround radius $r_{ta}\sim 10\,{\rm Mpc}$ of a massive cluster and avoiding the central region within $\sim 0.5 r_{ta}$. We found that in the resulting $\sim5-10$~Mpc annulus, the probability of finding localized patches with ${\rm EM_{\it l}}\gtrsim 10^3\,{\rm Mpc}$ is $\sim 10$\%. In terms of $\tau_T$, the excess associated with these patches is $\sim 2\times10^{-5}$. The annulus-averaged value of the ${\rm EM_{\it l}}$ is a factor of a few lower. These levels are within reach of future high spectral resolution missions like \textit{LEM}\citep[][]{2022arXiv221109827K}, although this critically depends on the metallicity of the actual metallicity of the WHIM gas, which might not be tightly constrained in the simulations (Section~\ref{sec:where}). 

A similar strategy of probing the volume inside $r_{ta}$ but still far from the virial or splashback radii of a massive cluster, could be used to study other types of objects in "representative" volumes of the Universe at well-defined redshifts (=redshift of the cluster). At least the objects for which the overdensity of matter of order 2-3 does not strongly affect their properties, but simply increases their space density compared to the field.

The level of contamination of the eWHIM50 signal by regions with larger overdensities has been estimated by comparing in projected images the contribution of the $(1+\delta)\lesssim 50$ gas to the ${\rm EM_{\it l}}$ with the total ${\rm EM_{\it l}}$ provided by the gas at any overdensity. The condition that the low-density gas contributes more than half of the total ${\rm EM_{\it l}}$ excludes much of the region within 5~Mpc from the cluster center, but less than $\sim50$\% of the area in the 5-10~Mpc annulus (and $\sim 20$\% of the projected image of the entire cluster box, see Fig.~\ref{fig:eml_masked}). This suggests that it should be possible to find uncontaminated regions in real data, although a practical recipe for doing this selection has to be identified (Section~\ref{sec:cont}).  

If the abundance of metals in the eWHIM50 gas (located well beyond the virial radii of halos) is at the level of $Z/Z_\odot\sim 10^{-2}$, the prospects of detecting this gas with X-ray spectrometers are rather pessimistic. The abundance of metals does not affect the dispersion measure and/or SZ signals as possible proxies for the presence of the gas, although in this case, the lack of a precise redshift "tag" complicates the detection of individual structures.

Considering the Coma cluster that was observed by \textit{SRG}/eROSITA during the CalPV phase, combining them with shallower all-sky survey data for background estimation,  we concluded that in the central region ($\lesssim 40'$) the current level of the calibration uncertainties precludes the robust characterization of the soft emission potentially associated with the eWHIM50 signal on top of the bright cluster emission. In the outer regions ($40-120'$), the impact of the systematic uncertainties is less prominent. There, 3 sigma upper limits are about one order of magnitude higher than the typically expected levels for the eWHIM50 gas  (Section~\ref{sec:coma}).  

Finally, we note that here we are specifically targeting typical filaments/overdense regions. This does not exclude the possibility of finding rare objects, e.g. exceptionally massive filaments oriented along the line of sight, that have much stronger integral WHIM signal. On contrary, the approach proposed here can be applied to the surroundings of any massive cluster, in particular, those located at the redshifts optimizing detection prospects versus bright Galactic foreground emission. We show that e.g. proposed soft X-ray microcalorimetric mission \textit{LEM} might be able to detect WHIM signal from a turnaround radius of a massive cluster in the redshift window around $z\approx0.08$, given that of the WHIM gas is enriched to $\sim$0.1 of the solar value.

\section*{Acknowledgments}
We are grateful to our reviewer for useful comments and suggestions.
This work is partly based on observations with the eROSITA telescope onboard \textit{SRG} space observatory. The \textit{SRG} observatory was built by Roskosmos in the interests of the Russian Academy of Sciences represented by its Space Research Institute (IKI) in the framework of the Russian Federal Space Program, with the participation of the Deutsches Zentrum für Luft- und Raumfahrt (DLR). The eROSITA X-ray telescope was built by a consortium of German Institutes led by MPE, and supported by DLR. The \textit{SRG} spacecraft was designed, built, launched, and is operated by the Lavochkin Association and its subcontractors. The science data are downlinked via the Deep Space Network Antennae in Bear Lakes, Ussurijsk, and Baikonur, funded by Roskosmos. The eROSITA data used in this work were converted to calibrated event lists using the eSASS software system developed by the German eROSITA Consortium and analysed using proprietary data reduction software developed by the Russian eROSITA Consortium.

IK and KD acknowledge support by the COMPLEX project from the European Research Council (ERC) under the European Union’s Horizon 2020 research and innovation program grant agreement ERC-2019-AdG 882679. KD acknowledges support by the Deutsche Forschungsgemeinschaft (DFG, German Research Foundation) under Germany’s Excellence Strategy - EXC-2094 - 390783311. The calculations for the hydrodynamical simulations were carried out at the Leibniz Supercomputer Center (LRZ) under the project pr83li and we are especially grateful for the support through the Computational Center for Particle and Astrophysics (C2PAP). 

\section*{Data availability}
X-ray data analysed in this article were used by permission of the Russian \textit{SRG}/eROSITA consortium. The data will become publicly available as a part of the corresponding \textit{SRG}/eROSITA data release along with the appropriate calibration information. All other data are publicly available and can be accessed at the corresponding public archive servers.



\bibliographystyle{mnras}
\bibliography{ref} 

\begin{thebibliography}{}
\makeatletter
\relax
\def\mn@urlcharsother{\let\do\@makeother \do\$\do\&\do\#\do\^\do\_\do\%\do\~}
\def\mn@doi{\begingroup\mn@urlcharsother \@ifnextchar [ {\mn@doi@}
  {\mn@doi@[]}}
\def\mn@doi@[#1]#2{\def\@tempa{#1}\ifx\@tempa\@empty \href
  {http://dx.doi.org/#2} {doi:#2}\else \href {http://dx.doi.org/#2} {#1}\fi
  \endgroup}
\def\mn@eprint#1#2{\mn@eprint@#1:#2::\@nil}
\def\mn@eprint@arXiv#1{\href {http://arxiv.org/abs/#1} {{\tt arXiv:#1}}}
\def\mn@eprint@dblp#1{\href {http://dblp.uni-trier.de/rec/bibtex/#1.xml}
  {dblp:#1}}
\def\mn@eprint@#1:#2:#3:#4\@nil{\def\@tempa {#1}\def\@tempb {#2}\def\@tempc
  {#3}\ifx \@tempc \@empty \let \@tempc \@tempb \let \@tempb \@tempa \fi \ifx
  \@tempb \@empty \def\@tempb {arXiv}\fi \@ifundefined
  {mn@eprint@\@tempb}{\@tempb:\@tempc}{\expandafter \expandafter \csname
  mn@eprint@\@tempb\endcsname \expandafter{\@tempc}}}

\bibitem[\protect\citeauthoryear{{Anders} \& {Grevesse}}{{Anders} \&
  {Grevesse}}{1989}]{1989GeCoA..53..197A}
{Anders} E.,  {Grevesse} N.,  1989, \mn@doi [\gca]
  {10.1016/0016-7037(89)90286-X}, \href
  {https://ui.adsabs.harvard.edu/abs/1989GeCoA..53..197A} {53, 197}

\bibitem[\protect\citeauthoryear{{Angelinelli}, {Ettori}, {Dolag}, {Vazza}  \&
  {Ragagnin}}{{Angelinelli} et~al.}{2022}]{2022A&A...663L...6A}
{Angelinelli} M.,  {Ettori} S.,  {Dolag} K.,  {Vazza} F.,   {Ragagnin} A.,
  2022, \mn@doi [\aap] {10.1051/0004-6361/202244068}, \href
  {https://ui.adsabs.harvard.edu/abs/2022A&A...663L...6A} {663, L6}

\bibitem[\protect\citeauthoryear{{Arnaud}}{{Arnaud}}{1996}]{1996ASPC..101...17A}
{Arnaud} K.~A.,  1996, in {Jacoby} G.~H.,  {Barnes} J.,  eds,  Astronomical
  Society of the Pacific Conference Series Vol. 101, Astronomical Data Analysis
  Software and Systems V. p.~17

\bibitem[\protect\citeauthoryear{{Beck} et~al.,}{{Beck}
  et~al.}{2016}]{2016MNRAS.455.2110B}
{Beck} A.~M.,  et~al., 2016, \mn@doi [\mnras] {10.1093/mnras/stv2443}, \href
  {https://ui.adsabs.harvard.edu/abs/2016MNRAS.455.2110B} {455, 2110}

\bibitem[\protect\citeauthoryear{{Bertone}, {Schaye}, {Dalla Vecchia}, {Booth},
  {Theuns}  \& {Wiersma}}{{Bertone} et~al.}{2010}]{2010MNRAS.407..544B}
{Bertone} S.,  {Schaye} J.,  {Dalla Vecchia} C.,  {Booth} C.~M.,  {Theuns} T.,
   {Wiersma} R. P.~C.,  2010, \mn@doi [\mnras]
  {10.1111/j.1365-2966.2010.16932.x}, \href
  {https://ui.adsabs.harvard.edu/abs/2010MNRAS.407..544B} {407, 544}

\bibitem[\protect\citeauthoryear{{Bertschinger}}{{Bertschinger}}{1985}]{1985ApJS...58...39B}
{Bertschinger} E.,  1985, \mn@doi [\apjs] {10.1086/191028}, \href
  {https://ui.adsabs.harvard.edu/abs/1985ApJS...58...39B} {58, 39}

\bibitem[\protect\citeauthoryear{{Biffi} et~al.,}{{Biffi}
  et~al.}{2017}]{2017MNRAS.468..531B}
{Biffi} V.,  et~al., 2017, \mn@doi [\mnras] {10.1093/mnras/stx444}, \href
  {https://ui.adsabs.harvard.edu/abs/2017MNRAS.468..531B} {468, 531}

\bibitem[\protect\citeauthoryear{{Bonamente}, {Mirakhor}, {Lieu}  \&
  {Walker}}{{Bonamente} et~al.}{2022}]{2022MNRAS.514..416B}
{Bonamente} M.,  {Mirakhor} M.,  {Lieu} R.,   {Walker} S.,  2022, \mn@doi
  [\mnras] {10.1093/mnras/stac1318}, \href
  {https://ui.adsabs.harvard.edu/abs/2022MNRAS.514..416B} {514, 416}

\bibitem[\protect\citeauthoryear{{Bregman} et~al.,}{{Bregman}
  et~al.}{2019}]{2019BAAS...51c.450B}
{Bregman} J.,  et~al., 2019, \mn@doi [\baas] {10.48550/arXiv.1903.11630}, \href
  {https://ui.adsabs.harvard.edu/abs/2019BAAS...51c.450B} {51, 450}

\bibitem[\protect\citeauthoryear{{Cen} \& {Ostriker}}{{Cen} \&
  {Ostriker}}{1999}]{1999ApJ...514....1C}
{Cen} R.,  {Ostriker} J.~P.,  1999, \mn@doi [\apj] {10.1086/306949}, \href
  {https://ui.adsabs.harvard.edu/abs/1999ApJ...514....1C} {514, 1}

\bibitem[\protect\citeauthoryear{{Churazov}, {Gilfanov}, {Forman}  \&
  {Jones}}{{Churazov} et~al.}{1996}]{1996ApJ...471..673C}
{Churazov} E.,  {Gilfanov} M.,  {Forman} W.,   {Jones} C.,  1996, \mn@doi
  [\apj] {10.1086/177997}, \href
  {https://ui.adsabs.harvard.edu/abs/1996ApJ...471..673C} {471, 673}

\bibitem[\protect\citeauthoryear{{Churazov}, {Haehnelt}, {Kotov}  \&
  {Sunyaev}}{{Churazov} et~al.}{2001}]{2001MNRAS.323...93C}
{Churazov} E.,  {Haehnelt} M.,  {Kotov} O.,   {Sunyaev} R.,  2001, \mn@doi
  [\mnras] {10.1046/j.1365-8711.2001.04090.x}, \href
  {https://ui.adsabs.harvard.edu/abs/2001MNRAS.323...93C} {323, 93}

\bibitem[\protect\citeauthoryear{{Churazov}, {Khabibullin}, {Lyskova},
  {Sunyaev}  \& {Bykov}}{{Churazov} et~al.}{2021}]{2021A&A...651A..41C}
{Churazov} E.,  {Khabibullin} I.,  {Lyskova} N.,  {Sunyaev} R.,   {Bykov}
  A.~M.,  2021, \mn@doi [\aap] {10.1051/0004-6361/202040197}, \href
  {https://ui.adsabs.harvard.edu/abs/2021A&A...651A..41C} {651, A41{~(Paper
  I)}}

\bibitem[\protect\citeauthoryear{{Churazov}, {Khabibullin}, {Bykov}, {Lyskova}
  \& {Sunyaev}}{{Churazov} et~al.}{2022}]{2022arXiv220507511C}
{Churazov} E.,  {Khabibullin} I.,  {Bykov} A.~M.,  {Lyskova} N.,   {Sunyaev}
  R.,  2022, \mn@doi [arXiv e-prints] {10.48550/arXiv.2205.07511}, \href
  {https://ui.adsabs.harvard.edu/abs/2022arXiv220507511C} {p. arXiv:2205.07511}

\bibitem[\protect\citeauthoryear{{Dav{\'e}} et~al.,}{{Dav{\'e}}
  et~al.}{2001}]{2001ApJ...552..473D}
{Dav{\'e}} R.,  et~al., 2001, \mn@doi [\apj] {10.1086/320548}, \href
  {https://ui.adsabs.harvard.edu/abs/2001ApJ...552..473D} {552, 473}

\bibitem[\protect\citeauthoryear{{Diemer} \& {Kravtsov}}{{Diemer} \&
  {Kravtsov}}{2014}]{2014ApJ...789....1D}
{Diemer} B.,  {Kravtsov} A.~V.,  2014, \mn@doi [\apj]
  {10.1088/0004-637X/789/1/1}, \href
  {https://ui.adsabs.harvard.edu/abs/2014ApJ...789....1D} {789, 1}

\bibitem[\protect\citeauthoryear{{Dolag}, {Komatsu}  \& {Sunyaev}}{{Dolag}
  et~al.}{2016}]{2016MNRAS.463.1797D}
{Dolag} K.,  {Komatsu} E.,   {Sunyaev} R.,  2016, \mn@doi [\mnras]
  {10.1093/mnras/stw2035}, \href
  {https://ui.adsabs.harvard.edu/abs/2016MNRAS.463.1797D} {463, 1797}

\bibitem[\protect\citeauthoryear{{Dolag}, {Mevius}  \& {Remus}}{{Dolag}
  et~al.}{2017}]{2017Galax...5...35D}
{Dolag} K.,  {Mevius} E.,   {Remus} R.-S.,  2017, \mn@doi [Galaxies]
  {10.3390/galaxies5030035}, \href
  {https://ui.adsabs.harvard.edu/abs/2017Galax...5...35D} {5, 35}

\bibitem[\protect\citeauthoryear{{Ferland} et~al.,}{{Ferland}
  et~al.}{2017}]{2017RMxAA..53..385F}
{Ferland} G.~J.,  et~al., 2017, \mn@doi [\rmxaa] {10.48550/arXiv.1705.10877},
  \href {https://ui.adsabs.harvard.edu/abs/2017RMxAA..53..385F} {53, 385}

\bibitem[\protect\citeauthoryear{{Fillmore} \& {Goldreich}}{{Fillmore} \&
  {Goldreich}}{1984}]{1984ApJ...281....1F}
{Fillmore} J.~A.,  {Goldreich} P.,  1984, \mn@doi [\apj] {10.1086/162070},
  \href {https://ui.adsabs.harvard.edu/abs/1984ApJ...281....1F} {281, 1}

\bibitem[\protect\citeauthoryear{{Gal{\'a}rraga-Espinosa}, {Aghanim}, {Langer}
  \& {Tanimura}}{{Gal{\'a}rraga-Espinosa} et~al.}{2021}]{2021A&A...649A.117G}
{Gal{\'a}rraga-Espinosa} D.,  {Aghanim} N.,  {Langer} M.,   {Tanimura} H.,
  2021, \mn@doi [\aap] {10.1051/0004-6361/202039781}, \href
  {https://ui.adsabs.harvard.edu/abs/2021A&A...649A.117G} {649, A117}

\bibitem[\protect\citeauthoryear{{Gilfanov}, {Sunyaev}  \&
  {Churazov}}{{Gilfanov} et~al.}{1987}]{1987SvAL...13....3G}
{Gilfanov} M.~R.,  {Sunyaev} R.~A.,   {Churazov} E.~M.,  1987, Soviet Astronomy
  Letters, \href {https://ui.adsabs.harvard.edu/abs/1987SvAL...13....3G} {13,
  3}

\bibitem[\protect\citeauthoryear{{HI4PI Collaboration} et~al.,}{{HI4PI
  Collaboration} et~al.}{2016}]{2016A&A...594A.116H}
{HI4PI Collaboration} et~al., 2016, \mn@doi [\aap]
  {10.1051/0004-6361/201629178}, \href
  {https://ui.adsabs.harvard.edu/abs/2016A&A...594A.116H} {594, A116}

\bibitem[\protect\citeauthoryear{{Hansen}, {Hassani}, {Lombriser}  \&
  {Kunz}}{{Hansen} et~al.}{2020}]{2020JCAP...01..048H}
{Hansen} S.~H.,  {Hassani} F.,  {Lombriser} L.,   {Kunz} M.,  2020, \mn@doi
  [\jcap] {10.1088/1475-7516/2020/01/048}, \href
  {https://ui.adsabs.harvard.edu/abs/2020JCAP...01..048H} {2020, 048}

\bibitem[\protect\citeauthoryear{{Hickox} \& {Markevitch}}{{Hickox} \&
  {Markevitch}}{2007}]{2007ApJ...661L.117H}
{Hickox} R.~C.,  {Markevitch} M.,  2007, \mn@doi [\apjl] {10.1086/519003},
  \href {https://ui.adsabs.harvard.edu/abs/2007ApJ...661L.117H} {661, L117}

\bibitem[\protect\citeauthoryear{{Hirschmann}, {Dolag}, {Saro}, {Bachmann},
  {Borgani}  \& {Burkert}}{{Hirschmann} et~al.}{2014}]{2014MNRAS.442.2304H}
{Hirschmann} M.,  {Dolag} K.,  {Saro} A.,  {Bachmann} L.,  {Borgani} S.,
  {Burkert} A.,  2014, \mn@doi [\mnras] {10.1093/mnras/stu1023}, \href
  {https://ui.adsabs.harvard.edu/abs/2014MNRAS.442.2304H} {442, 2304}

\bibitem[\protect\citeauthoryear{{Khabibullin} \& {Churazov}}{{Khabibullin} \&
  {Churazov}}{2019}]{2019MNRAS.482.4972K}
{Khabibullin} I.,  {Churazov} E.,  2019, \mn@doi [\mnras]
  {10.1093/mnras/sty2992}, \href
  {https://ui.adsabs.harvard.edu/abs/2019MNRAS.482.4972K} {482, 4972}

\bibitem[\protect\citeauthoryear{{Korkidis}, {Pavlidou}, {Tassis}, {Ntormousi},
  {Tomaras}  \& {Kovlakas}}{{Korkidis} et~al.}{2020}]{2020A&A...639A.122K}
{Korkidis} G.,  {Pavlidou} V.,  {Tassis} K.,  {Ntormousi} E.,  {Tomaras} T.~N.,
    {Kovlakas} K.,  2020, \mn@doi [\aap] {10.1051/0004-6361/201937337}, \href
  {https://ui.adsabs.harvard.edu/abs/2020A&A...639A.122K} {639, A122}

\bibitem[\protect\citeauthoryear{{Kraft} et~al.,}{{Kraft}
  et~al.}{2022}]{2022arXiv221109827K}
{Kraft} R.,  et~al., 2022, \mn@doi [arXiv e-prints]
  {10.48550/arXiv.2211.09827}, \href
  {https://ui.adsabs.harvard.edu/abs/2022arXiv221109827K} {p. arXiv:2211.09827}

\bibitem[\protect\citeauthoryear{{Kronberg}, {Lesch}  \& {Hopp}}{{Kronberg}
  et~al.}{1999}]{1999ApJ...511...56K}
{Kronberg} P.~P.,  {Lesch} H.,   {Hopp} U.,  1999, \mn@doi [\apj]
  {10.1086/306662}, \href
  {https://ui.adsabs.harvard.edu/abs/1999ApJ...511...56K} {511, 56}

\bibitem[\protect\citeauthoryear{{Liedahl}, {Osterheld}  \&
  {Goldstein}}{{Liedahl} et~al.}{1995}]{1995ApJ...438L.115L}
{Liedahl} D.~A.,  {Osterheld} A.~L.,   {Goldstein} W.~H.,  1995, \mn@doi
  [\apjl] {10.1086/187729}, \href
  {https://ui.adsabs.harvard.edu/abs/1995ApJ...438L.115L} {438, L115}

\bibitem[\protect\citeauthoryear{{Lieu}, {Mittaz}, {Bowyer}, {Breen},
  {Lockman}, {Murphy}  \& {Hwang}}{{Lieu} et~al.}{1996}]{1996Sci...274.1335L}
{Lieu} R.,  {Mittaz} J. P.~D.,  {Bowyer} S.,  {Breen} J.~O.,  {Lockman} F.~J.,
  {Murphy} E.~M.,   {Hwang} C.-Y.,  1996, \mn@doi [Science]
  {10.1126/science.274.5291.1335}, \href
  {https://ui.adsabs.harvard.edu/abs/1996Sci...274.1335L} {274, 1335}

\bibitem[\protect\citeauthoryear{{Lieu}, {Ip}, {Axford}  \& {Bonamente}}{{Lieu}
  et~al.}{1999}]{1999ApJ...510L..25L}
{Lieu} R.,  {Ip} W.~H.,  {Axford} W.~I.,   {Bonamente} M.,  1999, \mn@doi
  [\apjl] {10.1086/311790}, \href
  {https://ui.adsabs.harvard.edu/abs/1999ApJ...510L..25L} {510, L25}

\bibitem[\protect\citeauthoryear{{Lodders}}{{Lodders}}{2003}]{2003ApJ...591.1220L}
{Lodders} K.,  2003, \mn@doi [\apj] {10.1086/375492}, \href
  {https://ui.adsabs.harvard.edu/abs/2003ApJ...591.1220L} {591, 1220}

\bibitem[\protect\citeauthoryear{{Luo} et~al.,}{{Luo}
  et~al.}{2017}]{2017ApJS..228....2L}
{Luo} B.,  et~al., 2017, \mn@doi [\apjs] {10.3847/1538-4365/228/1/2}, \href
  {https://ui.adsabs.harvard.edu/abs/2017ApJS..228....2L} {228, 2}

\bibitem[\protect\citeauthoryear{{Markevitch}}{{Markevitch}}{1999}]{1999ApJ...522L..13M}
{Markevitch} M.,  1999, \mn@doi [\apjl] {10.1086/312217}, \href
  {https://ui.adsabs.harvard.edu/abs/1999ApJ...522L..13M} {522, L13}

\bibitem[\protect\citeauthoryear{{Martizzi} et~al.,}{{Martizzi}
  et~al.}{2019}]{2019MNRAS.486.3766M}
{Martizzi} D.,  et~al., 2019, \mn@doi [\mnras] {10.1093/mnras/stz1106}, \href
  {https://ui.adsabs.harvard.edu/abs/2019MNRAS.486.3766M} {486, 3766}

\bibitem[\protect\citeauthoryear{{Meisner} \& {Finkbeiner}}{{Meisner} \&
  {Finkbeiner}}{2015}]{2015ApJ...798...88M}
{Meisner} A.~M.,  {Finkbeiner} D.~P.,  2015, \mn@doi [\apj]
  {10.1088/0004-637X/798/2/88}, \href
  {https://ui.adsabs.harvard.edu/abs/2015ApJ...798...88M} {798, 88}

\bibitem[\protect\citeauthoryear{{Mewe}, {Gronenschild}  \& {van den
  Oord}}{{Mewe} et~al.}{1985}]{1985A&AS...62..197M}
{Mewe} R.,  {Gronenschild} E.~H.~B.~M.,   {van den Oord} G.~H.~J.,  1985,
  \aaps, \href {https://ui.adsabs.harvard.edu/abs/1985A&AS...62..197M} {62,
  197}

\bibitem[\protect\citeauthoryear{{Nandra} et~al.,}{{Nandra}
  et~al.}{2013}]{2013arXiv1306.2307N}
{Nandra} K.,  et~al., 2013, \mn@doi [arXiv e-prints]
  {10.48550/arXiv.1306.2307}, \href
  {https://ui.adsabs.harvard.edu/abs/2013arXiv1306.2307N} {p. arXiv:1306.2307}

\bibitem[\protect\citeauthoryear{{Nicastro}, {Fang}  \& {Mathur}}{{Nicastro}
  et~al.}{2022}]{2022arXiv220315666N}
{Nicastro} F.,  {Fang} T.,   {Mathur} S.,  2022, \mn@doi [arXiv e-prints]
  {10.48550/arXiv.2203.15666}, \href
  {https://ui.adsabs.harvard.edu/abs/2022arXiv220315666N} {p. arXiv:2203.15666}

\bibitem[\protect\citeauthoryear{{O'Neil}, {Barnes}, {Vogelsberger}  \&
  {Diemer}}{{O'Neil} et~al.}{2021}]{2021MNRAS.504.4649O}
{O'Neil} S.,  {Barnes} D.~J.,  {Vogelsberger} M.,   {Diemer} B.,  2021, \mn@doi
  [\mnras] {10.1093/mnras/stab1221}, \href
  {https://ui.adsabs.harvard.edu/abs/2021MNRAS.504.4649O} {504, 4649}

\bibitem[\protect\citeauthoryear{{Paerels}, {Kaastra}, {Ohashi}, {Richter},
  {Bykov}  \& {Nevalainen}}{{Paerels} et~al.}{2008}]{2008SSRv..134..405P}
{Paerels} F.,  {Kaastra} J.,  {Ohashi} T.,  {Richter} P.,  {Bykov} A.,
  {Nevalainen} J.,  2008, \mn@doi [\ssr] {10.1007/s11214-008-9323-6}, \href
  {https://ui.adsabs.harvard.edu/abs/2008SSRv..134..405P} {134, 405}

\bibitem[\protect\citeauthoryear{{Parimbelli}, {Branchini}, {Viel},
  {Villaescusa-Navarro}  \& {ZuHone}}{{Parimbelli}
  et~al.}{2022}]{2022arXiv220900657P}
{Parimbelli} G.,  {Branchini} E.,  {Viel} M.,  {Villaescusa-Navarro} F.,
  {ZuHone} J.,  2022, \mn@doi [arXiv e-prints] {10.48550/arXiv.2209.00657},
  \href {https://ui.adsabs.harvard.edu/abs/2022arXiv220900657P} {p.
  arXiv:2209.00657}

\bibitem[\protect\citeauthoryear{{Pavlinsky} et~al.,}{{Pavlinsky}
  et~al.}{2021}]{2021A&A...650A..42P}
{Pavlinsky} M.,  et~al., 2021, \mn@doi [\aap] {10.1051/0004-6361/202040265},
  \href {https://ui.adsabs.harvard.edu/abs/2021A&A...650A..42P} {650, A42}

\bibitem[\protect\citeauthoryear{{P{\'e}roux} \& {Howk}}{{P{\'e}roux} \&
  {Howk}}{2020}]{2020ARA&A..58..363P}
{P{\'e}roux} C.,  {Howk} J.~C.,  2020, \mn@doi [\araa]
  {10.1146/annurev-astro-021820-120014}, \href
  {https://ui.adsabs.harvard.edu/abs/2020ARA&A..58..363P} {58, 363}

\bibitem[\protect\citeauthoryear{{Planck Collaboration} et~al.,}{{Planck
  Collaboration} et~al.}{2013}]{2013A&A...554A.140P}
{Planck Collaboration} et~al., 2013, \mn@doi [\aap]
  {10.1051/0004-6361/201220247}, \href
  {https://ui.adsabs.harvard.edu/abs/2013A&A...554A.140P} {554, A140}

\bibitem[\protect\citeauthoryear{{Planck Collaboration} et~al.,}{{Planck
  Collaboration} et~al.}{2016}]{2016A&A...594A..13P}
{Planck Collaboration} et~al., 2016, \mn@doi [\aap]
  {10.1051/0004-6361/201525830}, \href
  {https://ui.adsabs.harvard.edu/abs/2016A&A...594A..13P} {594, A13}

\bibitem[\protect\citeauthoryear{{Planelles}, {Mimica}, {Quilis}  \&
  {Cuesta-Mart{\'\i}nez}}{{Planelles} et~al.}{2018}]{2018MNRAS.476.4629P}
{Planelles} S.,  {Mimica} P.,  {Quilis} V.,   {Cuesta-Mart{\'\i}nez} C.,  2018,
  \mn@doi [\mnras] {10.1093/mnras/sty527}, \href
  {https://ui.adsabs.harvard.edu/abs/2018MNRAS.476.4629P} {476, 4629}

\bibitem[\protect\citeauthoryear{{Predehl} et~al.,}{{Predehl}
  et~al.}{2021}]{2021A&A...647A...1P}
{Predehl} P.,  et~al., 2021, \mn@doi [\aap] {10.1051/0004-6361/202039313},
  \href {https://ui.adsabs.harvard.edu/abs/2021A&A...647A...1P} {647, A1}

\bibitem[\protect\citeauthoryear{{Rahmati}, {Schaye}, {Crain}, {Oppenheimer},
  {Schaller}  \& {Theuns}}{{Rahmati} et~al.}{2016}]{2016MNRAS.459..310R}
{Rahmati} A.,  {Schaye} J.,  {Crain} R.~A.,  {Oppenheimer} B.~D.,  {Schaller}
  M.,   {Theuns} T.,  2016, \mn@doi [\mnras] {10.1093/mnras/stw453}, \href
  {https://ui.adsabs.harvard.edu/abs/2016MNRAS.459..310R} {459, 310}

\bibitem[\protect\citeauthoryear{{Richter}, {Paerels}  \& {Kaastra}}{{Richter}
  et~al.}{2008}]{2008SSRv..134...25R}
{Richter} P.,  {Paerels} F.~B.~S.,   {Kaastra} J.~S.,  2008, \mn@doi [\ssr]
  {10.1007/s11214-008-9325-4}, \href
  {https://ui.adsabs.harvard.edu/abs/2008SSRv..134...25R} {134, 25}

\bibitem[\protect\citeauthoryear{{Shi}}{{Shi}}{2016}]{2016MNRAS.459.3711S}
{Shi} X.,  2016, \mn@doi [\mnras] {10.1093/mnras/stw925}, \href
  {https://ui.adsabs.harvard.edu/abs/2016MNRAS.459.3711S} {459, 3711}

\bibitem[\protect\citeauthoryear{{Shull}, {Smith}  \& {Danforth}}{{Shull}
  et~al.}{2012}]{2012ApJ...759...23S}
{Shull} J.~M.,  {Smith} B.~D.,   {Danforth} C.~W.,  2012, \mn@doi [\apj]
  {10.1088/0004-637X/759/1/23}, \href
  {https://ui.adsabs.harvard.edu/abs/2012ApJ...759...23S} {759, 23}

\bibitem[\protect\citeauthoryear{{Smith}, {Hallman}, {Shull}  \&
  {O'Shea}}{{Smith} et~al.}{2011}]{2011ApJ...731....6S}
{Smith} B.~D.,  {Hallman} E.~J.,  {Shull} J.~M.,   {O'Shea} B.~W.,  2011,
  \mn@doi [\apj] {10.1088/0004-637X/731/1/6}, \href
  {https://ui.adsabs.harvard.edu/abs/2011ApJ...731....6S} {731, 6}

\bibitem[\protect\citeauthoryear{{Springel}}{{Springel}}{2005}]{2005MNRAS.364.1105S}
{Springel} V.,  2005, \mn@doi [\mnras] {10.1111/j.1365-2966.2005.09655.x},
  \href {https://ui.adsabs.harvard.edu/abs/2005MNRAS.364.1105S} {364, 1105}

\bibitem[\protect\citeauthoryear{{Springel} \& {Hernquist}}{{Springel} \&
  {Hernquist}}{2003}]{2003MNRAS.339..289S}
{Springel} V.,  {Hernquist} L.,  2003, \mn@doi [\mnras]
  {10.1046/j.1365-8711.2003.06206.x}, \href
  {https://ui.adsabs.harvard.edu/abs/2003MNRAS.339..289S} {339, 289}

\bibitem[\protect\citeauthoryear{{Sunyaev} et~al.,}{{Sunyaev}
  et~al.}{2021}]{2021A&A...656A.132S}
{Sunyaev} R.,  et~al., 2021, \mn@doi [\aap] {10.1051/0004-6361/202141179},
  \href {https://ui.adsabs.harvard.edu/abs/2021A&A...656A.132S} {656, A132}

\bibitem[\protect\citeauthoryear{{Takei} et~al.,}{{Takei}
  et~al.}{2011}]{2011ApJ...734...91T}
{Takei} Y.,  et~al., 2011, \mn@doi [\apj] {10.1088/0004-637X/734/2/91}, \href
  {https://ui.adsabs.harvard.edu/abs/2011ApJ...734...91T} {734, 91}

\bibitem[\protect\citeauthoryear{{Tanimura}, {Aghanim}, {Kolodzig}, {Douspis}
  \& {Malavasi}}{{Tanimura} et~al.}{2020}]{2020A&A...643L...2T}
{Tanimura} H.,  {Aghanim} N.,  {Kolodzig} A.,  {Douspis} M.,   {Malavasi} N.,
  2020, \mn@doi [\aap] {10.1051/0004-6361/202038521}, \href
  {https://ui.adsabs.harvard.edu/abs/2020A&A...643L...2T} {643, L2}

\bibitem[\protect\citeauthoryear{{Tanimura}, {Aghanim}, {Douspis}  \&
  {Malavasi}}{{Tanimura} et~al.}{2022}]{2022A&A...667A.161T}
{Tanimura} H.,  {Aghanim} N.,  {Douspis} M.,   {Malavasi} N.,  2022, \mn@doi
  [\aap] {10.1051/0004-6361/202244158}, \href
  {https://ui.adsabs.harvard.edu/abs/2022A&A...667A.161T} {667, A161}

\bibitem[\protect\citeauthoryear{{Tanoglidis}, {Pavlidou}  \&
  {Tomaras}}{{Tanoglidis} et~al.}{2016}]{2016arXiv160103740T}
{Tanoglidis} D.,  {Pavlidou} V.,   {Tomaras} T.,  2016, \mn@doi [arXiv
  e-prints] {10.48550/arXiv.1601.03740}, \href
  {https://ui.adsabs.harvard.edu/abs/2016arXiv160103740T} {p. arXiv:1601.03740}

\bibitem[\protect\citeauthoryear{{Tornatore}, {Borgani}, {Dolag}  \&
  {Matteucci}}{{Tornatore} et~al.}{2007}]{2007MNRAS.382.1050T}
{Tornatore} L.,  {Borgani} S.,  {Dolag} K.,   {Matteucci} F.,  2007, \mn@doi
  [\mnras] {10.1111/j.1365-2966.2007.12070.x}, \href
  {https://ui.adsabs.harvard.edu/abs/2007MNRAS.382.1050T} {382, 1050}

\bibitem[\protect\citeauthoryear{{Truemper}}{{Truemper}}{1982}]{1982AdSpR...2d.241T}
{Truemper} J.,  1982, \mn@doi [Advances in Space Research]
  {10.1016/0273-1177(82)90070-9}, \href
  {https://ui.adsabs.harvard.edu/abs/1982AdSpR...2d.241T} {2, 241}

\bibitem[\protect\citeauthoryear{{Tuominen} et~al.,}{{Tuominen}
  et~al.}{2021}]{2021A&A...646A.156T}
{Tuominen} T.,  et~al., 2021, \mn@doi [\aap] {10.1051/0004-6361/202039221},
  \href {https://ui.adsabs.harvard.edu/abs/2021A&A...646A.156T} {646, A156}

\bibitem[\protect\citeauthoryear{{Vazza}, {Ettori}, {Roncarelli},
  {Angelinelli}, {Br{\"u}ggen}  \& {Gheller}}{{Vazza}
  et~al.}{2019}]{2019A&A...627A...5V}
{Vazza} F.,  {Ettori} S.,  {Roncarelli} M.,  {Angelinelli} M.,  {Br{\"u}ggen}
  M.,   {Gheller} C.,  2019, \mn@doi [\aap] {10.1051/0004-6361/201935439},
  \href {https://ui.adsabs.harvard.edu/abs/2019A&A...627A...5V} {627, A5}

\bibitem[\protect\citeauthoryear{{Wijers}, {Schaye}, {Oppenheimer}, {Crain}  \&
  {Nicastro}}{{Wijers} et~al.}{2019}]{2019MNRAS.488.2947W}
{Wijers} N.~A.,  {Schaye} J.,  {Oppenheimer} B.~D.,  {Crain} R.~A.,
  {Nicastro} F.,  2019, \mn@doi [\mnras] {10.1093/mnras/stz1762}, \href
  {https://ui.adsabs.harvard.edu/abs/2019MNRAS.488.2947W} {488, 2947}

\bibitem[\protect\citeauthoryear{{Willingale}, {Starling}, {Beardmore},
  {Tanvir}  \& {O'Brien}}{{Willingale} et~al.}{2013}]{2013MNRAS.431..394W}
{Willingale} R.,  {Starling} R.~L.~C.,  {Beardmore} A.~P.,  {Tanvir} N.~R.,
  {O'Brien} P.~T.,  2013, \mn@doi [\mnras] {10.1093/mnras/stt175}, \href
  {https://ui.adsabs.harvard.edu/abs/2013MNRAS.431..394W} {431, 394}

\bibitem[\protect\citeauthoryear{{Yoshikawa} \& {Sasaki}}{{Yoshikawa} \&
  {Sasaki}}{2006}]{2006PASJ...58..641Y}
{Yoshikawa} K.,  {Sasaki} S.,  2006, \mn@doi [\pasj] {10.1093/pasj/58.4.641},
  \href {https://ui.adsabs.harvard.edu/abs/2006PASJ...58..641Y} {58, 641}

\bibitem[\protect\citeauthoryear{{Yoshikawa}, {Yamasaki}, {Suto}, {Ohashi},
  {Mitsuda}, {Tawara}  \& {Furuzawa}}{{Yoshikawa}
  et~al.}{2003}]{2003PASJ...55..879Y}
{Yoshikawa} K.,  {Yamasaki} N.~Y.,  {Suto} Y.,  {Ohashi} T.,  {Mitsuda} K.,
  {Tawara} Y.,   {Furuzawa} A.,  2003, \mn@doi [\pasj] {10.1093/pasj/55.5.879},
  \href {https://ui.adsabs.harvard.edu/abs/2003PASJ...55..879Y} {55, 879}

\bibitem[\protect\citeauthoryear{{Zhu}, {Zhang}  \& {Feng}}{{Zhu}
  et~al.}{2021}]{2021ApJ...920....2Z}
{Zhu} W.,  {Zhang} F.,   {Feng} L.-L.,  2021, \mn@doi [\apj]
  {10.3847/1538-4357/ac15f1}, \href
  {https://ui.adsabs.harvard.edu/abs/2021ApJ...920....2Z} {920, 2}

\bibitem[\protect\citeauthoryear{{Zhuravleva} et~al.,}{{Zhuravleva}
  et~al.}{2013}]{2013MNRAS.435.3111Z}
{Zhuravleva} I.,  et~al., 2013, \mn@doi [\mnras] {10.1093/mnras/stt1506}, \href
  {https://ui.adsabs.harvard.edu/abs/2013MNRAS.435.3111Z} {435, 3111}

\bibitem[\protect\citeauthoryear{{{\v{S}}tofanov{\'a}}, {Simionescu}, {Wijers},
  {Schaye}  \& {Kaastra}}{{{\v{S}}tofanov{\'a}}
  et~al.}{2022}]{2022MNRAS.515.3162S}
{{\v{S}}tofanov{\'a}} L.,  {Simionescu} A.,  {Wijers} N.~A.,  {Schaye} J.,
  {Kaastra} J.~S.,  2022, \mn@doi [\mnras] {10.1093/mnras/stac1854}, \href
  {https://ui.adsabs.harvard.edu/abs/2022MNRAS.515.3162S} {515, 3162}

\makeatother
\end{thebibliography}




\appendix

\section{WHIM spectral model}
\label{app:smodel}
\begin{figure}
\centering
\includegraphics[angle=0,trim=4cm 0cm 3cm 0cm,width=0.99\columnwidth]{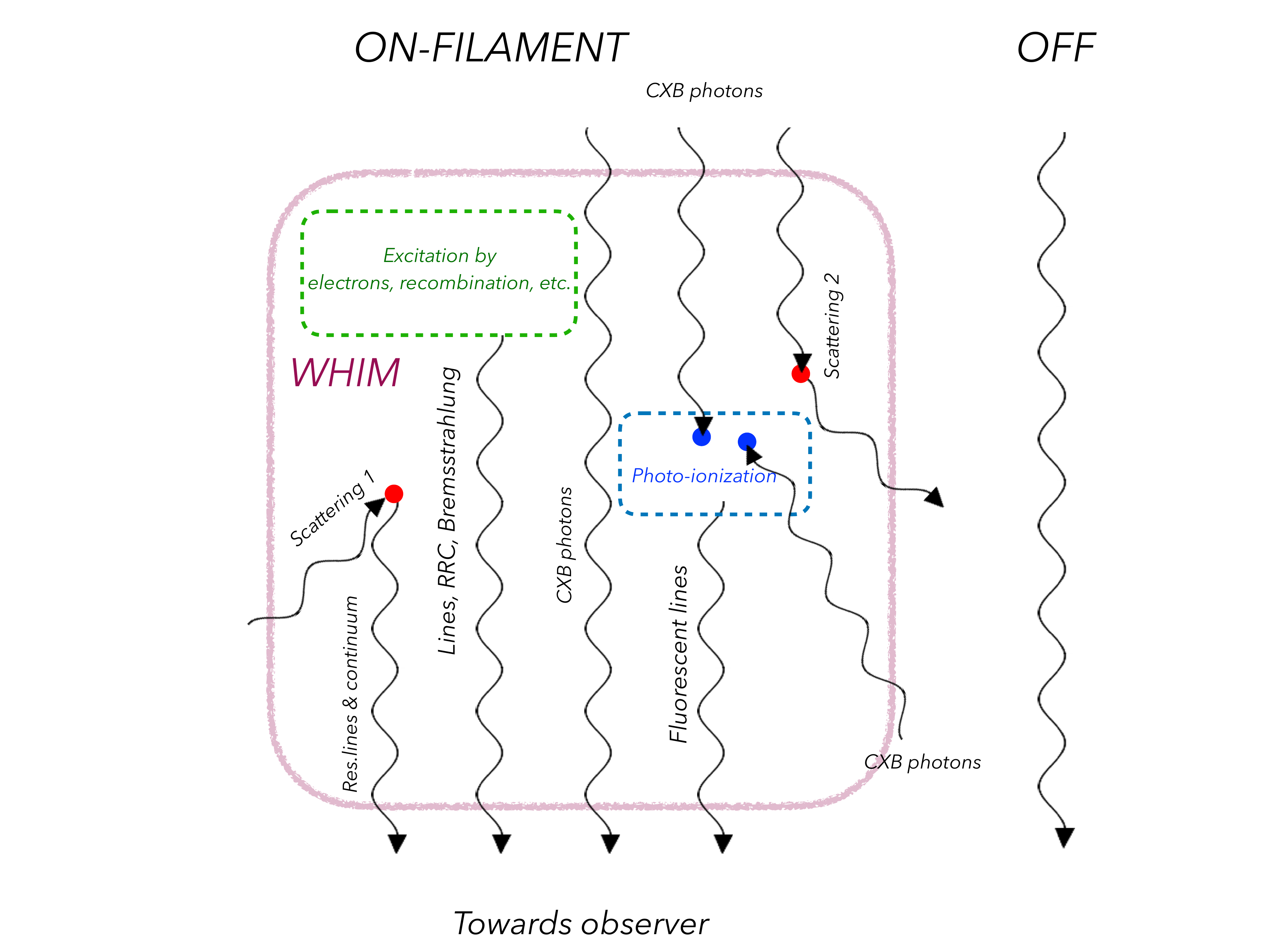}
\caption{Sketch showing (schematically) the most important processes that give rise to WHIM signatures in the X-ray spectra in the direction of a filament ("ON-FILAMENT"). 
Those include local production of X-ray photons (green box), resonant and Compton scattering (small red circles), and photo-electric absorption (small blue circles).  The region occupied by the WHIM patch is shown with a pink box. The red circles show the scattering of the CXB photons to the line of sight ("Scattering 1") and from the line of sight ("Scattering 2"), respectively. 
Outside the filament ("OFF" spectrum), the observer sees the CXB directly. This simplified picture assumes that all CXB sources are further away than the filament and ignores the foreground completely. Several other processes, e.g. secondary ionization by fast electrons that are produced by photo-ionization, are neglected. 
} 
\label{fig:sketch}
\end{figure}

\begin{figure*}
\centering
\includegraphics[angle=0,trim=1cm 5cm 0cm 2cm,width=0.66\columnwidth]{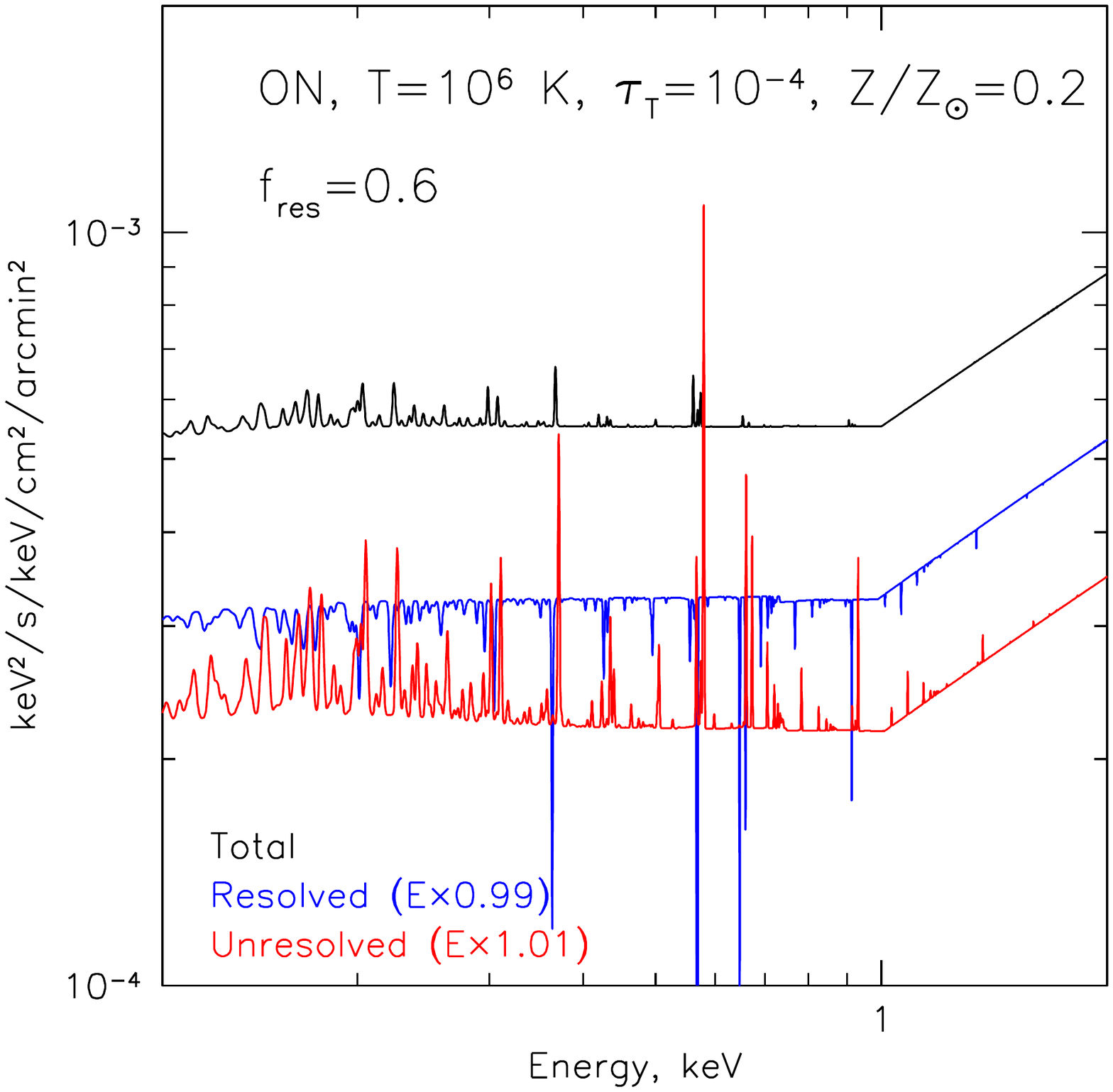}
\includegraphics[angle=0,trim=1cm 5cm 0cm 2cm,width=0.66\columnwidth]{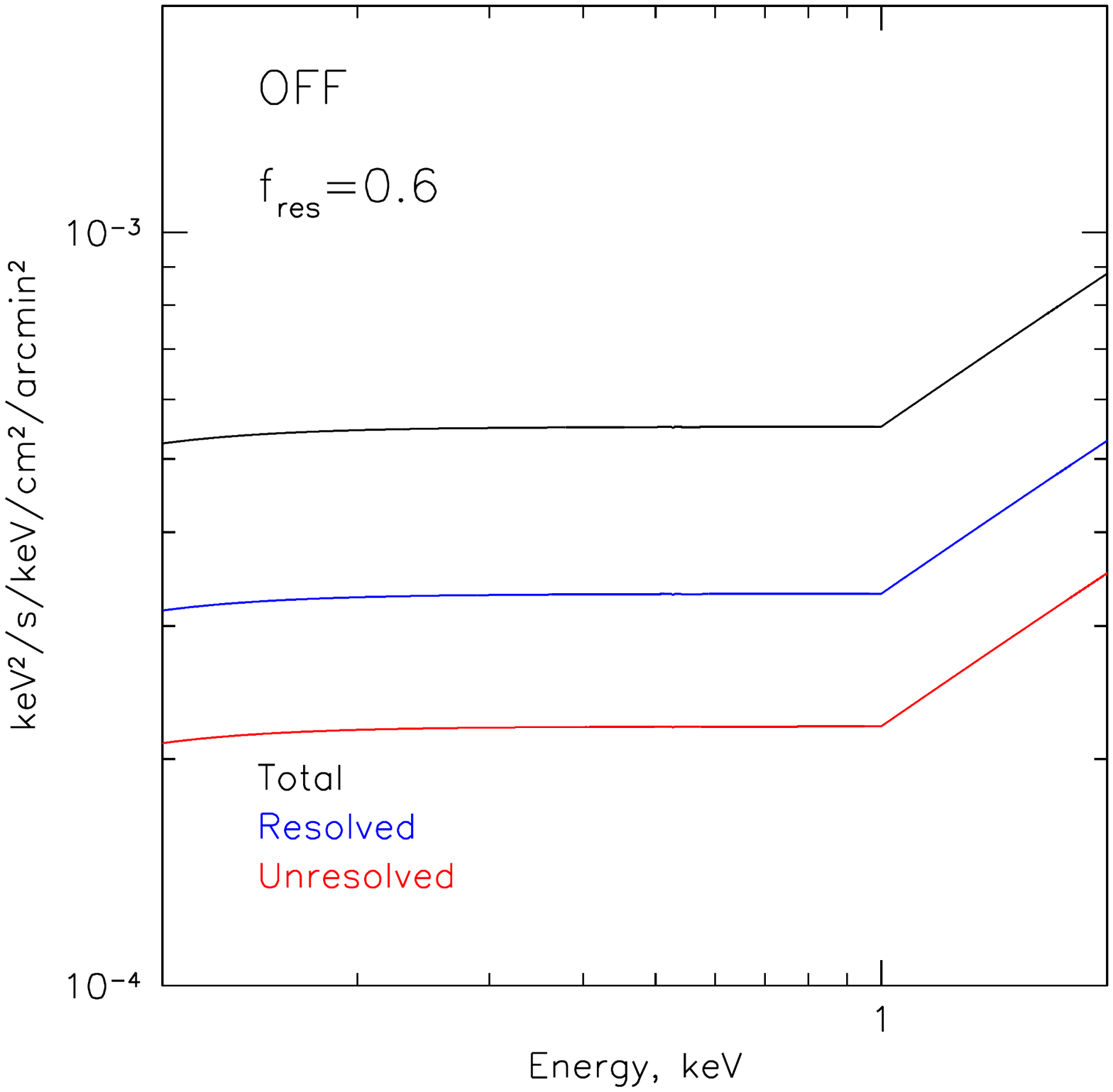}
\includegraphics[angle=0,trim=1cm 5cm 0cm 2cm,width=0.66\columnwidth]{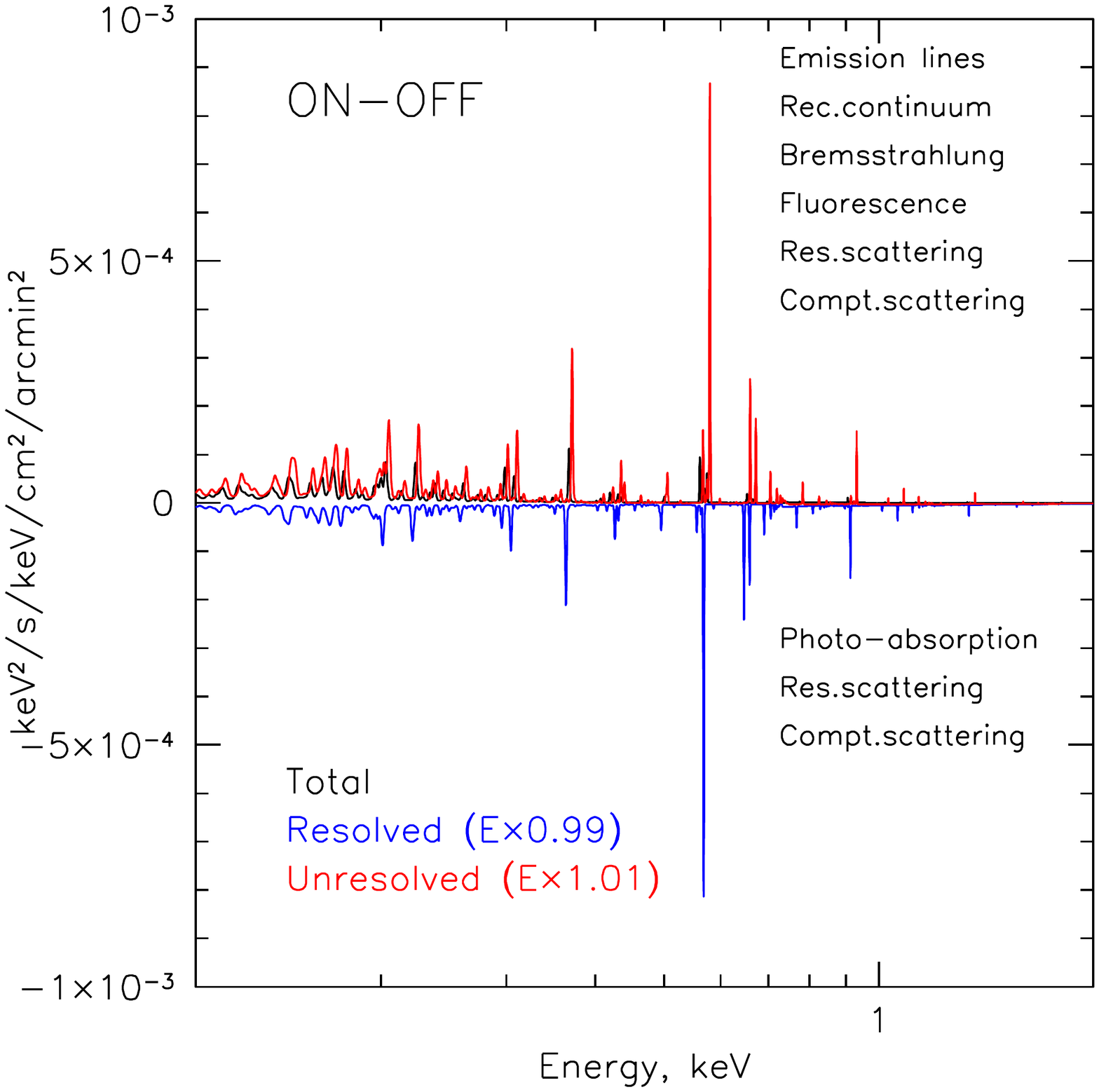}
\caption{Model of spectra that will be observed in the direction of a WHIM filament (ON, left), towards "an empty" region (OFF, middle), and the difference between ON and OFF (right). In each panel three spectra are shown: the total spectrum (black),  a collective spectrum of resolved X-ray sources (blue, a resolved fraction $f_{res}=0.6$ is assumed), and the remaining diffuse/unresolved flux (red). The energies of the resolved and unresolved components have been shifted by $\pm$1\% for the sake of clarity. The value $(1+\delta)=20$ is used in the model. The effects of the resonant lines saturation that might be important for oxygen lines for $\tau_T=10^{-4}$ are neglected. 
} 
\label{fig:onoff}
\end{figure*}

%
Fig.~\ref{fig:sketch} illustrates the WHIM spectrum formation adopted in this study \citep[see also Fig.1 in][for another version of the sketch]{2019MNRAS.482.4972K}.   
The "ON" and "OFF" regions correspond to the directions toward the WHIM filament and away from it, respectively. Corresponding model spectra are shown in Fig.~\ref{fig:onoff}. For display purposes, the spectra were convolved with a Gaussian, FWHM=2~eV. As in Fig.~\ref{fig:sketch}, the foreground emission is ignored for clarity. Accordingly, the "OFF" spectrum (middle panel) is a pure CXB (broken power law approximation), and the normalizations of the total, resolved, and unresolved spectra are related as $1:f_{res}:(1-f_{res})$, reflecting another simplifying assumption that shape of the resolved and unresolved CXB spectra are the same.  

For the "ON" direction (left panel in Fig.~\ref{fig:onoff}), the {\it total} spectrum contains shows the photons produced by electron excitation in the WHIM gas, some signatures of photo-electric absorption, fluorescent photons, and recombination continuum and lines. Conservative scattering processes (Compton and resonant) do not show up in the total spectrum \citep{2001MNRAS.323...93C,2019MNRAS.482.4972K}. The {\it resolved} spectrum on the contrary shows the absorption lines (resonant scattering from the line of sight) and photoelectric absorption. The attenuation by Compton scattering reduces the intensity of the resolved spectrum, but this effect is too small to see it in Fig.~\ref{fig:onoff}. For the sake of clarity, we have also assumed that the solid angle associated with resolved sources is very small and, therefore, the contribution of the diffuse components to the resolved spectrum can be neglected. In the {\it unresolved} spectrum the thermal emission, resonantly-scattered and fluorescent lines are present (red lines). 

Finally, the right panel of Fig.~\ref{fig:onoff} shows the difference between the ON and OFF spectra, i.e. pure signatures of the presence of the WHIM filament. For the {\it total} spectrum the "locally produced photons" are the most prominent for $T=10^6\,{\rm K}$ gas. In the {\it resolved} spectrum, the absorption lines are prominent, while in the  {\it unresolved} spectrum, the resonantly scattered lines add to local emission. The key process contributing to the excess emission is indicated on the top-right side, while the processes leading to negative/absorption structures are listed on the bottom-right side.

We now proceed with a slightly more elaborate model that accounts for the contribution of the Milky Way foreground and takes into account the effective area and the resolution of the {\it LEM} mission.

\begin{figure*}
\centering
\includegraphics[angle=0,trim=1cm 5cm 0cm 2cm,width=0.66\columnwidth]{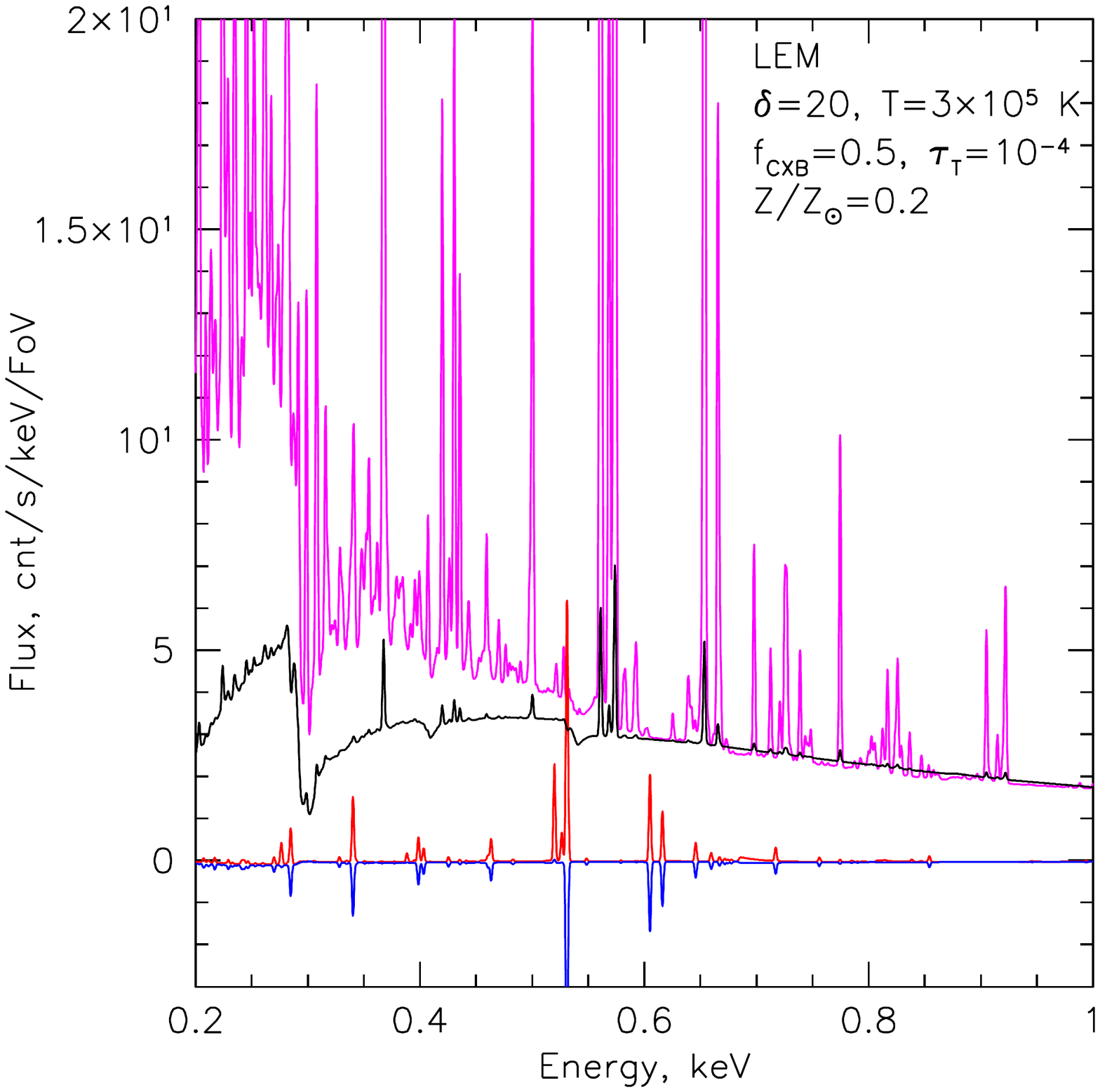}
\includegraphics[angle=0,trim=1cm 5cm 0cm 2cm,width=0.66\columnwidth]{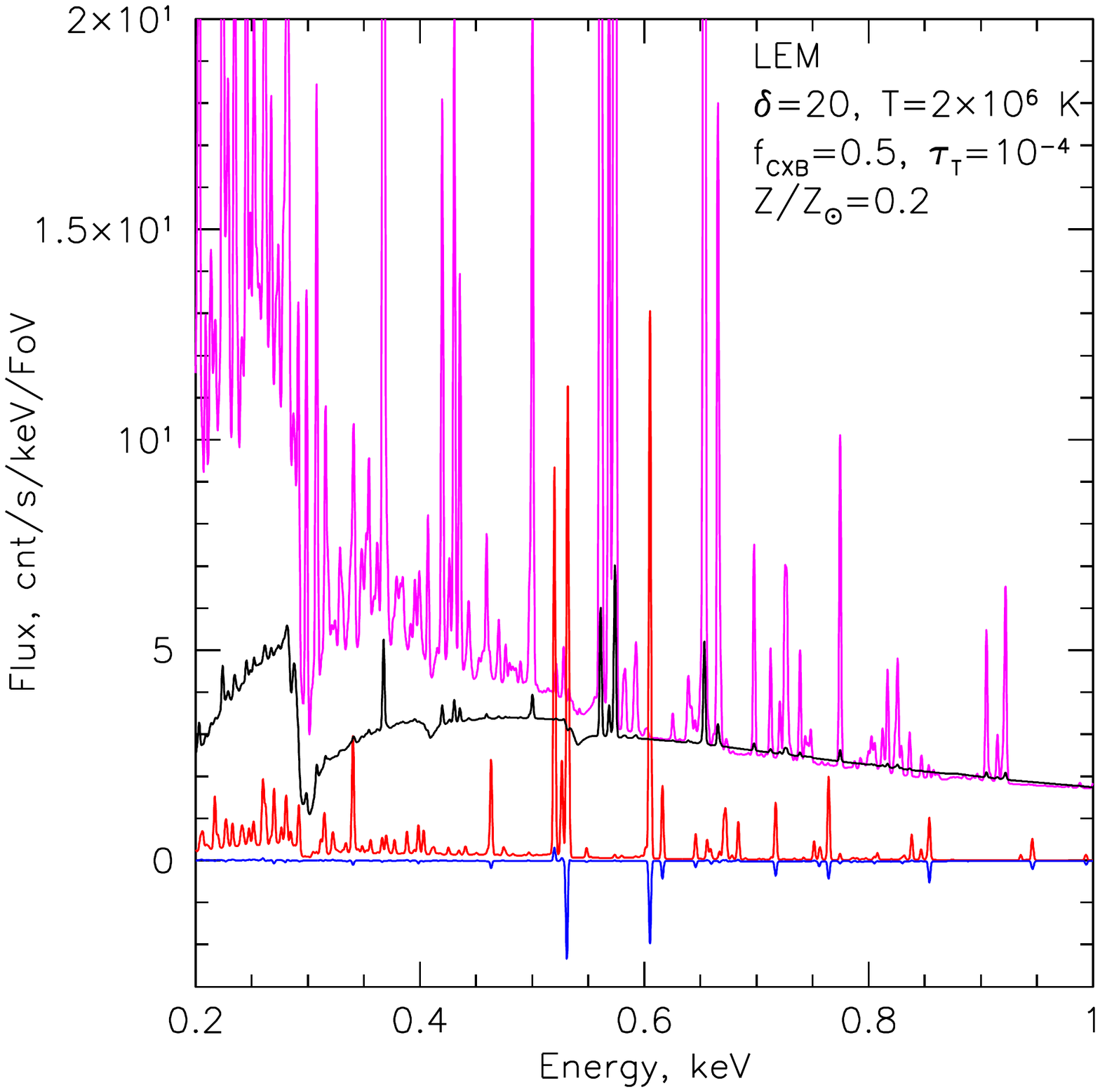}
\includegraphics[angle=0,trim=1cm 5cm 0cm 2cm,width=0.66\columnwidth]{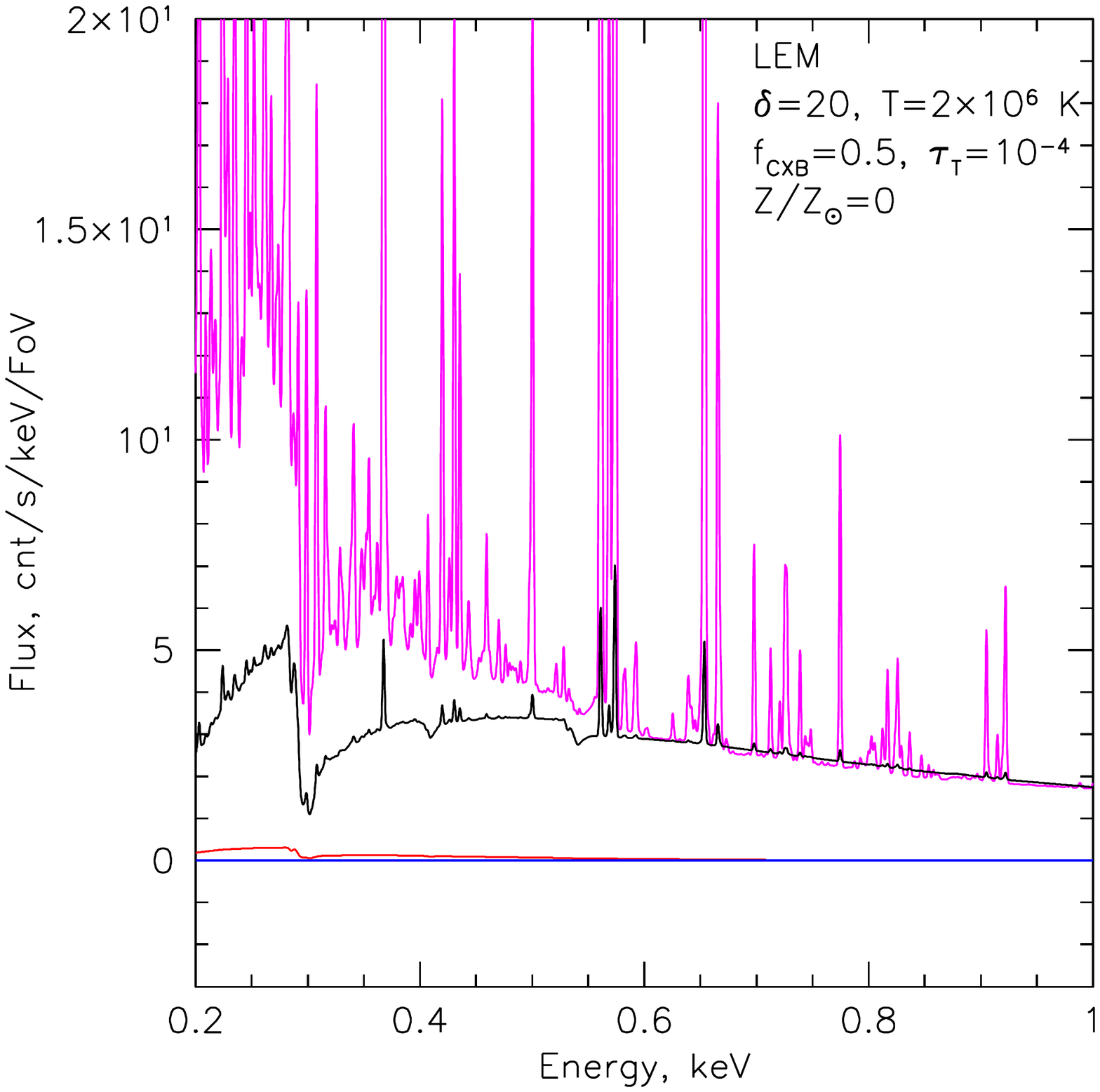}
\caption{Examples of expected WHIM signals for low and high-temperature regimes - left and middle, respectively, for metalicity $Z/Z_\odot=0.2$. For comparison, the right panel shows the spectra for the $Z/Z_\odot=0$ case. The top two magenta and black curves show the spectra, extracted from a "blank sky" region. The magenta curve correspond to the total emission (all backgrounds and foregrounds) from which 50\% of the sources have been excised (using $r=0.5'$ circular excise regions). The collective spectrum of the excised region is dominated by the CXB sources (black line). If WHIM is present at $z=0.08$ (filling the entire FoV), in the "excised" spectrum addition absorption lines predominantly associated with the resonant scattering (and continuum features) will appear - as shown by the bright blue line. The remaining "unresolved" spectrum (shown by the thick red line) will contain instead emission lines and locally produced photons.   For low temperatures (left panel), the absorption and emission spectra are almost identical, except for the sign. For higher temperatures (middle panel) thermal emission contribute significantly and the signal in the "unresolved" spectrum is larger. The right panel illustrates that the main WHIM signal is associated with metals. Once the metals are excluded, the only parts that remain are pure H+He bremsstrahlung and Thomson scattering.  
} 
\label{fig:resunres}
\end{figure*}

\begin{figure*}
\centering
\includegraphics[angle=0,trim=1cm 5cm 0cm 2cm,width=0.95\columnwidth]{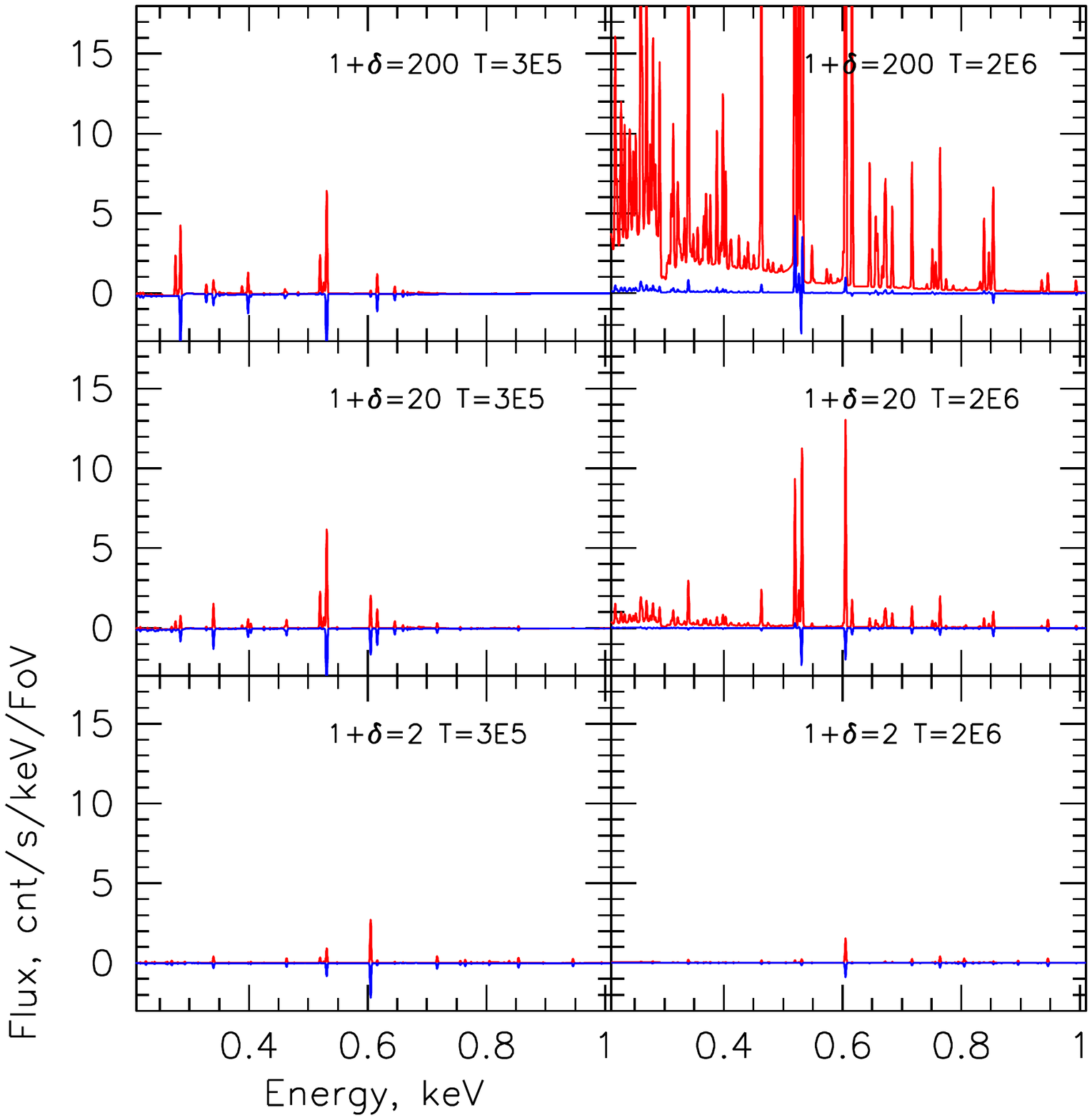}
\includegraphics[angle=0,trim=1cm 5cm 0cm 2cm,width=0.95\columnwidth]{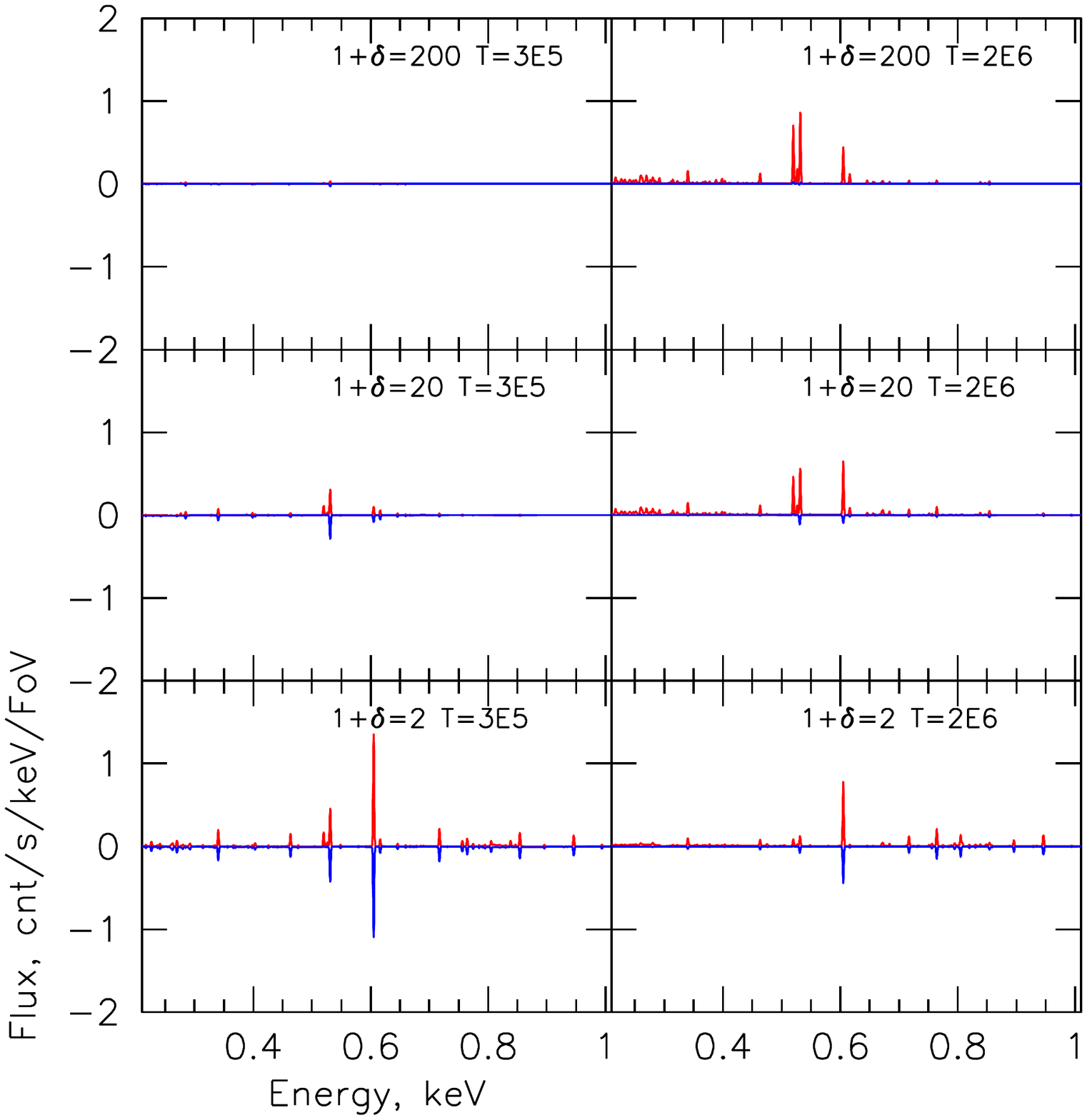}
\caption{WHIM spectral signatures (on-off) for several values of overdensity and temperature for gas at $z=0.08$. Similarly to Fig.~\ref{fig:resunres},  the blue curves show the absorption features in the combined spectrum of distant AGN that make 50\% of the CXB, while the red curve shows the WHIM signal in the remaining part of the image. 
In the left plot, the Thomson optical depth is kept the same in all panels; the two columns correspond to $T=3\times10^5\,{\rm K}$ and $T=2\times10^6\,{\rm K}$, while the rows correspond to densities $(1+\delta)$ of 2, 20 and 200, respectively.  The right plot shows the same set of temperatures and densities, but now the emission measure is kept the same across panels. The absolute normalizations are different for the left and the right plot, but they are the same for the sub-panels in each plot.  
\label{fig:dif2}}
\end{figure*}

\begin{figure}
\centering
\includegraphics[angle=0,trim=1cm 5cm 0cm 2cm,width=0.95\columnwidth]{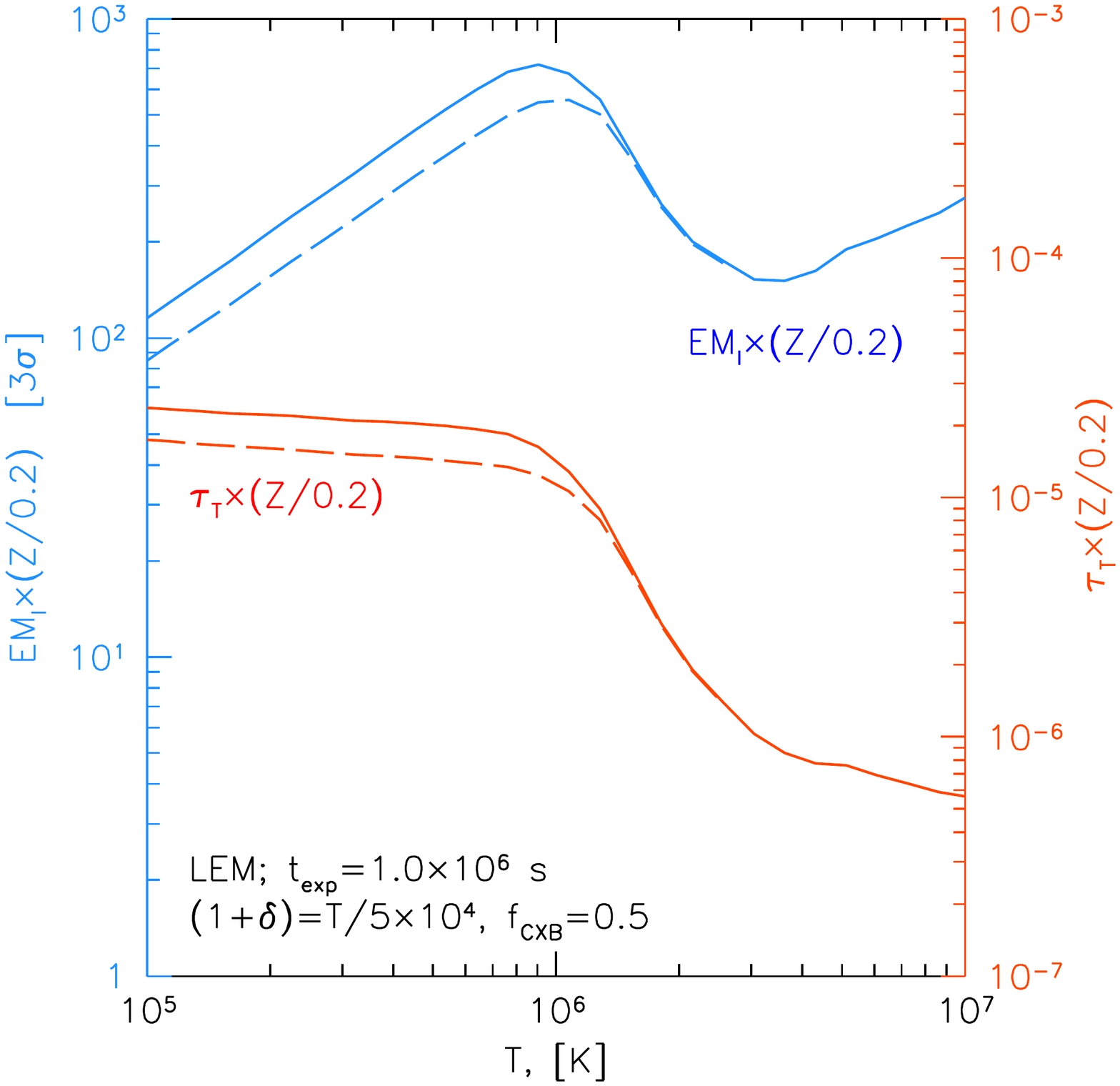}
\caption{LEM sensitivity for the "diagonal" WHIM, i.e. when the gas temperature and density are correlated as $T=5\times10^4\,{\rm K} \times (1+\delta)$. The "double" lines correspond to cases when only the "unresolved" spectrum is used to measure the WHIM signal (solid) or both "unresolved" and "resolved" spectra are considered (dashed). As expected from Fig.~\ref{fig:resunres}, at low temperatures, the signal (and the background) are almost equally split between the two spectra. Accordingly, the gain in sensitivity is $\sim \sqrt{2}$.  
\label{fig:diag}}
\end{figure}

\begin{figure*}
\centering
\includegraphics[angle=0,trim=0cm 0cm 0cm 0cm,width=1.95\columnwidth]{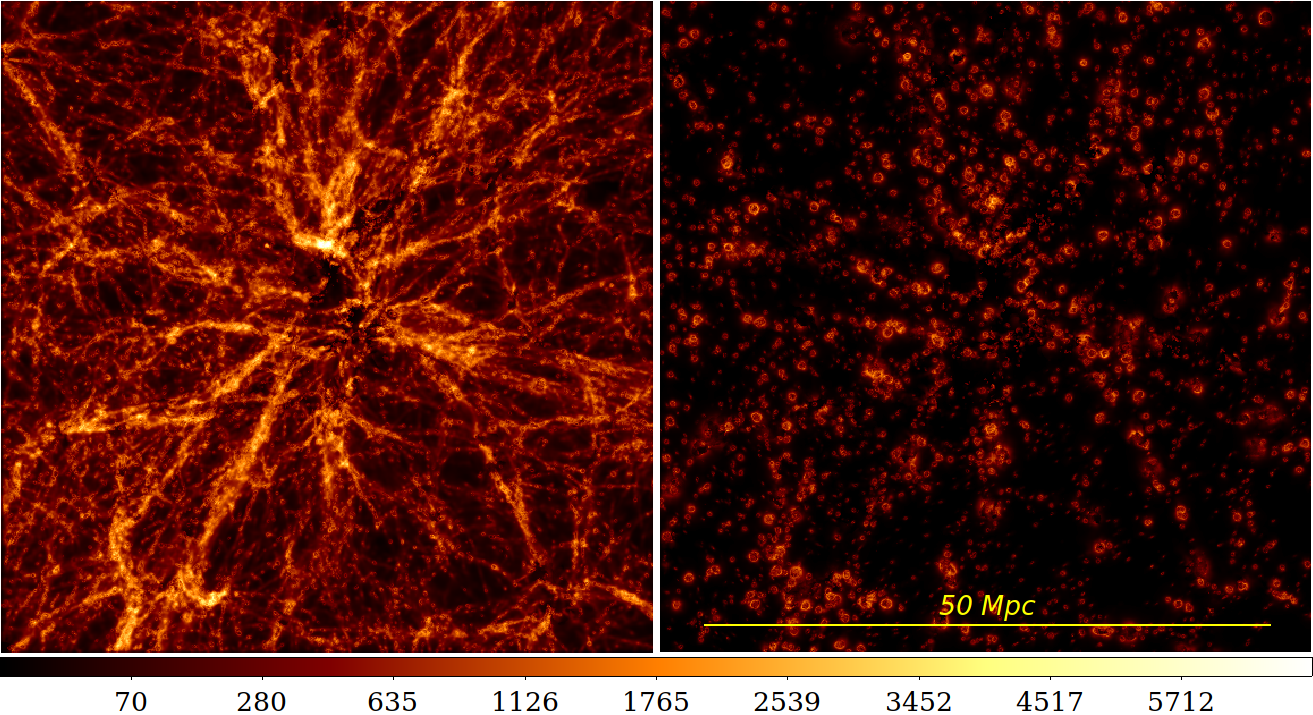}
\caption{ ${\rm EM_{\it l}}$ maps for the cluster box, calculated using particles with zero (left) and non-zero (right) metallicities. In both cases, the same density and temperature cuts (corresponding to eWHIM50) are used. While in the left plot, a filamentary structure is obvious, in the right panel the same structure is traced by round regions, which apparently correspond to the outskirts of individual small halos. An implication is that in the Magneticum simulations metals (more explicitly - particles enriched with metals) stay in the vicinity of individual halos rather than spread quasi-homogeneously through the volume of the filaments.   
\label{fig:mmetals}}
\end{figure*}

%

As discussed in Section \ref{sec:smodel}, the WHIM "on-off" signal can be conveniently factorized into three components
\begin{eqnarray}
S(E)=A\times \left (\frac{Z}{Z_\odot} \right)\times S(E,\delta,T), 
\end{eqnarray}
where $A$ characterize the {\it amount} of hydrogen in a patch of the WHIM, $Z/Z_\odot$ is the metallicity and $S(E,\delta,T)$ describes the shape of the spectrum for a given $\delta$ and $T$.  This prescription ignores the contributions of two continuum components that do not scale with the abundance of heavy elements, namely, the Thomson scattered continuum and a pure H+He thermal bremsstrahlung. In the above equation, $A$ is expressed via $\tau_T\,\times\,{\rm arcmin^2}$ (an equivalent formulation can be done in terms of ${\rm EM_{\it l}} \,\times\,{\rm arcmin^2}$, see  Eq.~\ref{eq:tau2eml}). The ${\rm arcmin^2}$ appears in the normalization factor since we are considering the diffuse source and mostly consider nearby objects so that the strength of the WHIM signal increased with the solid angle subtended by the source. When calculating the sensitivity (\S\ref{sec:smodel}) we assumed that the source fills the entire FoV of a given instrument. $Z_\odot$ stays for the Solar abundance \citep[e.g.][]{1989GeCoA..53..197A} and $S(E,\delta,T)$ is the shape of the WHIM signal. 

In reality, observational data (spatially resolved spectra of a sky region where the WHIM signal is present) can be split into two parts:
 $S_{r}(E)$ and $S_{u}(E)$. The former is the combined spectrum of bright distant compact sources (primarily AGN) that constitute a large part of the CXB that can be resolved in the data. This spectrum is simply a sum of spectra of small circles around these sources. We assume that a fraction $f_{CXB}\sim 0.5$ can be resolved this way. 
 
 Accordingly
\begin{eqnarray}
S_{r}(E)\approx f_{CXB} S_{CXB}(E) \times e^{-\tau_a(E)} + s_{dif}(E) +B(E)\approx \nonumber \\ f_{CXB} S_{CXB}(E) \times (1-\tau_a(E)) + s_{dif}(E)+B(E),
\end{eqnarray}
where $S_{CXB}(E)$ is the CXB spectrum\footnote{Strictly speaking - the spectrum of resolved CXB. We assume, however, that it has the same shape as the total CXB spectrum.} and $s_{dif}(E)$ is an additional (small) contribution of the diffuse emission to the circles around resolved sources. The $s_{dif}(E)$ is small as long the solid angle subtended by the circles is small. Finally, $B(E)$ is the sum of the instrumental background and the Milky Way foreground. 

Here $\tau_a(E)=\tau_T+\tau_{ph}(E)+\tau_{rs}(E)$ is the total optical depth due to the Thomson scattering, photoelectric absorption, and resonant scattering. Here and below we assume that $\tau_a(E)\ll 1$ at any energy and any second-order terms in $\tau$ can be neglected. In general, this spectrum will contain a number of spectral features including absorption edges and absorption lines. Studying such spectra is similar to the normal absorption studies \citep[see][for the recent review]{2022arXiv220315666N}, albeit instead of a single bright source, the spectrum combined from a large number of individual sources is used. 

The second component ($S_{u}(E)$) is the spectrum of the remaining (unresolved) spectrum of the WHIM region without excised regions
\begin{eqnarray}
S_{u}(E)\approx (1-f_{CXB}) S_{CXB}(E) \times e^{-\tau_a(E)} + \nonumber \\ (\tau_T+\tau_{rs}(E)) S_{CXB}(E) + S_{fl}(E) + S_{th}(E) +B(E),
\end{eqnarray}
where $S_{fl}(E)$ is the fluorescent emission and $S_{th}(E)$ is the "thermal" emission of the WHIM, consisting of recombination emission and emission caused by excitation of various transitions by electrons and the thermal bremsstrahlung (see \S\ref{sec:smodel}. The examples of $S_{r}(E)$ and $S_{u}(E)$ are shown in Fig.~\ref{fig:resunres} with light-blue and light-red lines, respectively. There we show the spectra convolved with the LEM spectral response. Since the solid angle for $S_{r}(E)$ is much smaller\footnote{In the 0.5-2~keV band, the number of CXB sources per square degree is $\sim 120 (f_{CXB}/0.5)^{3.2}$ and for $f_{CXB}=0.5$ and $30''$ radius of excised circles, less than 3\% of the solid angle is excised.} than for $S_{u}(E)$, all diffuse components are subdominant in  $S_{r}(E)$, while the CXB makes equal contributions to the spectra (for $f_{CXB}=0.5$).

In reality, when searching for the WHIM signatures, one has to compare the signal in the direction of the WHIM patch and the signal from the "blank" field, i.e. where the WHIM signal is absent (at least at the same redshift). Therefore, a more revealing is the difference between the on-WHIM and off-WHIM spectra, in which many contributions, e.g. the instrumental background and the Milky Way foreground as well as parts of the CXB signal are absent.  We denote these {\it difference} spectra as $S_{r,d}(E)$ and $S_{u,d}(E)$ for resolved and unresolved spectra, respectively. 

The {\it resolved difference} spectrum is
\begin{eqnarray}
S_{r,d}(E)=S_{r}(E)-f_{CXB} S_{CXB}(E)-s_{dif}(E)-B(E)\approx  \nonumber \\ -f_{CXB} S_{CXB}(E) \times \tau_a(E), 
\end{eqnarray}
i.e. it simply shows the missing flux in the spectrum of resolved CXB sources. A small term related to the absorption of the diffuse background behind the WHIM patch is neglected.

Similarly, the {\it unresolved difference} spectrum is 
\begin{eqnarray}
S_{u,d}(E)=S_{u}(E)-(1-f_{CXB}) S_{CXB}(E)-B(E)\approx \nonumber \\ -(1-f_{CXB})S_{CXB}(E)\tau_a(E) + \nonumber \\ \tau_{rs}(E) S_{CXB}(E) + S_{fl}(E) + S_{th}(E)= \nonumber \\
-(1-f_{CXB})S_{CXB}(E)\tau_{ph}(E)+\nonumber \\ f_{CXB}(\tau_T+\tau_{rs}(E)) S_{CXB}(E) + S_{fl}(E) + S_{th}(E).
\end{eqnarray}

Both $S_{r,d}(E)$ and $S_{r,d}(E)$ are shown in Fig.~\ref{fig:resunres} as the bright blue and red lines. To illustrate the dependence of the difference spectra on the density and temperature, Fig.~\ref{fig:dif2} shows a set of difference spectra for $T=3\times10^5$ and $2\times10^6$~K and densities $(1+\delta)=2, 20, 200$ for a fixed $\tau_T$ (left plot) and fixed ${\rm EM_{\it l}}$ (right plot). It is clear that at low temperatures and/or small overdensities, both spectra are dominated by the effects of the resonant scattering, i.e. the resolved spectrum shows absorption lines, while the unresolved spectrum shows the same lines in emission (provided that a significant part of the CXB is resolved). For high temperatures and substantial overdensities, the signal is dominated instead by the "locally" produced photons.  This means that the joint analysis of the resolved and unresolved spectra can boost the sensitivity by a factor $\sim \sqrt{2}\,$ for low temperatures and/or overdensities, while in the opposite limit, the unresolved spectra bear most of the information on the presence of the WHIM patch. This is further illustrated by Fig.~\ref{fig:diag}, where the sensitivity is calculated for the WHIM subsample when  the gas temperature and density are correlated as $T=4\times10^5\,{\rm K} \times (1+\delta)$ (see black dashed line in Fig.~\ref{fig:n_t_boxes}).
For this plot we assumed that the CXB resolved fraction is $\sim 0.5$, the CXB dominates the continuum, and the WHIM lines are shifted from the brightest lines of the Milky Way foreground. Should the angular resolution of an instrument be high enough so that a large fraction of the CXB can be excised from the data without much impact on the remaining diffuse emission, the sensitivity of the unresolved spectra to the WHIM emission could be higher. This is not the case for eROSITA or \textit{LEM} discussed here.

\subsection{Limitations of the model}
The spectral model used here has a number of limitations/assumptions that stem from attempts to keep the model simple and target the regime when the expected WHIM signal is very small. Here we list the most important of these limitations that one has to keep in mind.

\begin{itemize}
\item The model applies to $z\sim 0$.
\item Ionization fractions correspond to the equilibrium state (including photoionization).
\item The electron temperature is not affected by the photoionization of metals by CXB photons.
\item The abundance ratios among metals are fixed to the Solar photospheric values. 
\item Resonant lines are not saturated and treated by the model as Dirac delta functions of energy. In general, the optical depth $\tau(E)$ is assumed to be small at any energy, i.e. the Taylor expansion $e^{-\tau(E)}\approx 1-\tau(E)$ is valid for any $E$. 
\item Any second-order terms in $\tau$ are neglected.
\item A reduced list of transitions is used when the resonantly scattered emission is evaluated.
\end{itemize}

\section{Distribution of metals in the intergalactic medium (in Magneticum simulations)}
\label{app:metals}

Although the total amount of metals accumulated over the lifetime of the Universe can be predicted rather robustly, since it is closely connected with the overall star formation history, their split between different phases and corresponding spatial distribution is rather uncertain \citep[e.g.][]{2020ARA&A..58..363P}. In particular, this is determined by the efficiency of the metal transport from the sites of star formation (i.e. dark matter halos of various sizes) and their spread across intergalactic volumes (including mixing and diffusion). 

Similarly to the situation with the magnetic field in the IGM \citep[e.g.][]{1999ApJ...511...56K}, the former part is especially sensitive to the kinetic energy associated with the stellar and AGN-driven winds in the first galaxies at high redshift, $z>2$ \citep[e.g.][]{2017MNRAS.468..531B}. Various simulation sets differ in this regard substantially, resulting in different predictions for the efficiency and spatial distribution of this so-called early enrichment \citep[e.g.][]{2021A&A...646A.156T}.

The \texttt{Magneticum} model of the feedback is rather gentle, lacking extreme energetic outbursts and kinetic energy injections from supermassive black holes \citep[][]{2016MNRAS.463.1797D}. As a result, the spread of metals is limited to a few viral radii of the halos. Consequently, the diffuse intergalactic medium, and WHIM in particular, stays largely not enriched, although the average metallicity of all gas components stays constant at 0.2 \citep[][]{2022A&A...663L...6A}. We illustrate this by showing the projected emission measure distribution of the SPH particles in the "cluster" box with zero and non-zero metallicity in the left and right panels of Fig.\ref{fig:mmetals}, respectively.

Taken at the face value, this prediction would preclude the detection of this gas in X-rays (either in emission or in absorption) with any reasonable parameters of the future X-ray facilities. Namely, even a high effective area of the detector would not solve the problem given the inability to single out the signal corresponding to a particular redshift range of the "focusing" galaxy cluster or galaxy overdensity. The same is true for the thermal and kinematic SZ effects and other redshift-insensitive probes. Detection of the dispersion measure jump for a line-of-sight distributed sources (e.g. FRBs) might be used for that, requiring though the rather high density of the sources. Perhaps, more promising could be the detection of Thomson reflection of the light from extremely bright and moderately collimated lighthouses, namely that their illumination cones can be resolved spatially and cross an intergalactic filament almost perpendicularly \citep[e.g.][]{2001MNRAS.323...93C}.


\bsp	
\label{lastpage}
\end{document}